\documentclass[aps, pre, onecolumn, floatfix, groupedaddress]{revtex4-2}

\usepackage{standalone}
\usepackage[bottom]{footmisc}
\usepackage{lipsum}
\usepackage{float}
\usepackage{graphicx}
\usepackage{algorithm}
\usepackage{amssymb, amsmath, amsfonts}
\usepackage[caption=false]{subfig} % <- Preamble
\usepackage{pgfplots} % <- Preamble
\usepackage{xcolor}
\usepackage{multibib}
\usepackage{mathtools}
\usepackage{bm}
\usepackage{amsopn}
\usepackage{xr}
\usepackage{hyperref}
\usepackage[normalem]{ulem} % for crossing out text
\DeclareMathAlphabet{\mathdutchcal}{U}{dutchcal}{m}{n} \SetMathAlphabet{\mathdutchcal}{bold}{U}{dutchcal}{b}{n}
\DeclareMathAlphabet{\mathdutchbcal}{U}{dutchcal}{b}{n}

\makeatletter
\renewcommand*\env@matrix[1][\arraystretch]{%
\edef\arraystretch{#1}%
\hskip -\arraycolsep \let\@ifnextchar\new@ifnextchar \array{*\c@MaxMatrixCols c}}
\makeatother

\makeatletter
\def\bbordermatrix#1{\begingroup \m@th \@tempdima 4.75\p@ \setbox\z@\vbox{%
    \def\cr{\crcr\noalign{\kern2\p@\global\let\cr\endline}}%
    \ialign{$##$\hfil\kern2\p@\kern\@tempdima&\thinspace\hfil$##$\hfil &&\quad\hfil$##$\hfil\crcr
      \omit\strut\hfil\crcr\noalign{\kern-\baselineskip}%
      #1\crcr\omit\strut\cr}}%
  \setbox\tw@\vbox{\unvcopy\z@\global\setbox\@ne\lastbox}%
  \setbox\tw@\hbox{\unhbox\@ne\unskip\global\setbox\@ne\lastbox}%
  \setbox\tw@\hbox{$\kern\wd\@ne\kern-\@tempdima\left[\kern-\wd\@ne \global\setbox\@ne\vbox{\box\@ne\kern2\p@}%
    \vcenter{\kern-\ht\@ne\unvbox\z@\kern-\baselineskip}\,\right]$}%
  \null\;\vbox{\kern\ht\@ne\box\tw@}\endgroup}
\makeatother

\begin{document}
\raggedbottom
% Use the \preprint command to place your local institutional report number in the upper righthand corner of the title
% page in preprint mode. Multiple \preprint commands are allowed. Use the 'preprintnumbers' class option to override
% journal defaults to display numbers if necessary \preprint{}

%Title of paper
\title{Persistent and anti-Persistent Motion in Bounded and Unbounded Space: Resolution of the First-Passage Problem}

% repeat the \author .. \affiliation  etc. as needed \email, \thanks, \homepage, \altaffiliation all apply to the
% current author. Explanatory text should go in the []'s, actual e-mail address or url should go in the {}'s for \email
% and \homepage. Please use the appropriate macro foreach each type of information

% \affiliation command applies to all authors since the last \affiliation command. The \affiliation command should
% follow the other information \affiliation can be followed by \email, \homepage, \thanks as well.
\author{Daniel Marris}
\email[]{daniel.marris@bristol.ac.uk}
\affiliation{School of Engineering Mathematics and Technology, University of Bristol, BS8 1TW}
\author{Luca Giuggioli}
\email[]{luca.giuggioli@bristol.ac.uk}
\affiliation{School of Engineering Mathematics and Technology, University of Bristol, BS8 1TW}

\date{\today}

\begin{abstract}
 The presence of temporal correlations in random movement trajectories is a widespread phenomenon across biological,
 chemical and physical systems. The ubiquity of persistent and anti-persistent motion in
 many natural and synthetic systems has led to a large literature on the modelling of temporally correlated movement
 paths. Despite the substantial body of work, little progress has been made to determine the dynamical properties of
 various transport related quantities, including the first-passage or first-hitting probability to one or multiple
 absorbing targets when space is bounded. To bridge this knowledge gap we generalise the renewal theory of
 first-passage and splitting probabilities to correlated discrete variables. We do so in arbitrary dimensions on a
 lattice for the so-called correlated or persistent random walk, the one step non-Markovian extension
 of the simple lattice random walk in bounded and unbounded space. We focus on bounded domains and consider both
 persistent and anti-persistent motion in hypercubic lattices as well as the hexagonal lattice. The discrete
 formalism allows us to extend the notion of the first-passage to that of the directional first-passage, whereby the
 walker must reach the target from a prescribed direction for a hitting event to occur.  As an application to
 spatio-temporal observations of correlated moving cells that may be either repelled or attracted to hard surfaces,
 we compare the first-passage statistics to a target within a reflecting domain depending on whether an interaction
 with the reflective interface invokes a reversal of the movement direction or not. With strong persistence we
 observe multi-modality in the first-passage distribution in the former case, which instead is greatly suppressed in
 the latter.
\end{abstract}

\maketitle
\section{Introduction}\label{sec: Intro} 
Advances in tracking sensors for animals \cite{nathan2022big}, cells
\cite{meijering2012methods} and particles \cite{chenouard2014objective} have made conspicuous the importance of
quantifying the spatio-temporal dynamics of movement. To this end, two fundamental frameworks, the lattice random walk (LRW)
\cite{rws_on_latticesII,weiss1994aspects,LucaPRX}, and its continuous counterpart, the Brownian walk
\cite{morters2010brownian}, have been widely employed to model and analyse Markovian trajectories, that is when the
dynamics arise from history-independent random processes.

Much of the versatility of the two frameworks, comes from the analytic description of diffusive processes, the
quintessential example of a Markovian system. The exact mathematical representation of diffusion has in fact contributed
to its widespread application across disciplines, from cell biology \cite{bressloff2014stochastic} and ecology
\cite{kenkre2021theory} to social sciences \cite{bejan2007constructal} and economics \cite{embrechts2013modelling}.
However, the increased resolution of modern tracking technologies has made apparent the approximate nature of the
diffusive paradigm and is highlighting the need to model accurately the non-Markovian statistics within movement paths.

One such example of non-Markovian features in spatio-temporal trajectories is persistence or anti-persistence in the
movement steps, that is by having, respectively, a higher tendency to continue in the same direction or to reverse it.
Although both movement statistics have been observed, persistent motion, where the particle has inertia and possesses momentum, is more common and has been observed at small
scales, e.g., bacteria undergoing run and tumble dynamics \cite{selmeczi2005cell}, as well as being seen at large scales, e.g., animals
foraging \cite{prasad2006searching} and pedestrians in dense crowds \cite{echeverria2021estimating}, while
anti-persistent motion has been reported in hemocytes \cite{korabel2022hemocytes}, cell extracts
\cite{speckner2021single}, telomeres of bone osteosarcoma cells \cite{benelli2022probing} through to the evolution of
the score difference in a game of basketball \cite{gabel2012random}. Persistent motion also plays a central role in the exciton
coherence question \cite{kenkre1983coherence, kenkre1981effect}.

The ubiquity of persistent motion in many natural and artificial phenomena has generated a wealth of literature on the
subject. Within the space-time continuous description much is known about the one-dimensional occupation probability
\cite{weiss1994aspects, rudnick2004elements, jose2022active, martens2012probability, malakar2018steady}. However, unlike
the case of Brownian walks, with correlation in the motion, the standard renewal theory approach
\cite{rws_on_latticesII, redner2001guide, metzler2014first} is not sufficient to determine the
temporal dependence of transport related quantities such as the first-passage or first-hitting statistics to one or
multiple absorbing targets. While advances have been made in bounded space, these studies are limited to one dimension \cite{bressloff2022encounter,
angelani2015run} or can only access the mean first-passage time \cite{guerin2016mean}.

The main issue with a continuous spatial description in higher dimensions relates to the need to track the previous
movement direction as an internal degree of freedom. In two or higher dimensions the turning angle (relative to the
previous movement direction) is a continuous variable and it thus requires an infinite number of internal states
\cite{larralde1997transport, masoliver2017three}. Moreover, as the absorption property of a target needs to be specified for each direction
in which it can be reached, formalising first-passage processes in continuous space in higher dimensions remains a
challenge.

By employing a discrete spatial description \cite{rws_on_latticesII, LucaPRX} it is possible to overcome this
limitation. To do that we use the so-called correlated random walk (CRW),  proposed for the first time in the 1920s
\cite{furth1920brownian, taylor1922diffusion}, as the natural non-Markovian extension to the LRW. For the CRW, instead
of prescribing a probability to move in a given direction as in the standard LRW, the movement steps are determined by
the probability that the walker \textit{continues} in the same direction as the previous step \cite{weiss1994aspects,
rudnick2004elements}. In other words, with one parameter one creates persistence, or positive correlation, if the walker
is more likely to continue and anti-persistence, or negative correlation, if backtracking is more likely.

Despite the long modelling history of persistent motion in the spatially discrete literature, the only major work on the
CRW first-passage dynamics has appeared very recently \cite{larralde2020first}. That study considers an unbounded
lattice both in one and two dimensions under the assumption that the walker starts by facing in one predetermined
direction. With this simplification the one-dimensional problem may be reduced to a known first-passage quotient
\cite{rws_on_latticesIII} and the exact first-passage dynamics to a single target is derived, while in two dimensions
the dynamics are obtained asymptotically. Despite this recent progress, determining the general first-passage time
dependence to one or multiple targets in finite space has remained an elusive task, with insights for the one target
case relying exclusively upon studies of the global mean first-passage time in periodically bounded space
\cite{tejedor2012optimizing}. 

In this paper, we generalise the discrete renewal equation and develop a general theory to determine the first-passage
probability to a finite number of targets for a CRW in bounded and unbounded domains of arbitrary dimensions. Focusing
on bounded space, we first find the analytic representation of the CRW occupation probability when space is bounded by
periodic and reflective boundaries in hypercubic lattices and by periodic boundaries in a six-sided hexagonal lattice.
By modelling the special properties of the target site as an impurity or defect problem \cite{montroll1955effect, szabo1984localized,giuggioli2022spatio,
kenkre2021memory} we derive a general form of the first-passage probability and elucidate that with
persistence, even in one-dimensional domains, the first-passage probability can be multi-modal. Furthermore, we extend
the notion of first-passage time to the \textit{directional} first-passage time, which we define as the probability that
the walker reaches a target site from a specific direction for the first time. Monte Carlo simulations are
then used to validate our theoretical findings.

As an application to flagellated bacterium that may be attracted or repelled by solid walls \cite{shum2015hydrodynamic},
we present an iterative process to extract occupation and first-passage probabilities with arbitrary boundary
conditions. With this method we study the effect of two different reflecting boundary conditions, namely whether the
boundary invokes a change of direction on various transport statistics, and show that is the former, which aids a
searching walker at intermediate times. 

The contents of the rest of the paper is laid out as follows. Section \ref{sec: Unbounded} treats the formulation of the
unbounded problem. The bounded propagators or Green's function of the occupation probability in one dimensional domains
are derived in Sec. \ref{sec: Bounded}, while their higher dimensional counterparts, together with the case of the
hexagonal lattice, which we use to model chiral persistent motion, are treated in Sec. \ref{sec: 2D}. In Sec. \ref{sec:
Absorbing} we turn our attention to the placement of absorbing traps and discuss the relationship between their location
and the imposition of absorbing boundary conditions on the lattice. It is here that we generalise the discrete renewal
equation to correlated motion. In Sec. \ref{sec: MFPT}, we derive the first moments of these distributions, then in Sec.
\ref{sec: numerical}, we display the iterative method for calculating transport statistics for cases where the analytic
solutions are not known. Concluding remarks and possible future avenues are then discussed in Sec. \ref{sec:
conclusion}.
\begin{figure}[h]
 \centering
 \includegraphics[width = 0.7\textwidth]{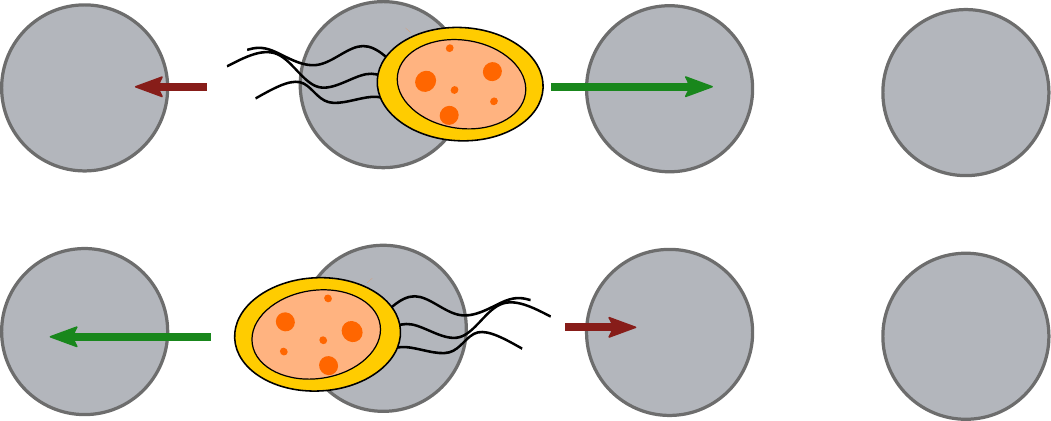}
 \caption{(Colour Online). A cartoon depicting persistent motion, for simplicity drawn over a one-dimensonal lattice,
 with each circle representing a lattice site. We consider some entity, displayed here as a bacterium, travelling
 with momentum. If it is travelling to the right, as shown in the upper track, its momentum will keep it travelling
 to the right (green arrow) unless a random fluctuation (red arrow) causes it to change direction. The equivalent
 case is then true with the direction flipped once it starts moving to the left, which is shown on the lower track.
 The smaller size of the reversing arrows stems from the reduced chance of that occurring compared to continuing to
 move in the same direction.}
 \label{fig: cartoon}
\end{figure}

\section{Unbounded Propagator}\label{sec: Unbounded} 
\begin{figure}[h]
    \centering
    \includegraphics[width = \textwidth]{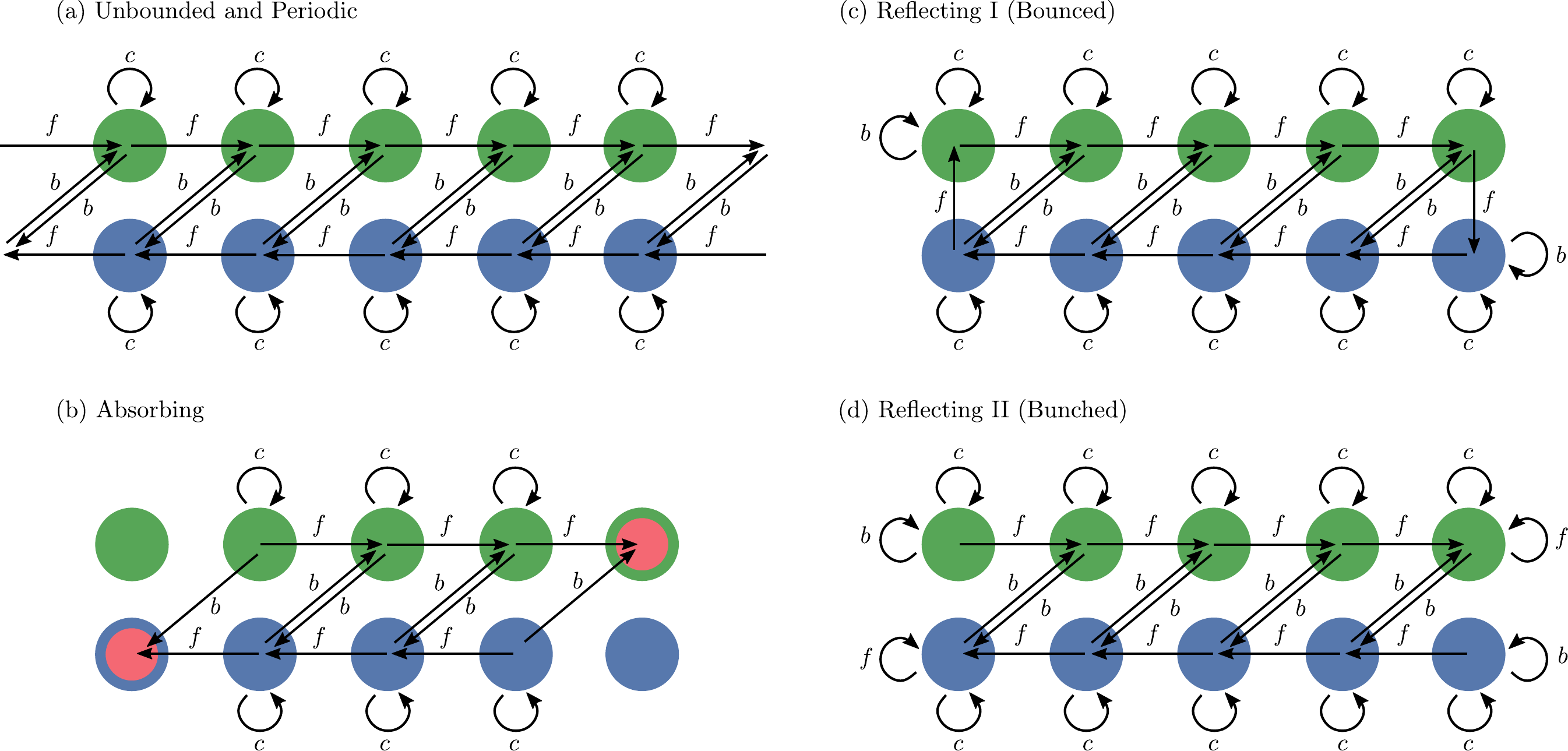}
    \caption{ (Colour Online). Schematics of the two internal state representation of a persistent or anti-persistent
    movement process in one dimension. A spatial site consists of two states, $m = 1, 2$ shown,
    respectively, in green (upper circle) and blue (lower circle). Allowed steps are drawn with arrows and the
    parameters of their corresponding jump probabilities are shown next to the corresponding arrows. In (a), the
    periodic and unbounded lattices are displayed together, because the transition probabilities are identical, with the 
    only difference arising as the wrapping between sites $1$ and $N$ in the periodic case. In panel (b) the absorbing state is drawn in
    red, while panels (c) and (d) represent the two types of reflection, which we label `bounced' and `bunched' due to
    the different interaction at the boundary. In the former, as a walker tries to move out of the domain, it reverses
    its movement direction, while in the latter it does not.}
    \label{fig: lattice_schematic}
   \end{figure}

Although the CRW is a one-step non-Markov process, its dynamics can be represented by a set of coupled second-order Markov equations
\cite{van1998remarks}, as amply shown  across the literature both in continuous \cite{masoliver1992solutions,
malakar2018steady, landman1979stochastic2} and discrete time \cite{larralde2020first}. For the latter, in one dimension it is given by 
\begin{align}
\label{eq: Master_eq}
&Q(n, 1, t\!+\!1) \! = \!fQ(n\!-\!1,1, t)\! + \!bQ(n\!-\!1,2, t) \!+\! cQ(n,1,t), \\ \nonumber
&Q(n, 2, t\!+\!1) \! = \!bQ(n\!+\!1,1, t)\! + \!fQ(n\!+\!1,2, t) \!+\! cQ(n,2,t),
\end{align}
where $Q(n,m, t)$ is the occupation probability at time $t$ to be at site $n$ and state $m$. The state (or internal
degree of freedom) represents the direction the walker may enter the site, namely a walker in state 1 must have arrived
from the left, while a walker in state 2 from the right (see schematic in Fig. \ref{fig: lattice_schematic}). The
parameter $f$ is the probability of continuing forward in the direction last travelled while $b$ is the probability of
backtracking away from the previous direction travelled. $c = 1-(f+b)$ is the so-called lazy or sojourn parameter
whereby $0 < f+b \leq 1$, with $f+b =1$ corresponding to a walker that moves at every time step. When $f > b$ the walker is in
the persistent regime, shown pictorially in Fig. \ref{fig: cartoon}, while for $f < b$ the walker is subject to anti-persistence. When $f=b$, one recovers the
diffusive dynamics modelled by the LRW. The effect of the lazy parameter is akin to a rescaling of the velocity in the
continuous limit of the CRW, the telegrapher's equation (see supplementary material \cite{supp_mat} for details).

To solve Eq. (\ref{eq: Master_eq}), it is convenient to write it in matrix form as
\begin{equation}
\bm{Q}(n, t+1) = \mathbb{A}\cdot\bm{Q}(n-1, t) + \mathbb{B}\cdot\bm{Q}(n+1, t) + \mathbb{C}\cdot\bm{Q}(n, t),
\label{eq: matrix_master_eq}
\end{equation}
where $\bm{Q}(n, t) = [Q(n, 1, t), Q(n, 2, t)]^{\intercal}$, $\mathbb{A} = \bigl[\begin{smallmatrix}f & b\\ 0 &
0\end{smallmatrix}\bigr]$, $\mathbb{B} = \bigl[\begin{smallmatrix}0 & 0\\ b & f\end{smallmatrix}\bigr]$, and $\mathbb{C} =
\bigl[\begin{smallmatrix}c & 0\\ 0 & c\end{smallmatrix}\bigr]$ with the dot operator representing matrix multiplication. 

We assume that the walker enters its initial site from the left with some arbitrary probability $\alpha_1$ and from the
right with $\alpha_2 = 1-\alpha_1$, such that the initial condition is written as $\bm{Q}(n, 0) = \delta_{n,
n_0}\bm{U_{\bm{m}_0}}$ where $\bm{U_{\bm{m}_0}} = \left[\alpha_1, \alpha_2\right]^{\intercal}$, i.e., the probability of
starting in state $m_{0_1} = \alpha_1$ while the probability of starting in state $m_{0_2} = \alpha_2$,
and the propagation is symmetric if $\alpha_1 = \alpha_2$. Using known procedures for random walks with internal degrees
of freedom \cite{barry_hughes_book,rws_on_latticesIII} (outlined in Sec. II of the supplementary Material
\cite{supp_mat} for convenience), the generating function, $\widetilde{f}(z)=\sum_{t=0}^{\infty}f(t)z^t$, solution of
Eq. (\ref{eq: matrix_master_eq}), the so-called lattice Green's function or unbounded propagator, can be obtained
analytically as a column vector of the occupation probability in each internal state:
\begin{equation}
\widetilde{\bm{Q}}_{n_0}(n, z) = \frac{1}{2\pi}\int_{-\pi}^{\pi} e^{-i\xi(n-n_0)}\big[ \mathbb{I}-z\bm{\lambda}(\xi)\big]^{-1}\!\cdot\bm{U_{\bm{m}_0}}\text{d}\xi,
\label{eq: internal_states_LGF}
\end{equation}
where $\mathbb{I}$ is the identity matrix and 
\begin{equation}
\bm{\lambda}(\xi) = \left(\begin{array}{cc}
  fe^{i\xi}+c & be^{i\xi} \\
  be^{-i\xi} &  fe^{-i\xi}+c
\end{array}\right),
\label{eq: 1D_sf}
\end{equation}
is the so-called structure function with $\xi$ the Fourier variable. The $(i,j)$-th element of $
\bm{\lambda}(\xi)$ governs the movement from state $j$ to state $i$, for example, the element $\bm{\lambda}_{1,2}(\xi)$ encodes
moving from state 2 to state 1, which is achieved only via a backtracking, which increases the coordinate $n$.
$\bm{\lambda}(0)$ allows to verify the normalisation $\left(\sum_{i=1}^{2}\bm{\lambda}_{i,j}(0) = 1, j \in
\{1,2\}\right)$, as expected from a stochastic matrix. The associated expression in arbitrary dimensions may be found in
ref. \cite{supp_mat}, where we show a generalisation of the known unbounded propagator \cite{ernst1988random}.

The integral in Eq. (\ref{eq: internal_states_LGF}) can be evaluated exactly (see \cite{supp_mat}), and after summing
over the internal states $\widetilde{Q}(n,
z) = \sum_{i=1}^{2}\widetilde{\bm{Q}}(n, z)_i$, one obtains, as a generating function, the unbounded occupation
probability for the entire site as 
\begin{align}
  \label{eqn:sol_1d}
   \widetilde{Q}_{n_0}(n,z)=\frac{1}{(1-2cz)\sqrt{\left(1+\delta z^2\right)^2-4\varepsilon^2z^2}} \left[\frac{1-cz}{r(z)^{|n-n_0|}} + \frac{\alpha_1 (b-f)z}{r(z)^{|n-n_0+1|}} + \frac{\alpha_2(b-f)z}{r(z)^{|n-n_0-1|}}\right],
\end{align}
where $r(z) = \frac{1+\delta z^2}{2\varepsilon z}\left(1+\sqrt{1-\frac{4\varepsilon^2z^2}{[1+\delta
z^2]^2}}\right)$, $\varepsilon = f(1-cz)(1-2cz)^{-1}$ and $\delta = (f^2-b^2+c^2)(1-2cz)^{-1}$, and where the appearance
of the higher order $z$ terms in Eq. (\ref{eqn:sol_1d}) makes conspicuous that the system contains internal degrees of freedom.
Using the relation $f+b+c=1$, the scaling term in Eq. (\ref{eqn:sol_1d}) can be rewritten as
$\left[(1-z)[1-z(1-2b)][1-z(1-2f)]\{1-z[1-2(f+b)]\}\right]^{-1/2}$. In the limit $z\rightarrow 1$, as the
remaining terms of Eq. (\ref{eqn:sol_1d}) give the constant $2b$, we obtain $\widetilde{Q}_{n_0}(n,z)\sim
b/\sqrt{2fb(f+b)(1-z)}$, alternatively the time dependence through Tauberian theorems is $Q_{n_0}(n,t)\sim b/\sqrt{2\pi
fb(f+b)t}$ for $t\rightarrow \infty$. As expected, this scaling is the same as the one of the diffusive LRW, and to
which it reduces when $f=b=q/2$. It is in fact a straightforward exercise to show that when $f=b$, irrespective of the
values of $\alpha_1$ and $\alpha_2$, Eq. (\ref{eqn:sol_1d}) reduces to the known solution of the symmetric unbounded
walk \cite{LucaPRX}. 

\section{Bounded Propagators}\label{sec: Bounded} 
We now consider a CRW confined to a bounded domain of size $N$ ($n \in [1, N]$) and seek the dynamics of the occupation
probability in the case of periodic and reflecting boundaries. Below we use the superscript notation $p$ and $r$,
respectively, to distinguish them and will use the letter $\bm{P}$ to denote a bounded propagator compared to 
$\bm{Q}$ for the infinite case.

\subsection{Periodic Boundary}\label{sec: periodic} For the periodic domain, the boundary condition is imposed in the
same way as for the standard P{\'o}lya walk \cite{rws_on_latticesII, LucaPRX} making it straightforward to find the
periodic propagator via the method of images (see \cite{supp_mat} Sec. III) as
\begin{equation}
\bm{P}^{(p)}_{n_0}(n, t)\! =\! \frac{1}{N}\!\sum_{\kappa=0}^{N-1}e^{\frac{-2\pi i\kappa(n-n_0)}{N}}\bm{\lambda}\left(\frac{2\pi\kappa}{N}\right)^t\!\cdot\bm{U_{\bm{m}_0}},
\label{eq: matrix_periodic}
\end{equation}
where $\bm{\lambda}(\cdot)^t$ denotes the matrix power. Diagonalising $\bm{\lambda}\left(\frac{2\pi\kappa}{N}\right)$ to take the matrix power explicitly and performing a
summation over the states, one finds the site occupation probability $P^{(p)}_{n_0}(n, t)=\sum_{i=1}^2\bm{P}^{(p)}_{n_0}(n, t)_i$, and when $\alpha_1 = \alpha_2 =
\frac{1}{2}$ we obtain
\begin{equation}
P^{(p)}_{n_0}(n, t)\! =\! \frac{1}{2N}\!\sum_{\kappa=0}^{N-1}\!\cos\left(\frac{2\pi\kappa(n-n_0)}{N}\right)\!\Big[f(\kappa, t)+ h(\kappa, t)\Big], 
\label{eq: scalar_periodic}
\end{equation}
where 
\begin{equation}
\begin{aligned}
f(\kappa, t) \!=\! \lambda_{+}^{t}\! +\! \lambda_{-}^{t},\qquad h(\kappa, t)\! = \!\frac{b\cos\!\left(\frac{2\pi\kappa}{N}\right)\!\left[\lambda_{+}^{t}\! - \!\lambda_{-}^{t}\right]}{\sqrt{f^2\!\cos^2\!\left(\frac{2\pi\kappa}{N}\right) \!+\! b^2\! - \!f^2}}, \\
\end{aligned}
\end{equation}
and 
\begin{equation} 
\lambda_{\pm} = c + f\cos\left(\frac{2\pi\kappa}{N}\right) \pm
\sqrt{f^2\cos^2\left(\frac{2\pi\kappa}{N}\right) + b^2 - f^2},
\label{eq: lambdapm}
\end{equation}
are the eigenvalues of the matrix structure function $\bm{\lambda}(\frac{2\pi\kappa}{N})$. The non-symmetric case is
less compact and we give it explicitly in Eq. (S23) of the supplementary material
\cite{supp_mat}.

In the diffusive limit, where $f = b= \frac{q}{2}$,  Eq. (\ref{eq: matrix_periodic}), reduces to the expression for the
propagator for the LRW on a one-dimensional periodic lattice \cite{LucaPRX}, while in the $f = 1$, limit one has $P^{(p)}_{n_0}(n, t) =
N^{-1}\sum_{\kappa=0}^{N-1}\left\{\alpha_1\cos\left[2\pi\kappa(n-n_0+t)/N\right]\right.
+\left.\alpha_2\cos\left[2\pi\kappa(n-n_0-t)/N\right]\right\}$,
%\begin{equation} \begin{aligned} P^{(p)}_{n_0}(n, t) =
% \frac{1}{N}\sum_{\kappa=0}^{N-1}&\alpha_1\cos\left(\frac{2\pi\kappa(n-n_0+t)}{N}\right) + \\
% &\alpha_2\cos\left(\frac{2\pi\kappa(n-n_0-t)}{N}\right), \end{aligned} \label{eq: scalar_periodic_ballistic}
% \end{equation}
which shows a superposition of two weighted ballistic waves travelling in opposite directions. In the other limit $(b =
1)$, corresponding to the perfectly anti-persistent case, the walker constantly hops between the initial condition,
which it visits on even time steps, and the lattice sites on either side at odd
time steps, as one may see from the resulting expression: $P^{(p)}_{n_0}(n, t) = (2N)^{-1}\sum_{\kappa=0}^{N-1}\exp[-2\pi
i \kappa  \left(n-n_{0}\right)/N]\big\{[1+(-1)^{t}]-[-1+(-1)^{t}][\alpha_1\exp(2\pi i \kappa/N) + \alpha_2\exp(-2\pi i
\kappa/N)]\big\}$.
%\begin{equation}
%\begin{aligned}
%&P^{(p)}_{n_0}(n, t) = \frac{1}{2N}\sum_{\kappa=0}^{N-1}e^{\frac{2\pi i \kappa  \left(n-n_{0}\right)}{N}}\times \\
%& \left[\left(1+\left(-1\right)^{t}\right)-\left(-1+\left(-1\right)^{t}\right)\left(\alpha_1e^{-\frac{2\pi i \kappa}{N}} + \alpha_2e^{\frac{2\pi i \kappa}{N}}\right)\right].
% \end{aligned}
% \label{eq: periodic_b1}
%\end{equation} 

\subsection{Reflective Barrier}\label{sec: reflective} 
When the correlated particle is confined within a reflecting box, there are two classes of boundary interactions that
one might consider as displayed in panel (c) and (d) of Fig. \ref{fig: lattice_schematic}. More specifically, one must
choose whether the direction of persistence is flipped upon hitting the boundary or not. When the persistence direction
is not flipped then an accumulation of particles appears at the boundary, which is reflected in the increase in the
magnitude of the steady state probability at the boundary \cite{malakar2018steady, bechinger2016active}. We call this
accumulation a `bunching'. In contrast, when the boundary induces a change of direction, which we call a `bounce', there
is no accumulation. It is the `bouncing' case we consider here and we find the propagator via two different methods,
viz. the method of images \cite{LucaPRX} and the so-called \textit{squeezing} method \cite{kalay2012effects}, which
allow, respectively, for two variations on this boundary condition. We treat the `bunched' case later in Sec. \ref{sec:
numerical} and provide a comparison of the transport statistics with the `bounced' case. 

\subsubsection*{The Method of Images}
The method of images is a dimensionally scalable method that allows to include the effects of interfaces or boundary
conditions on the walker dynamics from the knowledge of the dynamics in the absence of boundaries
\cite{keller1953scope}. To account for the change in movement direction 
caused by a reflection, we ensure that the bounded
walker changes state upon interacting with the boundary. To achieve this, one must specify boundary conditions for each internal
state. More specifically, we require that $P^{(r)}(1, \{2, 1\}, t) = P^{(r)}(0, \{1, 2\}, t)$, and $P^{(r)}(N, \{1, 2\},
t) = P^{(r)}(N+1, \{2, 1\}, t)$, where the compact notation $\{i, j\}$ implies pairing the left index $i$ (right index
$j$), on the LHS of the equation, to the left index (right index) on the RHS. In this way, by allowing the image of state two to interact with the dynamics on state
one, and vice versa, the persistent walker trying to leave the boundary will be reflected back and continue its journey
in the opposite direction (see  \ref{app: dD_reflective} for further discussion). With the boundary conditions
set, before applying the method of images, we need to ensure that the transport process is spatially symmetric, which we
accomplish by imposing $\alpha_1 = \alpha_2 = \frac{1}{2}$. 

By taking the state-level boundary conditions given above, and performing a summation over the internal states, one
finds the boundary condition for the entire lattice site. Upon doing so, one find the boundary condition to be
$P^{(r)}_{n_0}(1, t) = P^{(r)}_{n_0}(0, t)$, $P^{(r)}_{n_0}(N, t) = P^{(r)}_{n_0}(N+1, t)$, which, as expected
\cite{masoliver1993solution}, is the same boundary condition as the diffusive random walker. Since the image set for this
boundary condition has been derived recently \cite{LucaPRX}, one can find the reflective propagator as $P^{(r)}_{n_0}(n, t) = \sum_{i=1}^{2}\bm{P}^{(r)}_{n_0}(n, t)_i$
where
\begin{equation}
\bm{P}^{(r)}_{n_0}(n, t)= \frac{1}{N}\sum_{\kappa=0}^{N-1}\beta_{\kappa}\cos\left(\frac{\pi\kappa(n-\frac{1}{2})}{N}\right) \cos\left(\frac{\pi\kappa(n_0-\frac{1}{2})}{N}\right)\bm{\lambda}\left(\frac{\pi\kappa}{N}\right)^t\cdot\bm{U_{\bm{m}_0}}, 
\label{eq: matrix_reflective1}
\end{equation}
with $\beta_{\kappa} = 1$ when $\kappa = 0$, else $\beta_{\kappa} = 2$ (see Sec. IV of the
supplementary material \cite{supp_mat} for a full derivation).

From Eq. (\ref{eq: matrix_reflective1}), and similarly in Eq. (\ref{eq: matrix_periodic}), one may notice that, in
contrast to the unbounded dynamics in Eq. (\ref{eqn:sol_1d}), within periodic and reflecting domains we are able to keep
separate the spatial and temporal components of the propagator. In doing so, one realises that, as expected, the
respective spatial dependence in the symmetric $\alpha_1 = \alpha_2$ case coincides with the one obtained for the
so-called lazy P{\'o}lya's walk in finite domains \cite{LucaPRX} and the structure function pertains in both solutions
as the boundary condition does not impact the individual jump probabilities in the bulk of the domain. Since the
structure of the two propagators is analogous, one can also expand Eq. (\ref{eq: matrix_reflective1}) via the same procedure
as the periodic case to find explicitly $P^{(r)}_{n_0}(n, t)$. The semi-bounded bounced propagator can also be found via
the method of images and, for completeness, we have presented it in the supplementary material \cite{supp_mat}.

\subsubsection*{The Squeezing Method}

We exploit an alternate procedure to bound the propagator in a reflective box, which uses only the properties of the
periodic propagator \cite{kalay2012effects, montroll1956random,beylkin2008fast}. While being less scalable to higher dimensions, when compared to
the method of images, it is simple to construct analytically in one-dimensional systems. This technique is
particularly convenient to analyse the case when  $\alpha_1 \neq \alpha_2$, since the use of an image set in a periodic
geometry wraps upon itself, which avoids the interplay between the two independent image sets used to derive Eq. (\ref{eq: matrix_reflective1}).

We follow ref. \cite{kalay2012effects} and \textit{squeeze} a torus into a reflecting lattice. More specifically, we
transform a periodic domain with $M \in 2\mathbb{Z}$ sites into a reflective one with $N = (M+2)/2$ sites
($N\geq 2$) by summing the probabilities of the sites a distance $l$ on
either side of the $1^{\text{st}}$ and $N^{\text{th}}$ site (see Fig. 1b in \cite{kalay2012effects}
or \ref{app: dD_reflective} for a schematic
representation). In doing so, we impose the boundary condition as implemented
originally by Chandrasekhar for the LRW \cite{chandrasekhar1943stochastic}, which states that the walker attempting to escape from
$N$ is bounced back to $N-1$ with certainty, and similarly for site $n=1$. The reflective propagator with
this condition
is \cite{kalay2012effects}
\begin{equation}
\bm{P}_{n_0}^{(r_s)}(n, t) = \bm{P}^{(p)}_{n_0}(n, t) + \mu(n)\bm{P}^{(p)}_{n_0}(M+2-n, t),
\label{eq: 1d_reflective_kalay}
\end{equation}
where we indicate with the superscript $r_s$ that the reflecting domain has been accounted by squeezing a periodic one,
and where $\mu(n) = 0$ when $n = 1$ or $N$, and $\mu(n)=1$, otherwise.  We note that by construction, in this instance,
one should not think of the states as being representative of the direction of motion previously travelled, since the
baseline propagator on the torus, from which Eq. (\ref{eq: 1d_reflective_kalay}) is built, considers trajectories that may 
enter each lattice site from either direction.
\begin{figure*}[t]
 \centering
 \includegraphics[width=0.8\textwidth]{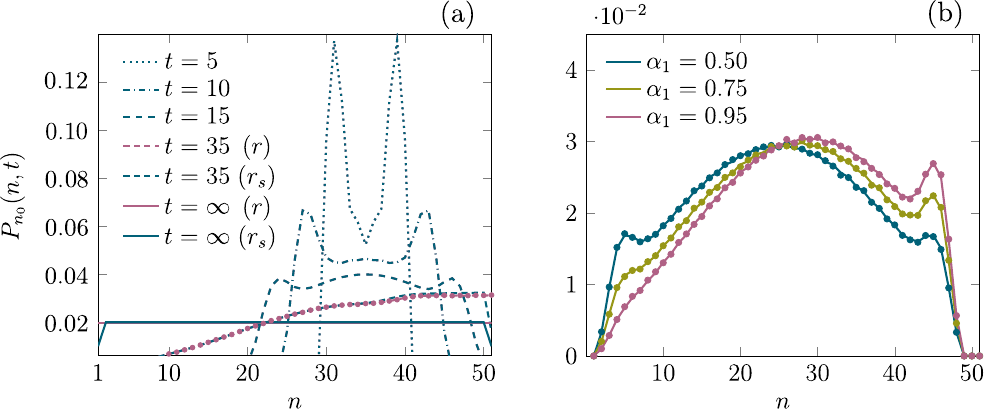}
 \caption{(Colour Online). The occupation probability in a one-dimensional reflecting domain of length $N=51$ found
 via the method of images, Eq. (\ref{eq: matrix_reflective1}), and the squeezing technique, Eq. (\ref{eq:
 1d_reflective}). Panel (a) shows the evolution of the occupation probability of both propagators, with the method of
 images $(r)$ solution plotted in purple and the squeezed solution $(r_s)$ displayed in blue, for the symmetric $\alpha_1 = \alpha_2 = 0.5$ case. For
 clarity, as there is no differences between the dynamics of the two propagators before boundary interaction, we omit
 in the legend the entries for the method of images solution at short times. The dynamics seen are with $f =0.7$, $b = 0.1$
 and $n_0 = 35$. Panel (b) depicts the occupation
 probability of the squeezed propagator at $t = 23$ with $n_0 = 25$, $f = 0.8$, $b = 0.1$. Here we examine the effect 
 of changing $\alpha_1$ and $\alpha_2 =
 1-\alpha_1$. The dots on some curves across the two panels represent the ensemble average of $10^6$ stochastic
 simulations.}
 \label{fig: reflect_occu}
\end{figure*}

Using Eqs. (\ref{eq:
matrix_periodic}) and (\ref{eq: 1d_reflective_kalay}), and after some algebra,  the reflective propagator is given explicitly by
\begin{equation}
\begin{aligned}
\bm{P}^{(r_s)}_{n_0}(n,t) = \frac{1}{2N-2}&\sum_{\kappa= 0}^{2N-3}\bigg[\exp\left(\frac{-\pi i\kappa(n-n_0)}{N-1}\right)  \\ 
&+\mu(n)\exp\left(\frac{-\pi i\kappa(2N-n-n_0)}{N-1}\right) \bigg]\bm{\lambda}\left(\frac{\pi\kappa}{N-1}\right)^t\cdot\bm{U_{\bm{m}_0}}.
\end{aligned}
\label{eq: 1d_reflective}
\end{equation}
In Fig. \ref{fig: reflect_occu}(a), we plot the reflecting propagator found via both methods (Eqs. (\ref{eq:
matrix_reflective1}) and (\ref{eq: 1d_reflective})). For times small enough that no boundary interaction has occurred,
one cannot distinguish between the two representations, as expected. However, after boundary interaction, the occupation
probability differs slightly due to the no-waiting scheme at the boundary that Eq. (\ref{eq: 1d_reflective}) obeys.
Since these sites can be viewed as `slippy', i.e., the chance of leaving the site is higher than the rest of the domain,
the steady state distribution shows a bench-like structure, with these boundary sites having a lower probability of
being occupied. In fact, it can be shown (see  \ref{app: s_func}) that when either $f$ or $b$ are not identically one,
$P^{(r_s)}_{n_{0}}(n, t \to \infty) = \frac{1}{(N-1)}$ ($n \neq \{1, N\}$) while $P^{(r_s)}_{n_{0}}(1,t\to \infty) =
P^{(r_s)}_{n_{0}}(N,t\to \infty) = \frac{1}{2(N-1)}$. 

In Fig. \ref{fig: reflect_occu}(b), we display the effect of the weightings over the internal states in the initial condition.
When $\alpha_1 \neq \alpha_2$, the propagation is clearly asymmetric, as the higher weighting of trajectories that
started right are likely to continue heading right.

\section{Propagators in Higher Dimensions}\label{sec: 2D}
As mentioned in the introduction, in continuous space modelling persistence in higher dimensions requires the inclusion
of an infinite number of internal variables \cite{larralde1997transport}.  A discrete representation, on the other hand,
requires only $Z$ internal variables, where $Z$ is the coordination number of the lattice. In dimensions larger than
one, we require a third parameter, $\ell$, to govern the probability of turning, which we define as a change of
movement direction in any direction except turning back. For simplicity, we first assume the probability of moving in
any lateral direction, to be uniform, although this can be relaxed with no additional mathematical burden, as we will
show below with the hexagonal lattice. Moreover, we demand $f + b +
\mathfrak{d}\ell + c^{(\mathfrak{d})}= 1$, where $\mathfrak{d}$ is the number of permissible lateral movements, defined
as $\mathfrak{d} = Z-2$ and $c^{(\mathfrak{d})}$ is the sojourn probability. The initial condition must also account for
the higher number of movement directions, meaning that $\bm{Q}_{\bm{n}_0}(\bm{n}, 0) = \prod_{i= 1}^{d}\left(\delta_{n_i,
n_{0_i}}\right)\bm{U}_{\bm{m}_{0}}$, where $\bm{U}_{\bm{m}_{0}} = [\alpha_1, ... , \alpha_{2d}]^{\intercal}$. 

\subsection{Hypercubic Lattices}\label{sec: square} 

We again start by writing a Master equation to govern the unbounded dynamics. In $d$ dimensions, this takes the 
form of $2d$ coupled equations, which we have written explicitly for the $d=2$ case in the supplementary material \cite{supp_mat}. 
Akin to the $d=1$ case, it is convenient to re-write the coupled equations in matrix form, and 
for arbitrary dimension $d$, we obtain
\begin{equation}
 Q(\bm{n}, t\!+\!1) = \sum_{i=1}^{d}\left\{\mathbb{A}_{2d}^{(i)}\cdot Q(\bm{n}\!-\!\bm{e}_i, t) \!+\! \mathbb{B}_{2d}^{(i)}\cdot Q(\bm{n}\!+\!\bm{e}_i, t)\right\} + \mathbb{C}_{2d}\cdot Q(\bm{n}, t),
 \label{eq: Dd_ME_matrix}
\end{equation} 
where $\bm{e}_i$ is a unit vector along dimension $i$.
$\mathbb{A}_{2d}^{(i)}$, $\mathbb{B}_{2d}^{(i)}$ are $2d \times 2d$ matrices that govern the movement
probabilities along either direction of the $i^{\text{th}}$ axis, which is encoded via 
elements along $(2i-1)^{\text{th}}$ and $2i^{\text{th}}$ rows, respectively, while all other rows in $\mathbb{A}_{2d}^{(i)}$ and $\mathbb{B}_{2d}^{(i)}$
are identically zero. $\mathbb{C}_{2d}$ governs the
sojourn probability and is of the form  $\mathbb{C}_{2d} = c^{(\mathfrak{d})}\mathbb{I}$.

The method of images allows us to find the periodic and reflective propagator in arbitrary dimensions. Following the supplementary
material \cite{supp_mat}, we find the propagator in arbitrary dimensions again as a summation of column vector of state probabilities 
$P^{(\gamma)}_{n_{01}, ... , n_{0d}}(n_1, ..., n_d, t) = \sum_{i=1}^{2d}\bm{P}^{(\gamma)}_{n_{01}, ... , n_{0d}}(n_1, ..., n_d, t)_i $,
where
\begin{align}
\bm{P}^{(\gamma)}_{n_{01}, ... , n_{0d}}(n_1, ..., n_d, t)\! =\!  \frac{1}{N^d}\! \sum_{\kappa_1=0}^{N_1-1}\! ...\! \sum_{\kappa_d=0}^{N_d-1}\left[\prod_{i=1}^{d}g_{\kappa_i}^{(\gamma)}(n_i, n_{0_i}\!) \right]
\bm{\lambda}\left(\pi \mathcal{N}^{(\gamma)}_{\kappa_1}, ..., \pi \mathcal{N}^{(\gamma)}_{\kappa_d}\right)^t\cdot\bm{U}_{\bm{m}_{0}},
\label{eq: high_dim}
\end{align}
and $\gamma = p$ or $r$. The function $g_{\kappa_i}^{(\gamma)}(n_i, n_{0_i})$ is the spatial dependence of the
periodic or reflective bounded domain, given by $g_{\kappa_j}^{(p)}(n_j, n_{0_j}) =\exp[-2\pi i\kappa_{j}(n_j -
n_{0_j})/N_j]$ and $g_{\kappa_j}^{(r)}(n_j, n_{0_j}) =
\beta_{\kappa_j}\cos[\pi\kappa_j(n_j-1)/(N_j)]\cos[\pi\kappa_j(n_{0_j}-1)/(N_j)]$, and where
$\mathcal{N}^{(p)}_{i}=2\kappa_{i}/N_i$, while $\mathcal{N}^{(r)}_{i} =\kappa_{i}/N_i$. We note that, similarly to the
$d=1$ case, the image method demands that when $\gamma = r$ the initial weighting over the states to be uniform.

The $2d\times 2d$ matrix $\bm{\lambda}\left(\pi \mathcal{N}^{(\gamma)}_{\kappa_1}, ..., \pi \mathcal{N}^{(\gamma)}_{\kappa_d}\right)$
maintains a simple structure regardless of the dimension and also allows straight forward calculation of the uniform
steady state for both $\gamma = p$ or $r$ (see  \ref{app: s_func}). Explicit analytical expressions of the
scalar propagator for the site occupation probability depends on the analytical determination of the eigensystem of
$\bm{\lambda}\left(\pi \mathcal{N}^{(\gamma)}_{\kappa_1}, ..., \pi \mathcal{N}^{(\gamma)}_{\kappa_d}\right)$. We discuss some of these
analytic cases starting from the fully correlated or anti-correlated limits. In those cases many elements of
$\bm{\lambda}\left(\pi \mathcal{N}^{(\gamma)}_{\kappa_1}, ..., \pi \mathcal{N}^{(\gamma)}_{\kappa_d}\right)$ reduce to zero. When
$b, \ell=0$, the matrix is diagonal, so it is trivial to take the matrix power explicitly and the propagator is simply
the sum of the matrix diagonal elements weighted according to $\bm{U}_{\bm{m}_{0}}$. Furthermore, when $f, c ,\ell = 0$, the
structure function reduces to a generalised permutation matrix, which one may also take to arbitrary power in
closed form. 

As $\sum_{t=0}^{\infty}z^t\bm{\lambda}(\cdot)^t = [\mathbb{I}-z\bm{\lambda}(\cdot)]^{-1}$ in employing generating
functions, it may become more convenient to take a matrix inverse than calculating the eigensystem to determine the
propagator in time. To illustrate this aspect, we consider the simplified cases discussed in
\cite{proesmans2020phase, tejedor2012optimizing}, whereby the authors make the assumption that the
probability of a lateral movement is equivalent to that of a backward movement, which corresponds to setting $b= \ell$
in our notation. We display the generating function of the occupation probability for this case in Sec. V of ref. \cite{supp_mat}.

Furthermore, carrying out the same procedure, but now setting $f=\ell$, one obtains a structure function matrix
where the eigenvalues can be calculated explicitly (see \cite{supp_mat}). Following this, the propagator for the entire
site probability can in fact be determined in the time domain as 
\begin{align}
    \widetilde{P}^{(\gamma)}_{n_{0_1}, n_{0_2}}(n_1, n_2, t)=\frac{1}{2N^2}\sum_{\kappa_1 = 0}^{N-1}\sum_{\kappa_2 = 0}^{N-1}\left[\prod_{i=1}^{2}g_{\kappa_i}^{(\gamma)}(n_i, n_{0_i})\right] 
     \left[f(\kappa_1,\kappa_2,t)+\frac{1-2\ell}{2\ell}g(\kappa_1,\kappa_2,t)\right],
     \label{eqn:2D_corr_final}
\end{align}
where
\begin{align}
 f(\kappa_1,\kappa_2,t)= \psi_{+}(\kappa_1, \kappa_2)^t +  \psi_{-}(\kappa_1, \kappa_2)^t,
\end{align}
\begin{align}
g(\kappa_1,\kappa_2,t)&=\frac{\sigma(\kappa_1, \kappa_2)\left[\psi_{+}(\kappa_1, \kappa_2)^t -  \psi_{-}(\kappa_1, \kappa_2)^t\right]}{\sqrt{\sigma(\kappa_1, \kappa_2)^2+1-4\ell}},
\end{align}
\begin{equation}
 \psi_{\pm}(\kappa_1, \kappa_2) = \sigma(\kappa_1, \kappa_2)\pm\sqrt{\sigma(\kappa_1, \kappa_2)^2+1-4\ell},
\end{equation}
and 
\begin{equation}
 \sigma(\kappa_1, \kappa_2) = \ell\left(\cos\left[\pi\mathcal{N}_{\kappa_1}^{(\gamma)}\right]+\cos\left[\pi \mathcal{N}_{\kappa_2}^{(\gamma)}\right]\right).
\end{equation}

\subsection{Hexagonal Lattice}\label{sec: hex} 
There has been some attempts in the past to study correlated motion in a hexagonal setting. The steady state of a run
and tumble model which can change direction by a rotation of $60^{\circ}$ was studied analytically in
\cite{smith2022exact}, while the path of a persistent photon with varying angles of incidence is studied as a
$12^{\text{th}}$ order Markov chain in a reflecting hexagonal geometry in \cite{miri2003persistent}, where long time
diffusion constants and moments are derived. Moreover, the CRW on a hexagonal lattice has been used in
\cite{prasad2006searching} to determine optimal search strategies for central place foragers.

When considering the hexagonal lattice, it is convenient to use Her's three axis coordinate system
\cite{her1995geometric} where each coordinate is represented by three linearly dependent coordinates
$(n_1, n_2, n_3)$ such that $n_1 + n_2 + n_3 = 0$. The finite domain is six-sided, with the number of lattice points
$\Omega = 3R^2 + 3R + 1$ governed by the circumradius of the hexagon $R$. As the coordination number in this case is
$Z=6$, we require $f + b + \sum_{j=1}^{4}\ell_j + c^{(4)}= 1$ and $\bm{U}_{{\bm{m}_0}}$ is a $6\times 1$ column vector. Each
lateral movement direction is defined in relation to the previous movement step where $\ell_1$ governs the probability
of a $60^{\circ}$ rotation anticlockwise, $\ell_2$ a $120^{\circ}$ anticlockwise rotation, $\ell_3$ a $120^{\circ}$
clockwise rotation and $\ell_4$ that of a $60^{\circ}$ clockwise rotation, see Fig. \ref{fig: hex_direction} (Appendix \ref{app: hex}) for pictorial
representation. With these parameters, we find the propagator for correlated motion in a periodic
hexagonal geometry as (see \ref{app: hex}) 
\begin{align}
\bm{P}^{(\mathcal{H})}_{\bm{n}_{0}}(\bm{n}, t) = \Bigg\{\frac{\bm{\lambda}^{(\mathcal{H})}(0,0)^t }{\Omega}  +& \frac{1}{\Omega}\sum_{r=0}^{R-1}\sum_{s=0}^{3r+2} e^{\frac{2\pi i\bm{\kappa}\cdot(\bm{n} -\bm{n}_{0})}{\Omega}} \bm{\lambda}^{(\mathcal{H})}\big(\bm{\kappa}(r,s)\big)^t \Bigg. \nonumber \\ &+
  e^{\frac{-2\pi i\bm{\kappa}\cdot(\bm{n} -\bm{n}_{0})}{\Omega}}\bm{\lambda}^{(\mathcal{H})}\big(-\bm{\kappa}(r,s)\big)^t \Bigg\}\cdot \bm{U}_{{\bm{m}_0}},
\label{eq: hex_periodic}
\end{align}
with the vector $\bm{\kappa} (r,s)= 2\pi[\kappa_1(r,s), \kappa_2(r, s)]/\Omega$ having components $\kappa_1(r,s) = R(s+1) + s - r$ and $\kappa_2(r,s) = R(2-s+3r) + r + 1$
\cite{marris2023exact} and with $\bm{\lambda}^{(\mathcal{H})}(\cdot, \cdot)$ defined in Eq. (\ref{eq: hex_structure_func}).

\begin{figure*}[h]
 \includegraphics[width = \textwidth]{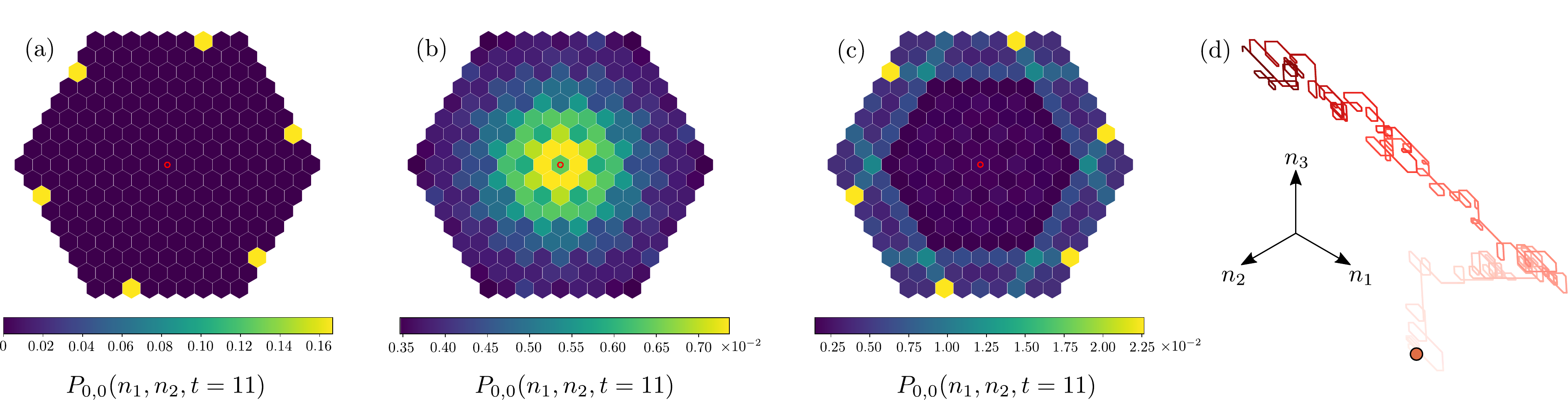}
 \caption{(Colour Online).  Correlated motion in hexagonal lattices. Panels (a)-(c) show the occupation probability
 from Eq. (\ref{eq: hex_periodic}) at time $t = 11$ in a hexagonal domain with circumradius $R = 8$ ($\Omega = 217$)
 and initial condition at the origin (shown via the red circles) with $\bm{U_{\bm{m}_0}} =
 \left(1,1,1,1,1,1\right)^{\intercal}/6$. The boundary condition is taken as the right-shift periodic boundary
 condition, and we provide further details of how this impacts the dynamics in \ref{app: hex}. In panel (a) the
 walker is ballistic ($f = 1$) and there is no steady state. There are instead probability pulses of magnitude
 $\frac{1}{6}$ that travel around the domain for all time. In panel (b) $f = 0.6$, $b = 0.06$ and $\ell_{\{1, 2, 3,
 4\}} = 0.085$, while in panel (c) $f=0.8$, $\ell_1 = 0.2$ ($b = \ell_{\{2,3,4\}} = 0$). Panel (d) depicts a sample trajectory, with the same
 movement probabilities as panel (c). The trajectory is over $2000$ timesteps and the colour of the trajectory
 indicates the time, with light red indicating short times and dark red indicating long times. The domain is of size
 $R = 700$, which is chosen so the walker does not interact with the boundary, and we omit the boundary for visual
 ease. The initial condition is shown by the orange circle at (0,0,0) and the coordinate axes for the hexagonal
 lattice are shown via the black arrows.}
 \label{fig: hex_occu}
\end{figure*}

In a periodic hexagon the non-orthogonality of the coordinates leads to a shift as the walker crosses the domain
boundaries \cite{marris2023exact}. In the context of this present work, if a ballistic walker traverses across the upper
boundary it will continue its path via the bottom edge, but shifted along one coordinate.  One can choose which way to
shift the walker, and we have arbitrarily chosen the right shift here. For such a case, in Fig. \ref{fig: hex_occu}(a),
we display the probability when the dynamics are ballistic, where one observes pulses of probability
with magnitude 1/6, which originally travelled from the origin to the six corners, have traversed the
boundary and are no longer on their original axis. Since each side of the hexagon is shifted in the same
way, the rotational symmetry remains, which is also seen in a less persistent case in panel (b).

An interesting feature of this boundary condition is that the ballistic walker in
the periodic hexagonal lattice is guaranteed to visit every site (see \ref{app: hex} for 
mathematical details). In contrast, in the square lattice, either periodic or reflecting, the ballistic walker is
confined to a genuine sublattice, i.e. only the coordinates $[(1, n_{0_2}), ..., (N,n_{0_2}), (n_{0_1}, 1), ...,
(n_{0_2},N)]$, which can be evinced as when $b$ and $\ell$ are identically zero, $\left[\mathbb{I} -
\bm{\lambda}\left(\pi \mathcal{N}^{(\gamma)}_{\kappa_1}, \pi \mathcal{N}^{(\gamma)}_{\kappa_2}\right)\right]^{-1}$ is singular away from
$(\kappa_1, \kappa_2) = (0,0)$, which occurs only when the walk is confined to a genuine sublattice
\cite{spitzer1976principles}. This has
interesting implications on the search time of the walkers in these lattices. Clearly, when the target is away from the
coordinates of the sublattice, the search time of the walker in the square lattice becomes increasingly longer as
persistence is increased \cite{tejedor2012optimizing} until the probability of a first-passage event ever taking place
becomes zero. However, on the hexagonal lattice, the ballistic walker will, with probability one, hit the target in $t
\leq \Omega$ timesteps making an increase of persistence a good search strategy with the periodic hexagonal boundary 
condition.

\subsubsection*{Chiral Motion}
By relaxing the assumption that all lateral movement directions are uniformly distributed, we introduce the possibility
of looping patterns into the movement. We present this as a discrete alternative to the
model of `chiral' persistent motion \cite{larralde1997transport, larralde2015three, masoliver1993some, hargus2021odd}, where an angular bias is introduced
into the turning angle, which is of interest as a model of a charged
particle subjected to soft scattering by aligned magnetic domains \cite{larralde1997transport} and particles driven by
microscopic torques such as active colloids \cite{kummel2013circular}, spermatozoa \cite{riedel2005self}, and
chlamydomonas \cite{fuchter2023three}. While chiral motion appears at first to lend itself to a continuous space
representation, it is known that even the two-dimensional case leads to an analytical intractable Fokker-Planck equation
\cite{larralde1997transport}, meaning occupation probabilities and further transport quantities are difficult to obtain.
On the other hand, in the discrete paradigm, to create looping, one merely alters the movement probabilities in
$\bm{\lambda}^{(\mathcal{H})}(\cdot, \cdot)$ allowing to obtain with ease the occupation probability.

We create this motion, with anti-clockwise loops, by first setting $c^{(4)}=0$ and then $f, \ell_1 > b,
\ell_{\{2,3,4\}}$, i.e. the probability of either continuing forward or turning $60^{\circ}$ is much larger than any
other movement direction. For this chiral example, using the parameters  $f=0.8$, $\ell_1 = 0.2$, we plot the occupation
probability in Fig. \ref{fig: hex_occu}(c) and a sample trajectory in Fig. \ref{fig: hex_occu}(d) in a large domain so
that no boundary interaction takes place. The single trajectory displays periods of persistent motion before making
loopy turns, which are anti-clockwise in most cases. To control these loops, one may either employ a larger $\ell_1$,
while decreasing $f$, so that their frequency increases, or one may increase $\ell_2$ at the expense of $\ell_1$, to
tighten loops. To create clockwise loops, one instead places the turning probability into $\ell_3$ or $\ell_4$ instead.

\section{Absorbing Traps and First-Passage Processes}\label{sec: Absorbing} 

In the presence of absorbing traps, the one-dimensional boundaries to be imposed on the equation governing the
occupation probability for the persistent random walk are known~\cite{masoliver1992solutions}. As shown
pictorially in Fig. \ref{fig: lattice_schematic}(b) for the one dimensional case, to constrain entirely the domain
between sites $n=1$ and $n=N$, the Master equation needs to satisfy the boundary
condition $P^{(a)}(N, 1, t)=P^{(a)}(1, 2, t)  =
0$, where now the superscript $a$ indicates the presence of absorbing boundaries.

In two and higher dimensions, the constraints to represent a fully absorbing domain can be understood by considering,
for simplicity, the square lattice. Here, with $Z=4$ the states 1,2,3 and 4 represent, respectively, a walker that has
moved West, East, North and South in the last movement step. In this case, the walker can reach the West of the domain
only by increasing its $n_1$ coordinate, thus reaching any site with coordinate $n_1=N$ in state 1, the north of the
domain by increasing its $n_2$ coordinate, thus reaching site $n_2=N$ in state 3, etc. As such, there is only one
movement direction the walker can undertake which results in an absorption at each boundary, leading to the boundary
condition $P^{(a)}(N, n_2,1, t)= P^{(a)}(1, n_2, 2, t)=P^{(a)}(n_1, N, 3, t) = P^{(a)}(n_1, 1, 4, t) = 0$. One may
extend this logic to any lattice, that is, to create a fully absorbing boundary traps need to be placed in only the one
state in each of the boundary sites that corresponds to the movement direction into that boundary.

On the other hand, in the cases when the traps are not complete along the domain boundaries, are partially absorbing, or more
generally, when traps are away from boundary sites, one needs to consider all the states in a site to make it fully
absorbing. To do that one must choose an appropriate absorption free propagator. While the method of images solution in
a periodic domain is amenable to placing traps, the reflective method of images solution is hardly useful when
extracting first-passage distributions. To explain why, recall, as shown in Sec. \ref{sec: reflective} (and
\ref{app: dD_reflective}), that the construction of the reflective propagator is reliant on the interplay between the
symmetric propagation in state one (two) and its oppposite image in state two (one). Hence, once a trap is placed on an
individual state, which stops movement along one direction, the dynamics with the trap in site one (two) is no longer
symmetric with its image counterpart representing movement along state two (one). To overcome this issue, in this
Section we use the squeezed propagator viz. Eq. (\ref{eq: matrix_reflective1}), for the reflective occupation
probability, which requires no symmetry between individual image sets, since it is built using periodic propagators.

We are interested in studying first-passage processes both when a single target consists of a subset of the $i=1,...,M$
internal degrees of freedom or all of them (the entire site), in the cases where the initial probability is shared
between any or all of the internal states in $\bm{n}_0$. Since the set of trajectories that originate in state
$\bm{n}_0, m_i$ are independent of those starting in $\bm{n}_0, m_j$ and those terminating in $\bm{s}, m_{\bm{s}_i}$ are
independent of those terminating in $\bm{s}, m_{\bm{s}_j}$ $(i\neq j)$, it suffices to consider them separately and sum
the result. Furthermore, when there are multiple targets, the summation is carried out over both the internal states and
the physical site locations.

To illustrate, consider the (unordered) set $\mathdutchcal{S} = \left\{(\bm{s}_i, m_{\bm{s}_i}), ...,
(\bm{s}_S, m_{\bm{s}_S}) \right\}$ comprising of $S$ localised targets. To begin, we write, in arbitrary dimensions, the Master equation for the state-level dynamics
\begin{align}
  P(\bm{n}, m, t+1) = \sum_{\bm{n}^{\prime}}\sum_{m^{\prime}=1}^{M}&\bigg[A(\bm{n}, m, \bm{n}^{\prime}, m^{\prime})P(\bm{n}^{\prime}, m^{\prime}, t) \nonumber
  \\ &+  \sum_{i=1}^{S}\delta_{\bm{n}, \bm{s}_i}\delta_{m, m_{\bm{s}_i}}(1-\rho_{\bm{s}_i,
  m_{\bm{s}_i}})A(\bm{s}_i, m_{\bm{s}_i}, \bm{n}^{\prime}, m^{\prime})P(\bm{n}^{\prime}, m^{\prime}, t)\bigg], 
  \label{eq: defective_ME}
\end{align}
where $A(\bm{n}, m, \bm{n'}, m')$ is the transition probability from state $\bm{n}', m'$ to state $\bm{n}, m$, and
$\rho_{\bm{s}_i, m_{\bm{s}_i}}$, $(0 < \rho_{\bm{s}_i, m_{\bm{s}_i}} \leq 1)$ governs the probability of getting
absorbed at defect $\bm{s}_i, m_{\bm{s}_i}$ where $\rho_{\bm{s}_i, m_{\bm{s}_i}} = 1$ represents perfect trapping
efficiency at that site. To proceed, we first consider the $\rho_{\bm{s}_i, m_{\bm{s}_i}} \neq 1$ case. 

Assuming $\sum_{i=1}^{S}\delta_{\bm{n}, \bm{s}_i}\delta_{m, m_{\bm{s}_i}}(1-\rho_{\bm{s}_i, m_{\bm{s}_i}})A(\bm{s}_i,
m_{\bm{s}_i}, \bm{n}^{\prime}, m^{\prime})P(\bm{n}^{\prime}, m^{\prime}, t)$, for any $m$, as a given known function the
formal solution of Eq. (\ref{eq: defective_ME}) is the convolution in space and time of the absorbing propagator with
the known terms \cite{giuggioli2022spatio} and proceeding in the $z$-domain (i.e., taking the generating function) we
find
\begin{align}
 \widetilde{P}^{(a)}(\bm{n}, m, &z) = \sum_{\bm{n}_0^{\prime}}\sum_{m_0^{\prime}=1}^{M}\widetilde{P}_{\bm{n}_0^{\prime}, m_0^{\prime}}(\bm{n}, m, z)P(\bm{n}, m, 0) \nonumber \\
 &- z \sum_{\bm{n}_0^{\prime}}\sum_{m_0^{\prime}=1}^{M}\widetilde{P}_{\bm{n}_0^{\prime}, m_0^{\prime}}(\bm{n}, m, z) \sum_{i=1}^{S}\rho_{\bm{s}_i, m_{\bm{s}_i}}\sum_{\bm{n}^{\prime}}\sum_{m^{\prime}=1}^{M}A(\bm{s}_i, m_{\bm{s}_i}, \bm{n}^{\prime}, m^{\prime})\widetilde{P}(\bm{n}^{\prime}, m^{\prime}, z).
 \label{eq: defect_sol0}
\end{align}
By rearranging, and substituting the $z$-transform of the second term on the RHS of Eq. (\ref{eq: defective_ME}) into
the second term of Eq. (\ref{eq: defect_sol0}) the generating function of the formal solution of Eq. (\ref{eq:
defective_ME}) is found as 
\begin{align}
\widetilde{P}^{(a)}(\bm{n}, m, z)&= \! \sum_{\bm{n}_0^{\prime}}\sum_{m_0^{\prime}=1}^{M}\!\bigg[\widetilde{P}_{\bm{n}_0^{\prime}, m_0^{\prime}}(\bm{n}, m, z)P(\bm{n}, m, 0) \nonumber \\ &+
\sum_{i=1}^{S}\frac{\rho_{\bm{s}_i, m_{\bm{s}_i}}}{\rho_{\bm{s}_i,m_{\bm{s}_i}}\!-\!1}\widetilde{P}_{\bm{s}_i, m_{\bm{s}_{i}}}(\bm{n}, m, \!z)\! \left[\widetilde{P}^{(a)}(\bm{s}_i,m_{\bm{s}_{i}}, \! z)\!-\!P^{(a)}(\bm{s}_i, m_{\bm{s}_{i}}, 0)\right]\! \bigg]\!,
\label{eq: defect_sol1}  
\end{align}
where we have used $\widetilde{P}_{\bm{n}_0, m_{0}}(\bm{n}, m, z)$ to denote any valid defect free propagator. 

An initial condition spatially localised over site $\bm{n}_0$ is given by $P^{(a)}(\bm{n}, 0) = \delta_{\bm{n},
\bm{n}_0}\sum_{m=1}^{M}P^{(a)}(\bm{n}, m, 0)$, where the contribution by each state is $P^{(a)}(\bm{n}, m, 0) =
\delta_{m, m_0}\alpha_{m_0}\left[(1-\rho_{\bm{s}_i,m_{\bm{s}_i}})\delta_{(\bm{n}_0, m_0) \in
\mathdutchcal{S}}+\delta_{(\bm{n}_0, m_0) \notin \mathdutchcal{S}}\right]$. Substitution of this initial condition into
Eq. (\ref{eq: defect_sol1}) leads directly to
\begin{align}
\widetilde{P}^{(a)}_{\bm{n}_0, m_0}(\bm{n}, m, z) &= \alpha_{m_0}\widetilde{P}_{\bm{n}_0, m_0}(\bm{n}, m, z) \nonumber \\ &+ \sum_{i=1}^{S}\frac{\rho_{\bm{s}_i,
m_{\bm{s}_i}}}{\rho_{\bm{s}_i,m_{\bm{s}_i}}-1}\alpha_{m_{\bm{s}_i}}\widetilde{P}_{\bm{s}_i, m_{\bm{s}_{i}}}(\bm{n}, m, z)\widetilde{P}^{(a)}_{\bm{n}_0, m_0}(\bm{s}_i,m_{\bm{s}_{i}}, z),
\label{eq: defect_sol2}
\end{align}
where the notation $\alpha_{m_{0_j}}P_{\bm{n}_0, m_{0_j}}(\bm{n}, m,
t) = \alpha_{m_{0_j}}\bm{e}_m^{\intercal}\cdot \bm{\mathcal{P}}_{\bm{n}_0}(\bm{n}, t)\cdot \bm{e}_{m_{0_j}}$, where $\bm{P}_{\bm{n}_0}(\bm{n}, t) = \bm{\mathcal{P}}_{\bm{n}_0}(\bm{n}, t)\cdot \bm{U}_{m_0}$, is the probability that the walker occupies site and state $(\bm{n}, m)$ given its initial position was $(\bm{n}_0, m_{0_j})$,
which is weighted by $\alpha_{m_{0_j}}$.

Following the standard defect technique procedure \cite{giuggioli2022spatio} i.e., solving via Cramer's rule and taking $\rho_{\bm{s}_i,m_{\bm{s}_i}} \to 1$
for all $(\bm{s}_i,m_{\bm{s}_i})$, we find the state-level propagator with absorption as
\begin{equation}
\widetilde{P}^{(a)}_{\bm{n}_0, m_0}(\bm{n},m, z) = \alpha_{m_0}\widetilde{P}_{\bm{n}_0, m_0}(\bm{n}, m, z) - \sum_{i = 1}^{S}\alpha_{m_{\bm{s}_i}}\widetilde{P}_{\bm{s}_i,m_{\bm{s}_{i}}}(\bm{n}, m, z)\frac{\det[\mathbb{H}^{(i)}(\bm{n}_0, m_0, z)]}{\det[\mathbb{H}(z)]},
\label{eq: defect_sol5}
\end{equation}
where $\mathbb{H}(z)_{l,k} = \alpha_{m_{\bm{s}_k}}\widetilde{P}_{\bm{s}_k, m_{\bm{s}_k}}(\bm{s}_l, m_{\bm{s}_l}, z)$,
$\mathbb{H}(z)_{k,k} = \alpha_{m_{\bm{s}_k}}\widetilde{P}_{\bm{s}_k, m_{\bm{s}_k}}(\bm{s}_k, m_{\bm{s}_k}, z)$ and
$\mathbb{H}^{(j)}(\bm{n}_0, m_0, z)$ is the same but with the $j^{\text{th}}$ column replaced with
$\alpha_{m_0}\left[\widetilde{P}_{\bm{n}_0, m_0}(\bm{s}_1, m_{\bm{s}_1}, z), ..., \widetilde{P}_{\bm{n}_0,
m_0}(\bm{s}_S, m_{\bm{s}_S}, z) \right]^{\intercal}$, i.e, the matricies are comprised of known state-level propagators.

As noted above, the occupation probability over the entire site may be found by performing a summation over all internal states and
using the fact that $\widetilde{P}^{(a)}_{\bm{n}_0}(\bm{n}, z) = \sum_{m=1}^{M}\sum_{j = 1}^{M}\widetilde{P}_{\bm{n}_0,
m_{0_j}}^{(a)}(\bm{n}, m, z) $, and $\widetilde{P}^{(\gamma)}_{\bm{n}_0}(\bm{n}, z) = \sum_{m=1}^{M}\sum_{j=
1}^{M}\alpha_{m_{0_j}}\widetilde{P}_{\bm{n}_0, m_{0_j}}^{(\gamma)}(\bm{n}, m, z) $, where there is no $\alpha_{m_0}$
multiplier on the absorbing case as it is already implemented via the initial condition in Eq. (\ref{eq: defect_sol2}),
one may perform the double summation over both sides of Eq. (\ref{eq: defect_sol5}) to arrive at
\begin{align}
 \widetilde{P}^{(a)}_{\bm{n}_0}(\bm{n}, z) = \widetilde{P}_{\bm{n}_0}(\bm{n}, z) - \sum_{m=1}^{M}\sum_{j=1}^{M}\sum_{i = 1}^{S}\alpha_{m_{\bm{s}_{i}}}\widetilde{P}_{\bm{s},m_{\bm{s}_{i}}}(\bm{n}, m, z)
 \frac{\det[\mathbb{H}^{(i)}(\bm{n}_0, m_{0_j}, z)]}{\det[\mathbb{H}(z)]}.
 \label{eq: correlated_defect}
\end{align}

To help elucidate the notation we note that while the site-state pair defining the trap location is unique, it is possible
to have multiple states within one site containing traps. For example, consider the case of a fully absorbing trap in
site $\bm{s}$ in a two-dimensional square lattice. Since a trap in state $m_{\bm{s}} =1$ of site $\bm{s}$ absorbs a
walker reaching $\bm{s}$ from the left (movement along the West direction) only,
traps must also be placed in sites $m_{\bm{s}}=2, 3$ and 4 to absorb a walker entering the site from any
direction. Hence, for this case $\mathdutchcal{S} = \left\{(\bm{s}, 1), (\bm{s}, 2), (\bm{s}, 3), (\bm{s}, 4) \right\}$. 

\subsection*{First-Passage and Splitting Dynamics}
\label{sec: first-passage} 
Knowledge of the occupation probability for the absorbing propagator allows us to study other transport
statistics such as the first exit or first-passage time \cite{redner2001guide, metzler2014first}. By performing a
spatial summation, we find
the generating function of the survival probability, namely $\widetilde{S}_{\bm{n}_0}(z)=\sum_{\bm{n}}\widetilde{P}^{(a)}_{\bm{n}_0}(\bm{n}, z)$,
and from the known relation $\widetilde{F}_{\bm{n}_0}(\bm{n}, z) = 1 -
(1-z)\widetilde{S}_{\bm{n}_0}(z)$, we obtain the generating function for the
first-passage probability to an arbitrary number of target states as
\begin{equation}
\widetilde{F}_{\bm{n}_0}(\mathdutchcal{S}, z) = \sum_{j=1}^{M}\sum_{i = 1}^{S}\alpha_{m_{\bm{s}_i}}\frac{\det[\mathbb{H}^{(i)}(\bm{n}_0, m_{0_j}, z)]}{\det[\mathbb{H}(z)]}.
\label{eq: multi_tar_FP}
\end{equation}
With multiple targets ($S>1$), the theory also allows the determination of the so-called splitting probability, 
that is the probability of reaching $\bm{s}_k$ and not any of the other targets in $\mathcal{S}$. Following ref. \cite{giuggioli2022spatio}, one finds 
\begin{equation}
 \widetilde{F}_{\bm{n}_0}(\bm{s}_k|\bm{s}_1; ...;\bm{s}_k-1;\bm{s}_k+1; ...; \bm{s}_S , z) = \alpha_{m_{\bm{s}_k}}\frac{\sum_{j=1}^{M}\det[\mathbb{H}^{(k)}(\bm{n}_0, m_{0_j}, z)]}{\det[\mathbb{H}(z)]}.
 \label{eq: multi_tar_splitting}
\end{equation}

Equations (\ref{eq: multi_tar_FP}) and (\ref{eq: multi_tar_splitting}) are the sought-after quantities. They are the
solution for the multi-target first-passage and splitting problems, respectively, for random walks with internal degrees
of freedom, of which the focus of this study, the CRW, is one such example. The expressions are general and accounts for
an arbitrary distribution over $\bm{U}_{{\bm{m}_0}}$ with any number of target states, which may be located in one or
multiple lattice sites.

Depending on the problem Eq. (\ref{eq: multi_tar_FP}) can be simplified to more compact expressions. The simplest case
is that of one target state, which is pertinent to the one-dimensional unbounded CRW. In such a scenario, if the target state is located at
site $s > n_0$, one may put the target at $(s, 1)$ since the walker may only arrive at $s$ from the left, as one can
evince by looking at Fig. \ref{fig: lattice_schematic}(b). With a single target state, Eq. (\ref{eq: multi_tar_FP}) can be written as
\begin{equation}
\widetilde{F}_{\bm{n}_0}(\bm{s}, m_{\bm{s}}, z) = \frac{\sum_{j=1}^{M}\alpha_{m_{0_j}}\widetilde{P}_{n_0, m_{0_j}}(\bm{s}, m_{\bm{s}}, z)}{\widetilde{P}_{\bm{s}, m_{\bm{s}}}(\bm{s}, m_{\bm{s}}, z)},
\label{eq: single_tar_FP}
\end{equation}
which can be further reduced to the known first-passage quotient for random walks with internal degrees of freedom $\widetilde{F}_{\bm{n}_0,
m_{0_k}}(\bm{s}, m_{\bm{s}}, z) = \frac{\widetilde{P}_{\bm{n}_0, m_{0_k}}(\bm{s}, m_{\bm{s}}, z)}{\widetilde{P}_{\bm{s},
m_{\bm{s}}}(\bm{s}, m_{\bm{s}}, z)}$ 
\cite{rws_on_latticesIII,larralde2020first} when the initial condition is localised at state $m_{0_k}$.

For the bounded one-dimensional lattice, where the walker may enter from either direction, the case with two target states at the same spatial site, e.g., $\mathcal{S} = \{(s, 1), (s,2)\}$, is
particularly relevant. Here, Eq. (\ref{eq: multi_tar_FP}) can be expanded out (see Sec. VII of ref. \cite{supp_mat}) to find 
\begin{small}
\begin{align}
 & \widetilde{F}_{n_0}(s, z)   = \nonumber \\ &\frac{\alpha_{1}\! \left(\!\widetilde{P}_{n_0, 1}(s, 1,z) \left[\widetilde{P}_{s, 2}(s, 2,z)\!-\!\widetilde{P}_{s, 1}(s, 2,z)\right]\!+\!\widetilde{P}_{n_0, 1}(s, 2,z)\! \left[\widetilde{P}_{s, 1}(s, 1,z)\!-\!\widetilde{P}_{s, 2}(s, 1,z)\right]\!\right)}{\widetilde{P}_{s, 1}(s, 1,z)\widetilde{P}_{s, 2}(s, 2,z)-\widetilde{P}_{s, 1}(s, 2,z)\widetilde{P}_{s, 2}(s, 1,z)\!} \nonumber \\ 
 &+\frac{\alpha_2 \!\left(\widetilde{P}_{n_0, 2}(s, 1,z) \left[\widetilde{P}_{s, 2}(s, 2,z)\!-\!\widetilde{P}_{s, 1}(s, 2,z)\right]\!+\!\widetilde{P}_{n_0, 2}(s, 2,z)\! \left[\widetilde{P}_{s, 1}(s, 1,z)\!-\!\widetilde{P}_{s, 2}(s, 1,z)\right]\right)}{\widetilde{P}_{s, 1}(s, 1,z)\widetilde{P}_{s, 2}(s, 2,z)-\widetilde{P}_{s, 1}(s, 2,z)\widetilde{P}_{s, 2}(s, 1,z)\!},
 \label{eq: one_whole_site_FP}
\end{align}
\end{small}
\begin{figure}
 \centering
 \includegraphics[width = \textwidth]{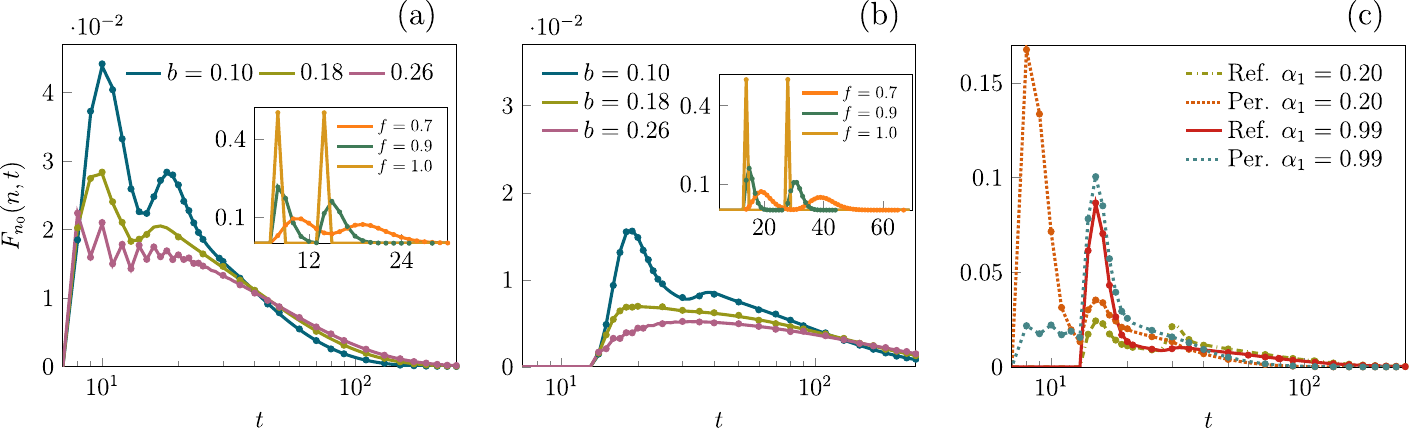}
 \caption{(Colour Online).  First-passage probability in a finite one dimensional domain of size $N = 22$ with
 initial condition $n_0 = 8$ and a target at $s=N$ obtained via a numerical $z$-inverse transform
 \cite{abate1992numerical, abate2000introduction} of Eq. (\ref{eq: one_whole_site_FP}), with the periodic propagator,
 Eq. (\ref{eq: matrix_periodic}), used for panel (a) and the squeezed reflective propagator, Eq. (\ref{eq:
 1d_reflective}), used for panel (b). We note here the use of the two state first-passage in the squeezed reflective
 propagator is needed as it is constructed via the periodic dynamics. In the main figure of panel (a) and (b) we use
 $\alpha_1 = \alpha_2 = 1/2$, $f=0.65$ and we vary $b$, while in the insets we set $b=0$ and vary $f$. In (c) we fix
 $f = 0.82$, $b = 0.08$ and study how the first-passage probability changes as we change $\alpha_1$. The lines are
 again obtained from Eq. (\ref{eq: one_whole_site_FP}) with the appropriate propagator. Across all three panels the
 dots represent the ensemble average of $10^6$ stochastic simulations.}
 \label{fig: 1D_fp}
\end{figure}
where we have lightened the notation with $\widetilde{F}_{n_0}(s, z) = \widetilde{F}_{n_0}(\{(s, 1), (s,2)\} , z)$,
$\alpha_{m_{0_1}} =\alpha_1$ and similarly for the second state weighting. We present an alternative derivation via a generalised
renewal equation and some special cases of Eqs. (\ref{eq: single_tar_FP}) and (\ref{eq: one_whole_site_FP}) in \ref{app: defect}.

A prominent feature of the first-passage probability for a bounded CRW is the appearance of multiple peaks. While in periodic domains bi-modal
\cite{sarvaharman2020closed,holhouse2023first,bonomo2021first} and multi-modal \cite{giuggioli2023multi} first-passage distributions for Markov
LRWs have been reported, in those cases the appearance of multiple peaks is a consequence of
the periodicity of the boundaries and the presence of a bias for which there are repeated occasions when a
fraction of the trajectories may hit a target. This is not the case for the CRW, as we
observe multiple peaks with periodic as well as reflecting domains.

With correlated motion we may in fact identify trajectories that travel both towards and away from the target. With the
existence of two peaks documented in Fig. \ref{fig: 1D_fp}, the first peak corresponds to those trajectories, which
travel persistently towards the target and the second peak represents those that travel away and later come back via,
either, a bouncing reflection or the periodic boundary. Since the peaks correspond to the two sets of trajectories, the
initial weighting controls which peak contains larger probability, while the overall level of persistence governs their
collective height. Furthermore, for a given domain size, the path that reaches the target via the periodic boundary is
always shorter than the path that bounces back via the reflective boundary. Hence, the time at which the second peak
occurs is always lower in periodic domains when compared to its reflective counterpart, which may also be seen by
comparing panels (a) and (b). 

In panel (c), the effect of the individual weighting on the height of the peaks is seen explicity, where we have placed
the target at $s = 22$ and the walker intially at $n_0 = 8$. With a smaller proportion of trajectories set to move in
the right direction at $t=0$ ($\alpha_1 = 0.2$), the periodic case has a large peak at short time, which correspond to
the larger fraction of trajectories that move towards the target via the left boundary. The smaller peak at later times
is due instead to those walker paths that first headed right. In contrast, with a reflecting boundary at $n=1$ we do not
see a peak until those trajectories that initially headed right reach the target, with a second smaller peak appearing
later once the trajectories initially travelling left are reflected back from $n=1$. In contrast, we observe a very
different dynamics when we weigh heavily the walker trajectories so that the majority starts moving right ($\alpha_1 =
0.99$). In such a case we see far less deviation between the two boundary conditions as most persistent trajectories do
not encounter the left boundary. We note also that the oscillatory behaviour at short times seen in panels (a) and (c)
is a simple consequence of a low sojourn probability as in $c = 0$ limit, one creates parity issues
\cite{kac1947random}. While here, we are away from this limit, with a low $c$ in short times
the chance of being on an even site at even times (given an even initial condition) is still higher. The oscillations dampen with time as the probability of
undertaking a sojourn (which breaks the parity) increases.

\begin{figure*}[h]
    \centering
    \includegraphics[width=0.9\textwidth]{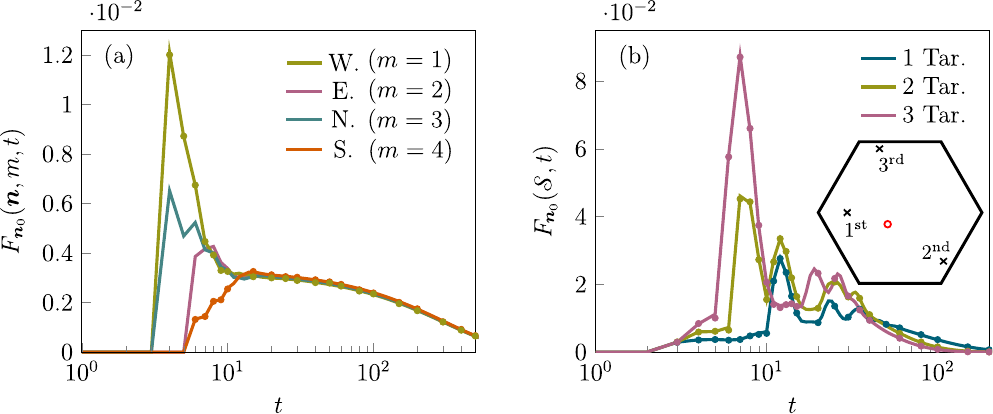}
    \caption{(Colour Online). First-passage probabilities of correlated random walks in two-dimensional periodic
    domains. In panel (a), through a numerical inversion of Eq. (\ref{eq: single_tar_FP}), we plot the directional
    first-passage probability for each of the four movement directions in a square periodic lattice of length $N=8$ to a
    single site at $\bm{s} = (7, 3)$. For example, in order to reach $(\bm{s}, m_{\bm_{s}}=1)$ the walker has to travel
    West, to reach $(\bm{s}, m_{\bm_{s}}=3)$ the walker has to travel North, etc. The initial condition is at $\bm{n}_0
    = (4, 2)$ with equal probability to move in any direction and with $f = 0.42$, $b = 0.04$ and $\ell = 0.18$. Panel
    (b) depicts results on a hexagonal lattice and shows, via a numerical inversion of Eq. (\ref{eq: multi_tar_FP}), the
    first-passage probability to a set $\mathdutchcal{S}$ of one, two, and three targets in a periodic hexagon of
    circumradius $R=4$ ($\Omega = 61$) at locations $\bm{s}_1 = (-3, 3, 0)$, $\bm{s}_2 = (4, -1, -3)$ and $\bm{s}_3 =
    (-3,-1, 4)$, which we add sequentially starting from $\bm{s}_1$. Since the walker can enter from any direction, all
    the internal states in $\bm{s}_i$ are targets. The initial condition in this case is just off centre at $\bm{n}_0 =
    (0, 1, -1)$ and $f = 0.80 $, $b = 0.02$, $\ell = 0.01$. The approximate location within the domain of the initial
    condition and target sites are shown via the schematic in the inset, with the initial condition a red open circle
    and the targets shown with crosses. Dots in (a) and (b) are the result of $5\cdot 10^{6}$ and $10^{6}$ stochastic
    simulations, respectively.}
   \label{fig: high_dim_FP}
   \end{figure*}
We now turn our attention to the periodic square lattice and study the probability that the walker reaches a site for
the first time moving in a specific direction, and we call this quantity the \textit{directional} first-passage
probability. This may be useful in a context where the target is only visible from a certain direction and is obtained
by placing a target in only one specific state in a lattice site. We show, in Fig. \ref{fig: high_dim_FP}(a), the cases
in which the walker reaches the target travelling in each of the four directions. We note that the directional
first-passage is a normalised probability distribution as it does not exclude trajectories that first reach the target
via an alternate direction. One may ask for the probability of reaching a site for the first time travelling in a specific
direction conditioned on the walker having not first reached for the first time from any other direction via the splitting 
probability in Eq. (\ref{eq: multi_tar_splitting}). 

With the initial condition at site $\bm{n}_0 = (4, 2)$, to reach $\bm{s}=(7,3)$ for the first time at $t=4$ or $t=5$,
the walker has to enter the site travelling either North or West. However, travelling West gives a higher directional
first-passage at these times because the walker travelling North into the target at $t=4$ must have made a left-hand
turn from $\bm{n}=(7,2)$ at the preceding time. In contrast, to access the site travelling West at an identical time,
the walker has three possible trajectories. After $t=4$, the West directional first-passage probability sees a sharp
decrease, which is a consequence of this timescale being dominated by unlikely trajectories that have had to make a
sojourn or an even number of backtracks between the intial condition and the target. Thus, it is here that the Eastern
directional first-passage becomes more likely as this timescale coincides with the time in which the Eastern travelling
persistent trajectories are able to reach the target. 

To analyse an example with multiple absorbing targets (Fig. \ref{fig: high_dim_FP}(b)), we consider the hexagonal
lattice and place the walker initially in the site $\bm{n}_0 = (0, 1, -1)$, just to the South-West of the centre of the
domain, with uniform initial weighting $\alpha_i=1/6$ for $i=1,...,6$. We place three targets, the first to the West of
the initial condition, the second to the South-East and the third to the North (see the caption of Fig. \ref{fig:
high_dim_FP} for the exact coordinates). Since the walker can enter any target via all six directions, the three target
case requires 18 target states. 

The sustained oscillations, which have only been seen in a small number of systems, including quantum walks
\cite{kulkarni2023first}, biased random walks in hexagonal domains \cite{giuggioli2023multi} and other non-Markov
systems \cite{verechtchaguina2006first} are the natural, higher-dimensional analogue of the bimodality seen in the
one-dimensional systems, where again, the peaks correspond to the timescales for which the walker moves away from the
target as well as those trajectories that miss the target and must travel round the entire domain. Naturally, as the
indirect trajectories take over and more trajectories reach the targets, the oscillations dampen and decay to
zero \cite{giuggioli2023multi}. 

While the multi-modal dynamics is present with one, two and three targets, the placement of targets has a strong effect
on the height and location of the peaks. In Fig. \ref{fig: high_dim_FP}(c), since the initial condition is slightly
off-centered, the high persistence of the walker means the first target (placed directly along the horizontal
circumradius) is likely bypassed leading to low search success until a trajectory, which first set off in the opposite
direction traverses the boundary. When a second and third target are added just away from the circumradius, the shift
imposed by the boundary condition aids the persistent walker in exploring new sections of the domain, which leads to the
emergence of the large peak in the two and three target cases. Finally, the third target then kills the second peak
since the placement of the third target blocks the route many persistent trajectories took to $\bm{s}_1$, subsequently
increasing the height of the first peak.  

\section{Summary Statistics and the Mean First-Passage Time}\label{sec: MFPT}
With the generating function of the first-passage time distribution to multiple targets, it becomes an algebraic
exercise to extract the $n^{\text{th}}$ moments, and in particular its first moment, the mean first-passage time (MFPT),
$\mathdutchcal{F}_{\bm{n_0}\to \{\mathdutchcal{S}\}} = \left.\frac{\partial \widetilde{F}_{\bm{n_0}}(\mathdutchcal{S},
z)}{\partial z}\right|_{z=1}$.  

When the initial condition is uniformly distributed across the site, the structure of Eq. (\ref{eq: multi_tar_FP}) is
very similar to the Markov first-passage expression for multiple targets \cite{giuggioli2022spatio}, and one
differentiates $\widetilde{F}_{n_0}(\mathdutchcal{S}, z) $ to find
\begin{equation}
\mathdutchcal{F}_{\bm{n_0}\to \{\mathdutchcal{S}\}} = \frac{\det[\mathbb{T}_{0}]}{\det[\mathbb{T}_{1}]-\det[\mathbb{T}]}.
\label{eq: MFPT}
\end{equation}

The matrices in Eq. (\ref{eq: MFPT}) are made up of the MFPT between localised sites $\mathdutchcal{F}_{\bm{n}_0,
m_{0}\to\bm{s}_i, m_{\bm{s}_i}}$. More precisely, $\mathbb{T}_{i,i} = 0$, $\mathbb{T}_{i,k} =
\mathdutchcal{F}_{\bm{s}_k, m_{\bm{s}_k}\to\bm{s}_i, m_{\bm{s}_i}}$ ($i\neq k$), $\mathbb{T}_{{0}_{i,k}} = \mathbb{T}_{i,k}
-(2d)^{-1}\sum_{j = 1}^{2d}\mathdutchcal{F}_{\bm{n}_0, m_{0_j}\to\bm{s}_i, m_{\bm{s}_i}}$ and $\mathbb{T}_{1_{i,k}} =
\mathbb{T}_{i,k} - 1$. To find these quantities one evaluates expressions of the form $ \mathdutchcal{F}_{\bm{n}_0,
m_{0}\to\bm{s}_i, m_{\bm{s}_i}} = \frac{\partial}{\partial z}\left.\frac{\widetilde{P}_{\bm{n}_0, m_0}(\bm{s},
m_{\bm{s}}, z)}{\widetilde{P}_{\bm{s}, m_{\bm{s}}}(\bm{s}, m_{\bm{s}}, z)}\right|_{z=1}, $ and we give details and
explicit quantities in Sec. VIII of ref. \cite{supp_mat}. 
\begin{figure*}[ht]
 \centering
 \includegraphics[width=\textwidth]{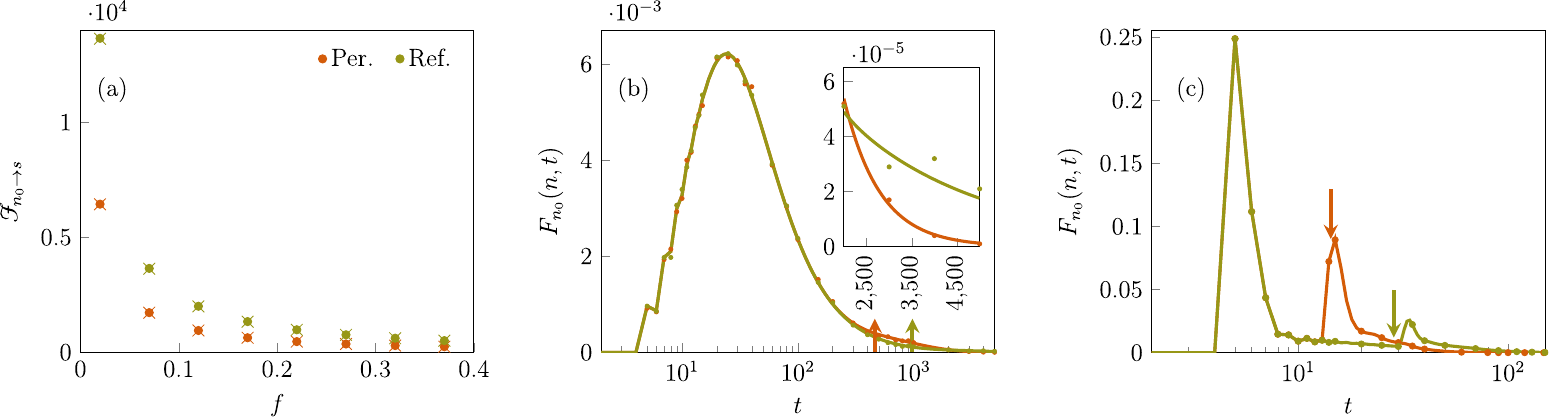}
 \caption{(Colour Online). First-passage probabilities to a single target site and their related mean first-passage
 times in one-dimensional lattices with periodic (orange) and reflecting (green) boundary conditions. The meaning of
 the colours, given explicitly in the legend of panel (a), extends over all three plots. In panel (a) we study the
 MFPT in an $N=35$ lattice for a walker starting at $n_0=30$ with equal weighting of the states in the
 anti-persistent regime ($b > f$) with the target at $N = s$, by fixing $b=0.6$ and varying $f$. Dots represent Eq.
 (\ref{eq: MFPT}) evaluated with Eq. (S74) of ref. \cite{supp_mat}, for the periodic lattice,
 and Eq. (S75) of ref. \cite{supp_mat}, for the reflecting lattice. Crosses represent the
 result of $10^6$ stochastic simulations. For the same domain and initial conditions, panel (b) shows the full
 first-passage probability when $f = 0.22$ and $b = 0.6$. The solid lines are a numerical inversion of Eq. (\ref{eq:
 single_tar_FP}), the dots are the result of $10^6$ stochastic simulations and the corresponding MFPTs are shown with
 the arrows. The inset of panel (b) is a blown-up look at the first-passage probabilities in the tails for $t \in
 [2000, 5000]$. Panel (c) displays the first-passage probability in the persistent regime ($f=0.86$, $b=0.06$) where
 $N = s = 19$ and $n_0 = 14$. Solid lines are generated from the numerical inversion of Eq. (\ref{eq:
 single_tar_FP}), the dots are obtained from $10^6$ stochastic simulations and the arrows indicate the MFPT.}
 \label{fig: MFPT}
\end{figure*}

In Fig. \ref{fig: MFPT}(a), we plot the MFPT for anti-persistent walkers $(f<b)$ as a function of
$f$, and a given $b$. The increase of $f$ has two effects on the walker's ability to spread, namely it decreases both
the chance of a sojourn and the probability of turning backwards. This leads to an exponential decrease in the mean
search time as the dynamics approaches the diffusive regime. Even though the target is placed only five lattice sites
away from the initial condition, the walker takes, on average, 13651 time steps to complete its search
in the strong anti-persistent regime $(f=0.02$). This is due to, in part, those trajectories that have moved far from
the target and subsequently have got `stuck' in cycles of constant backtracking. 

It is in panel (b) that one further sees the effect of those trajectories that get stuck far from the target as we
display the first-passage probability in another anti-persistent regime where ($f = 0.22$, $b = 0.6$). Since, the
initial condition is close to the target and the exploration is extremely slow, there is little difference between
periodic and reflecting domains at short and intermediate times. Differences surface at long times since in the periodic
domain those walkers that have explored the domain in its entirety may find a short route back to the target, while in
the reflective domain walkers are forced to travel back the whole length of the domain. It is these trajectories that
contribute to longer tails in the first-passage probability and subsequently the larger MFPT in the reflective case. We
note that, similarly to the persistent case, the local minima (and subsequent kinks) seen at short times is again due to
the low sojourn probability making the probability of a first-passage event at even times lower. 

Figure \ref{fig: MFPT}(c) draws attention to the importance of studying the entire first-hitting distribution in the
presence of persistent motion ($f>b$). When first-passage processes display multiple peaks in their time-dependent
probability, the MFPT may in fact provide misleading values of the time it takes to hit a target. An extreme example is
the ballistic limit, which has two distinct and narrowly defined hitting times given at, for example,
$\mathdutchcal{F}_{n_0 \to s}^{(r_s)} = s-1$ $(s>n_0)$, $\mathdutchcal{F}_{n_0 \to s}^{(r_s)} = N-s$ $(s<n_0)$ (for
$\alpha_1=\alpha_2$). In such a case the MFPT  would give a temporal value in between the two during which the
probability of a first-passage event is identically zero. As shown in panel (c), this aspect may also appear away from the ballistic limit, in the regime of high persistence.

We now make connection with ref. \cite{tejedor2012optimizing}, where they consider the global mean first-passage time
(GMFPT), $\mathdutchcal{G}_{\{\mathdutchcal{S}\}}$, of a persistent random walk. By performing a summation over $\bm{n}_0$
of Eq. (\ref{eq: MFPT}), and repeatedly using the multi-linearity property of the determinant, we find the GMFPT as 
\begin{equation}
 \mathdutchcal{G}_{\{\mathdutchcal{S}\}} =  \frac{\det[\mathbb{G}]}{\det[\mathbb{T}_{1}]-\det[\mathbb{T}]},
 \label{eq: GMFPT}
\end{equation}
where $\mathbb{G}_{i,j} = \mathbb{T}_{i,j} - (2dN)^{-1}\sum_{\bm{n}_0}\sum_{j = 1}^{2d}\mathdutchcal{F}_{\bm{n}_0,
m_{0_j}\to\bm{s}_i, m_{\bm{s}_i}}$, while $\mathbb{T}_{1}$ and $\mathbb{T}$ are defined above. In certain situations
such as the one target case in periodic domains treated
in \cite{tejedor2012optimizing}, the GMFPT is independent of the target location, however in general this is not the
case. In fact, $\mathdutchcal{G}_{\{\mathdutchcal{S}\}}$ is generally dependent on the location of the target(s), the
size of the domain and the movement parameters. 
\begin{figure}[h]
 \includegraphics[width = 0.9\textwidth]{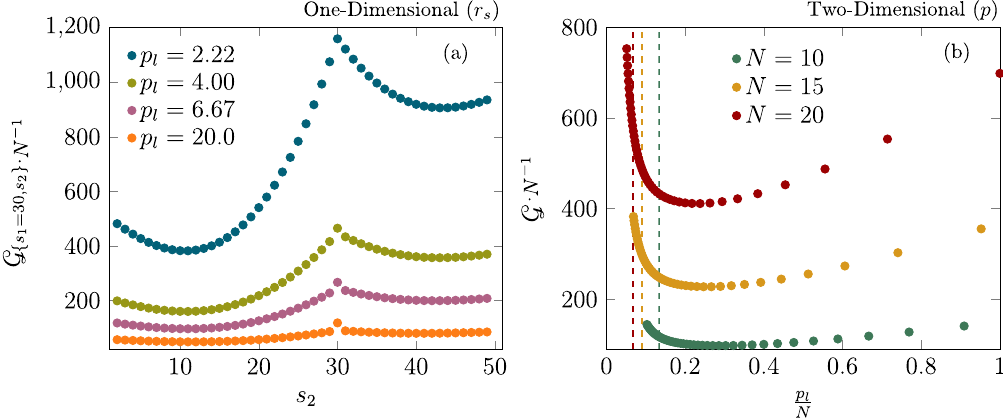}
 \caption{Global mean first-passage time, rescaled by $N$, plotted using Eq. (\ref{eq: GMFPT}). (a) By employing Eq. (S65) in Eq. (\ref{eq: GMFPT}), we show the case of two targets in a one-dimensional reflective domain
 with one fixed target at $s_1 = 30$ as a function of the placement of the second target $s_2$ for varying
 persistence lengths. (b) Using Eq. (S66) in Eq. (\ref{eq: GMFPT}) we treat the case of an $N\times N$ periodic
 domain and plot the GMFPT as a function of the rescaled persistence length for different domain sizes. In panel (a)
 when $s_1 = s_2$ the calculation reduces to a one target case, which explains the slight kink in the plots for this
 case. In (b) because of the periodicity of the lattice there is no dependence on the target location and we drop the
 subscript $\{\mathdutchcal{S}\}$ on the $y$-label. To vary $p_l$ in this case, we fix $b = 0.01$ and vary $f \in
 [0.02, 0.96]$ from low to high values, while simultaneously varying $\ell \in [0.01, 0.58]$ from high to low values
 to ensure normalisation with $c^{(2)} = 0.01$. The vertical dashed lines show $p_{l}/N$ values for the diffusive
 cases.}
 \label{fig: GMFPT}
\end{figure}  

To analyse this quantity further we consider the quantity of persistence length $p_l$, defined in ref.
\cite{tejedor2012optimizing} as the mean number of steps in a ballistic excursion. Since our formalisms allow for the
inclusion of a sojourn probability, we slightly redefine this quantity as the mean excursion time before a turn takes
place. More precisely, this is the mean of a geometric distribution i.e., $p_l =
\sum_{t=1}^{\infty}t\left[1-\left(f+c^{(\mathfrak{d})}\right)\right]\left(f+c^{(\mathfrak{d})}\right)^{(t-1)}$, which is
easily evaluated as $p_l = \left[1-\left(f+c^{(\mathfrak{d})}\right)\right]^{-1}$. Using the knowledge that the movement
parameters sum up to one (see Sec. \ref{sec: 2D}), one may rewrite this as
\begin{equation}
 p_l = \frac{1}{b + (Z-2)\ell}.
 \label{eq: persistence_length}
\end{equation}
In Eq. (\ref{eq: persistence_length}) as $p_l$ is defined through the coordination number of the lattice, $Z$, it means
that it is valid on all types of lattices. In the diffusive case ($b = \ell = q/Z$) $p_l = Z\left[q(Z-1)\right]^{-1}$,
where $0 < q \leq 1$ is the probability of changing lattice site, with $q=1$ a walker which always moves \cite{LucaPRX}.
In such a case on the square lattice, the movement probabilities are $q/4$ to move in any direction and $1-q$ to remain
on the current lattice site, implying $p_l = 4(3q)^{-1}$. 

In Fig. \ref{fig: GMFPT}, we plot, in panel (a), the GMFPT in a reflecting domain with two targets, with one fixed
target at $s_1 = 30$ and varying the position of the second $s_2$. In panel (b), we plot a 2D periodic domain with one
target and obtain minima as a function of $p_l$ as observed also in ref. \cite{tejedor2012optimizing}. To do so, we fix
a very low $b = 0.01$ and vary $p_l$ by changing both $f$ and $\ell$. 

In the one-dimensional reflection case, panel (a), an increase in the persistence length lowers the MFPT for all target
placement, which is expected due to an increase in persistence always aiding coverage in one-dimension. We also see strong
dependence on the placement of the second target with the higher GMFPTs occurring when $s_1$ and $s_2$ are close
together. We note the higher GMFPT when $s_2$ is to the right of $s_1$. To explain this feature, consider that when $n_0 < s_1 <
s_2$ or $s_2 < s_1 < n_0$, $s_2$ is inaccessible. With the static target at $s_1 > N/2$, the case where
$n_0 < s_1 < s_2$ leads to, on average, longer MFPTs and also occupies a larger proportion of the contributions to the
GMFPT than the $s_2 < s_1 < n_0$ case. If $s_1 = N/2$, one would see a symmetric plot around $s_2$.

In panel (b), we recover similar behaviour to the results reported in ref. \cite{tejedor2012optimizing}, whereby the
average search time may be minimised as a function of the persistence length. With larger domains, the minima occurs at
a smaller value of $p_l$, a feature which is expected as the average number of sideways steps required to reach the
target grows with the domain size. In all cases the minima falls within the persistent regime.

\section{Iterative Procedure and Comparison Between Reflective Interactions}\label{sec: numerical} 
In the preceding sections we have exploited the analytical expressions of propagator generating functions. For the cases
in which propagators are not known in closed form we present here an iterative procedure to find the first-passage
probability. While the procedure is valid in arbitrary dimensions, for notational ease we present it here
explicitly for the one-dimensional case, and use it as a tool to
make comparisons between the two forms of reflection mentioned in Sec. \ref{sec: reflective}. We differentiate
between the two reflections based on whether an interaction with the boundary invokes a reversal in the direction of movement, which we
call the `bouncing' case, or if the movement direction persists, which we call
the `bunching' case (see panels (c) and (d) in Fig. \ref{fig: lattice_schematic} for pictorial comparison). For an in
depth appreciation of the differences between the two boundary conditions we also compare here the occupation
probability and the mean square displacement (MSD) in the two cases. 

We begin by rewriting the Master equation, Eq. (\ref{eq: matrix_master_eq}), for a finite system using transition
matrices that describe the possible movement between lattice sites and link the different internal states. In other
words we represent spatial jumps for which the internal state does not change as well as those for which changes between
internal states occur. Noting that the probability across the entire site is given by
$\bm{P}_{n_0}(t) = \bm{P}_{n_0}(1,t) + \bm{P}_{n_0}(2,t)$, where $\bm{P}_{n_0}(1, t)$
is an $N \times 1$ column vector of the occupation probabilities of state 1 in each lattice site at time $t$, we have the Master equation
\begin{equation}
\begin{aligned}
 \bm{P}_{n_0}(1, t+1) &= \mathbb{D}\cdot \bm{P}_{n_0}(1, t) + \mathbb{E}\cdot \bm{P}_{n_0}(2, t), \\
 \bm{P}_{n_0}(2, t+1) &= \mathbb{F}\cdot \bm{P}_{n_0}(1, t) + \mathbb{G}\cdot \bm{P}_{n_0}(2, t),
\end{aligned}
\label{eq: iterative_ME}
\end{equation}
where $\mathbb{D}$ represents the transition probabilities from state 1 to state 1, $\mathbb{E}$ from state 2 to state
1, $\mathbb{F}$ from state 1 to state 2 and $\mathbb{G}$ encoding the transitions from state 2 to state 2. Due to the
finite space, each matrix is of size $N\times N$ and as each matrix only governs a subset of the movement possibilities,
they are sparse and have the following forms: $\mathbb{D}_{i,i} = c$, for $i \neq 1, N$, $\mathbb{D}_{i, i+1} = f$ and
$\mathbb{D}_{1,1} = \delta + c$, $\mathbb{D}_{N,N} = \sigma + c$; $\mathbb{E}_{i, i+1} = b$ and $\mathbb{E}_{1, 1} =
\omega$; $\mathbb{F}_{i, i-1} = b$ and $\mathbb{F}_{N, N} = \omega$ and $\mathbb{G}_{i, i-1} = f$ and $\mathbb{G}_{1,1}
= \sigma + c$, $\mathbb{G}_{N,N} = \delta + c$, where $\delta$, $\sigma$, $\omega$ are dependent on the chosen boundary
conditions. With initial conditions $P_{n_0}(n, 1, 0) = \alpha_1$, $P_{n_0}(n, 2, 0) = \alpha_2$, that is
$\bm{P}_{n_0}(1, 0) = \alpha_{1}\bm{e}_{n_0}$ and $\bm{P}_{n_0}(2, 0) = \alpha_{2}\bm{e}_{n_0}$, it is an iterative task
to find the occupation probability at time $t$.  
\begin{figure*}[b]
    \centering
   \includegraphics[width=0.9\textwidth, height=0.57\textwidth]{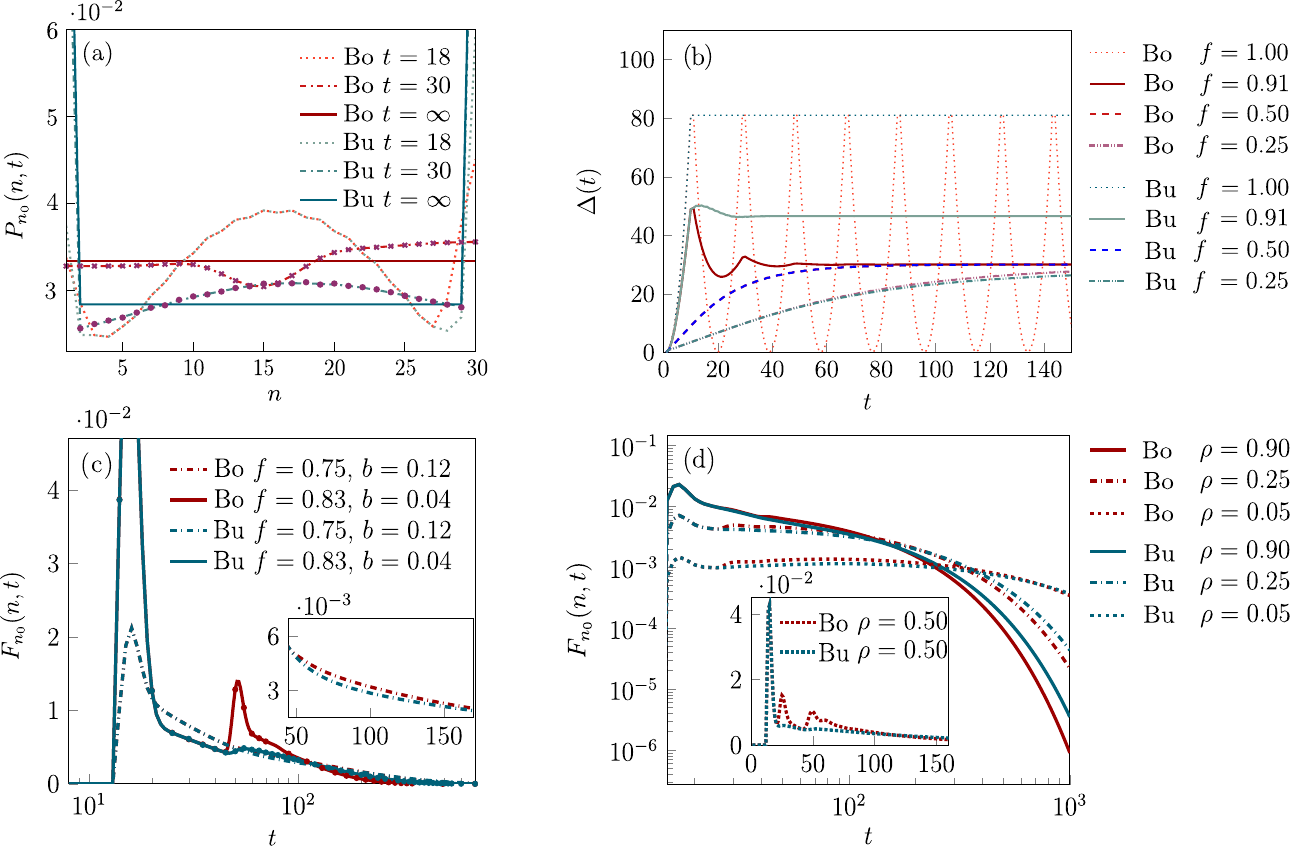}
    \caption{(Colour Online).  A comparison of the CRW dynamics between two forms of reflective boundaries, the `bunching'
    and `bouncing' interaction. Across all four panels, the blue-green colours represent the `bunching' boundary condition,
    while the red-purple colours represent the `bouncing' condition.  All lines are found via iteratively solving Eq.
    (\ref{eq: iterative_ME}), while crosses are obtained using the analytic propagator in Eq. (\ref{eq:
    matrix_reflective1}) and all dots are the result of $10^6$ stochastic simulations. Panel (a) displays the occupation
    probability at varying time steps with $f = 0.75$, $b = 0.12$ with $n_0 = 16$ and $N = 30$, and panel (b) shows the MSD
    for a variety of persistence levels with ($b=1-f$) in an $N = 19$ lattice with  $n_0 = 10$, that is with the initial
    condition in the middle of the domain. Panel (c) shows the first passage distribution to site $s=N=30$, while its inset is
    a close up of the region of discrepancy between the two boundary conditions when $f=0.75$, $b=0.12$. Panel (d) shows
    the first-passage probability to a partially absorbing target placed at $s = N-5$ ($N = 30$), with initial condition
    $n_0 = 12$ and $f=0.75$, $b=0.12$. The inset displays the same type of information, but the movement probabilities are
    now $f = 0.8$ and $b = 0.04$.}
    \label{fig: iter_analysis}
   \end{figure*}

For the `bunched' case, $\delta = b$, $\sigma = f$, $\omega = 0$, while for the `bouncing' case, $\delta = b$, $\omega =
f$ and $\sigma = 0$. To include absorbing traps one sets the outgoing probabilities of a trap state $(s, m_s)$ as
$(1-\rho_{s,m_s})A(s, m_s, n, m)$ where $\rho_{s,m_s}$ is the partial absorption probability at $(s, m_s)$ and with
$A(s, m_s, n, m)$ representing the elements of the transition matrix encoding movement out of $(s,m_s)$.  As an illustration, consider a trap
at site $s=N$ and state $m_s = 1$. In such a case we multiply the $N^{\text{th}}$ column of $\mathbb{D}$ and
$\mathbb{F}$ by $(1-\rho_{N,1})$. The first-passage is obtained via $F_{n_0}(n, t) =
S_{n_0}(t-1)-S_{n_0}(t)$ with $F_{n_0}(n, 0)=0$, where $S_{n_0}(t)=\sum_{n=1}^N P_{{n}_0}(n,t)$. 

In Fig. \ref{fig: iter_analysis}(a), we plot the occupation probability for the bouncing and bunching reflective
boundary conditions at varying time steps with uniform initial weighting $\alpha_1=\alpha_2=1/2$. Even at short times
one sees higher probability accumulating at the boundaries in the `bunched' case, with the asymmetry in heights due to
the initial condition being placed slightly to the right. At longer times, with the chance of more boundary interaction,
the differences become more stark, which pertains until we recover the expected steady states, i.e., uniform probability
for the `bouncing' case and increased probability at the boundaries for the `bunched' \cite{malakar2018steady}.

In panel (b), we compare the MSD, $\Delta(t) = \langle |n(t) - n_0|^2\rangle $, with $\langle \cdot \rangle$ denoting
the ensemble average, of a walker confined between either boundary. In the diffusive limit, both boundary conditions are
equivalent, which is confirmed by the identical MSD in this case. However, with increasing persistence one sees striking
differences between the MSD after boundary interactions have occurred. As predicted from the continuous space-time
equivalent analysis for the bouncing case \cite{giuggioli2012predicting}, with high positive correlation the MSD shows
oscillatory dynamics. In the ballistic limit, these oscillations persist for all times as the walker is guaranteed to
return to its initial condition periodically every $N$ timesteps. However, with non-zero backtracking probability, the
MSD displays oscillations at short times, which subsequently decay as the result of the different times at which trajectories return to
the starting location. Then, due to the uniform steady state when $b \neq 0$ the MSD eventually saturates to $N^2/12$. In contrast, the bunching
case sees a clear maxima before quickly lowering to a saturation value, which is dependent on the level of persistence
in the system. Since the shape of the steady state is dependent on the level of persistence in the system (where higher
persistence levels leads to larger values of the steady state at the boundary \cite{malakar2018steady}), the MSD
saturates at higher value for larger $f-b$ with the maximal displacement remaining for all times in the ballistic limit. 

In panels (c) and (d), we study the first-passage and as expected, as we increase the possibility of the walker
interacting with the boundary, more differences emerge in the first-passage dynamics. To illustrate,
in the fully absorbing case, seen in panel (c), differences in the dynamics emerge only at intermediate times when the
`bouncing' walkers display a much higher second peak compared to the `bunching' ones as they are allowed to reverse
their trajectory back towards the target. Instead, in panel (d), where the target is partially absorbing and away from
the boundary, the walker may now interact with both the left and right boundary and two timescales emerge where differences occur. The first
timescale is due to the trajectories overcoming the target and subsequently being reflected from the right boundary,
while the second results from the trajectories that are reflected from the left. Thus, it is the second timescale that
is present for all values of $\rho$ (including $\rho = 1$), while the first only emerges for a smaller absorption
probability. As we lower $\rho$ further and enter the reaction limited regime (when the timescale for a reaction to
occur is much higher than the transport time to reach the target \cite{kenkre2006exciton}) the two timescales lengthen
and begin to merge. For example, when $\rho = 0.05$, one sees a sustained region
where the `bouncing' case has a higher first-passage probability as the walker subject to this boundary condition
continuously passes over a target with low adsorption many times.

\section{Conclusion}\label{sec: conclusion}
First-passage processes in Markovian LRWs have been studied extensively over the last few years in a multitude of
settings \cite{das2022discrete, das2023dynamics,sarvaharman2022particle, sarvaharman2020closed, marris2023exact,
LucaPRX, benichou2014first, condamin2005first, godec2016first, godec2016universal}. On the other hand, for non-Markovian
LRWs a general formalism to derive the first-passage dynamics has been lacking. The simplest non-Markovian LRW
extension, the CRW, has received attention, but much of the analysis on first-passage statistics has relied either on
time-consuming computational methods \cite{prasad2006searching} or on the global MFPT~\cite{tejedor2012optimizing}, with very
few results known analytically. Here, we have constructed a general formalism to determine the CRW first-passage
probability with partially or fully absorbing sites. Furthermore, by deriving closed form expressions for the occupation
probability in some instances, we have been able to obtain the first-passage probability generating function
analytically. For those cases for which the occupation probability is not known in closed form, we have constructed an iterative
procedure to obtain the first-passage probability avoiding the use of computationally burdening stochastic
simulations. 

While there have been generalisations of renewal processes with continuous variables in the past, e.g., to non-Markovian dynamics
\cite{guerin2016mean} and ageing processes \cite{schultz2014aging}, they are limited by their analytic tractability, for
example, obtaining only moments or asymptotic expansions. Our study instead generalises the renewal formulation in
discrete space and time presented by Erd{\"o}s, Feller and Pollard in 1949 and allows us to link analytically the
occupation probability to the first-passage probability for persistent and anti-persistent walks with one or multiple
targets. Our formalisms have highlighted the natural appearance of multiple
modes in the first-passage dynamics when CRWs with strong persistence occur within finite, periodic and reflecting
domains. We have also quantified the \textit{directional} first-passage statistics, that is the first-hitting dynamics based on the
direction from which a walker reach a target. 

We have obtained analytically the MFPT to one or multiple targets in periodic and reflecting domains, and found a
general expression for the GMFPT, which was previously only known for periodic domains
\cite{tejedor2012optimizing}. By comparing the first-passage dynamics with its respective MFPT, we find that the
multi-modality of the first-passage may lead to cases where the mean actually falls between two modes in a temporal
range with a very low chance for a first-passage event to occur. 

The theory presented here is rather flexible. It can be extended to biased-correlated
random walks \cite{godoy1992reflection, rossetto2018one, garcia2007solution}, may cope with the inclusion of resetting
\cite{das2022discrete,bonomo2021first} and inert spatial heterogeneities \cite{sarvaharman2022particle} such as
permeable barriers \cite{kay2022diffusion,bressloff2022probabilistic, kosztolowicz2001random} or different media
\cite{das2023dynamics}, and may also be used to account for the dynamics in random environments
\cite{szasz1984persistent}. We wish also to draw the readers' attention to its applicability to other processes that are
conveniently modelled using random walks with internal states such as walks over non-Bravais lattices
\cite{marris2023exact,rws_on_latticesIII}, double-diffusivity \cite{hill1980discrete, hill1985general}, dimer migration
on a crystalline surface \cite{landman1979stochastic, landman1979stochastic2} and chromatographic processes
\cite{weiss1994aspects}.

Our findings may also open up the possibility to analyse, the dynamics of active particles in high spatial dimensions
\cite{proesmans2020phase}, encounter statistics \cite{giuggioli2022spatio} of persistent walkers, exclusion processes
\cite{zhang2019persistent, teomy2019transport, gavagnin2018modeling}, cover times \cite{regnier2023universal}, record
statistics \cite{regnier2023record} and walks with longer range memory such as the so-called alzheimers
\cite{cressoni2007amnestically} and the elephant random walks \cite{schutz2004elephants,kenkre2007analytic}. Finally,
owing to the connection between the two-step CRW and quantum random walks \cite{boettcher2013renormalization}, this work
in the classical paradigm could help provide tools for its quantum counterpart. 

\section*{Acknowledgements}
The authors acknowledge useful discussions with Toby Kay, Hern{\'a}n Larralde, Ralf Metzler, and Seeralan Sarvaharman and would like to
thank the Isaac Newton Institute for Mathematical Sciences for support and hospitality during the programme
\textit{Mathematics of movement: an interdisciplinary approach to mutual challenges in animal ecology and cell biology}
when part of the work on this paper was undertaken.  This work was supported by: EPSRC grant number EP/R014604/1 and was
carried out using the computational facilities of the Advanced Computing Research Centre, University of Bristol -
http://www.bris.ac.uk/acrc/. DM acknowledges funding from an Engineering and Physical Sciences Research Council (EPSRC)
DTP studentship, while LG acknowledges funding from Biotechnology and Biological Sciences Research Council (BBSRC) Grant
No. BB/T012196/1 and the Natural Environment Research Council (NERC) Grant No. NE/W00545X/1.
\section*{Data Availability}
The data that support the findings of this study are available within the article and any supplementary files.
\appendix

\section{Reflective Boundary Conditions}\label{app: dD_reflective}

\begin{figure}[H]
 \centering
 \includegraphics[width = 0.9\textwidth]{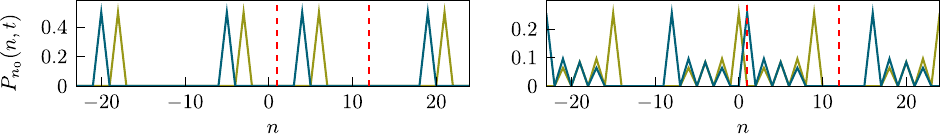}
 \caption{(Colour Online). The interplay between the bounded domain (shown between the two red dotted lines) and the
 images in three neighbouring domains at timestep $t=1$ (left) and $t =4$ (right) found by plotting the scalar expansion of
 Eq. (\ref{eq: matrix_reflective1}). The occupation probability of state
 one is shown in green while that of state two in blue. In both panels $f = 0.8$ and $b =
 0.2$.}
 \label{fig: reflect_images}
\end{figure}

To employ the method of images, we note the following. Since a walker moving right (left) must enter into state $m=1$,
($m=2$), at the right (left) boundary, it is only the contribution in state $m=1$ ($m=2$) that needs to be matched by an
image. Then to model the `bouncing', we invoke a change of state to allow propagation back in the opposing direction by allowing
the image of the other state to take over. In Fig. \ref{fig: reflect_images}, we show this interaction where in the bottom
panel, in the succeeding time step, the trajectories that continue left over boundary will be replaced by those in state
$m=1$, indicating the change in direction.

This boundary conditions are generalisable to higher dimensions, e.g. the case $d=2$ requires
\begin{equation}
\begin{aligned}
  P^{(r)}(1, n_2, \{2,1\}, t) &= P^{(r)}(0, n_2, \{1,2\}, t), \\
  P^{(r)}(N_1, n_2, \{1, 2\}, t) &= P^{(r)}(N_1+1, n_2, \{2,1\}, t), \\
  P^{(r)}(n_1, 1, \{4,3\}, t) &= P^{(r)}(n_1, 0, \{3,4\}, t),\\
  P^{(r)}(n_1, N_2, \{3,4\}, t) &= P^{(r)}(n_1, N_2+1, \{4,3\}, t),
\end{aligned}
\end{equation}
whose sum over the entire site gives, $P^{(r)}(n_1, \{1, N_2\}, t) = P^{(r)}(n_1, \{0, N_2+1\}, t)$ and $P^{(r)}(\{1,
N_1 \}, n_2, t) = P^{(r)}(\{0, N_1+1 \}, n_2, t)$. This allows one to see that the boundary conditions along either
axis are independent of the other, which allows the easy use of the methods of images (see details in the supplementary
Material \cite{supp_mat}).
\begin{figure}[h]
 \centering
 \includegraphics[width = 0.7\textwidth]{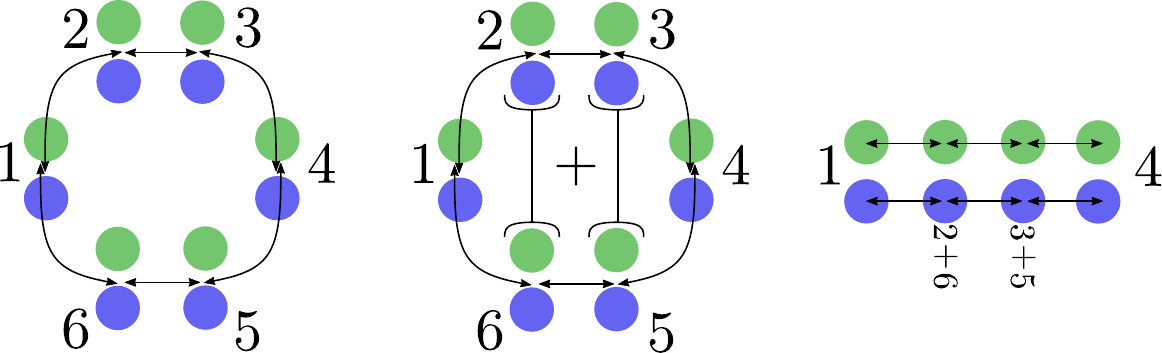}
 \caption{(Colour Online). In the style of Fig. 1b in ref. \cite{kalay2012effects}, a schematic depicting the method
 to turn a periodic domain into a reflecting domain by \textit{squeezing} the internal sites, while the boundary
 points (sites 1 and 4) remain unaffected by the transformation. For visual clarity we omit arrows depicting the
 state level dynamics, that is the directions from which a walker may enter a site, depicted in Fig. \ref{fig:
 lattice_schematic} of the main text. As shown in the right-hand panel, it is clear that after the transformation the
 walker may enter both internal states from either direction.}
 \label{fig: kalay_schematic}
\end{figure}

To employ the \textit{squeezing} method, we transform a periodic domain into one with reflecting boundaries. The
method, first introduced in ref. \cite{kalay2012effects} for the standard LRW case can be applied also to the CRW and the
essence of the method can be evinced from Fig. \ref{fig: kalay_schematic}. As shown, if the walker occupies site 1 of
the periodic lattice, its only options to move off the site are to travel to either site 2 or site 6. Thus, if we combine sites 2 and 6
together, that is we \textit{squeeze} the lattice, to create a new site (labelled $2+6$ in the far right panel), we have
precisely the reflecting boundary condition considered by Chandrasekhar \cite{chandrasekhar1943stochastic}. 

\section{The Structure Function}\label{app: s_func}
The structure (or characteristic) function is the discrete Fourier transform of the individual step 
probabilities of the walk \cite{klafter2011first,rws_on_latticesII}, which allows for easy inspection 
of the transport properties of the walk \cite{spitzer1976principles}. In systems with no internal degrees
of freedom the structure function is scalar, while the inclusion of internal states makes necessary the generalisation 
to matricial functions of size $\mathfrak{n}\times \mathfrak{n}$, where $\mathfrak{n}$ is the number of internal states. 

Presently, the matrix $\bm{\lambda}\left(\pi \mathcal{N}^{(\gamma)}_{\kappa_1}, ..., \pi \mathcal{N}^{(\gamma)}_{\kappa_d}\right)$
is of size $Z\times Z$ and has a common structure for all the lattices we have considered here. Namely, the
probability of remaining at the same state, i.e. persisting in the direction of the previous movement step or
sojourning, is encoded in the diagonal elements, the probability of backtracking is given on the upper and lower
diagonal in alternating rows, while all the remaining elements relate to the option of turning laterally. These features
are evident in Eq. (\ref{eq: 1D_sf}) for the $d=1$ case, and are seen explicitly also for the
two-dimensional hypercubic lattice when we write out the matrix
\begin{equation}
\bm{\lambda}\left(\pi \mathcal{N}^{(\gamma)}_{\kappa_1}, \pi \mathcal{N}^{(\gamma)}_{\kappa_2}\right) \!=\! \begin{bmatrix}
  fe^{i\pi \mathcal{N}^{(\gamma)}_{\kappa_1}} + c^{(2)} & be^{i\pi \mathcal{N}^{(\gamma)}_{\kappa_1}} & \ell e^{i\pi \mathcal{N}^{(\gamma)}_{\kappa_1}} & \ell e^{i\pi \mathcal{N}^{(\gamma)}_{\kappa_1}} \\
  be^{-i\pi \mathcal{N}^{(\gamma)}_{\kappa_1}} & fe^{-i\pi \mathcal{N}^{(\gamma)}_{\kappa_1}}+c^{(2)} & \ell e^{-i\pi \mathcal{N}^{(\gamma)}_{\kappa_1}} & \ell e^{-i\pi \mathcal{N}^{(\gamma)}_{\kappa_1}} \\
  \ell e^{i\pi \mathcal{N}^{(\gamma)}_{\kappa_2}}& \ell e^{i\pi \mathcal{N}^{(\gamma)}_{\kappa_2}} & fe^{i\pi \mathcal{N}^{(\gamma)}_{\kappa_2}}+c^{(2)} & be^{i\pi \mathcal{N}^{(\gamma)}_{\kappa_2}}\\
  \ell e^{-i\pi \mathcal{N}^{(\gamma)}_{\kappa_2}} & \ell e^{-i\pi \mathcal{N}^{(\gamma)}_{\kappa_2}} & be^{-i\pi \mathcal{N}^{(\gamma)}_{\kappa_2}} & fe^{-i\pi \mathcal{N}^{(\gamma)}_{\kappa_2}}+c^{(2)}
\end{bmatrix}\!.
\label{eq: 2D_sf}
\end{equation}
As mentioned in the main text (Sec \ref{sec: Unbounded}), the $(i, j)$-th element of $\bm{\lambda}\left(\pi
\mathcal{N}^{(\gamma)}_{\kappa_1}, ..., \pi \mathcal{N}^{(\gamma)}_{\kappa_d}\right)$ represents the movement that may occur from
state $m_j$ to $m_i$.  The elements of one row, which have identical exponents in the exponentials,
represent the one (unique) permissible movement direction of the walker entering the state
$m_i$. For example, $i\pi \mathcal{N}^{(\gamma)}_{\kappa_1}$ in the first row indicates that to enter state $m_1$, the walker must
increase its $n_1$ coordinate, that is by travelling West, $-i\pi
\mathcal{N}^{(\gamma)}_{\kappa_1}$ in the second row shows that to enter $m_2$ a decrease in the $n_1$ coordinate occurs by
moving East, etc.

From the structure function one may also gain information about
the steady state of the system.  When $f, b \neq 1 $, for hypercubic lattices and hexagonal lattices, in the limit $t \to
\infty$, one has, respectively, $\bm{\lambda}\left(\pi \mathcal{N}^{(\gamma)}_{\kappa_1}, ..., \pi
\mathcal{N}^{(\gamma)}_{\kappa_d}\right)^t = \bm{0}$ and
$\bm{\lambda}^{(\mathcal{H})}\!\big(\bm{\kappa}\big)^t=\bm{0}$ when $\bm{\kappa}\not=0$, while $\lim_{t\to
\infty}\bm{\lambda}\left(0, ..., 0\right)^t=(2d)^{-1}\mathbb{J}$, and
$\bm{\lambda}^{(\mathcal{H})}\!\big(0,0\big)^t=(6)^{-1}\mathbb{J}$, where $\mathbb{J}$ is the all ones matrix.
Therefore, $\bm{P}^{(\gamma)}_{n_{01}, ... , n_{0d}}(n_1, ..., n_d, t\to \infty) = (2d)^{-1}N^{-d}[1, ...,
1]^{\intercal}$ and $\bm{P}^{(\mathcal{H})}_{\bm{n}_{0}}(\bm{n}, t\to
\infty)=(6\Omega)^{-1}[1, ..., 1]^{\intercal}$, which means that
$P^{(\gamma)}_{n_{01}, ... , n_{0d}}(n_1, ..., n_d, t\to \infty) = N^{-d}$ for $\gamma \in \{p, r \}$
and $P^{(\mathcal{H})}_{n_{01}, n_{02}}(n_1, n_2, t\to \infty) = \Omega^{-1}$ as expected from an
irreducible (ergodic) aperiodic finite Markov chain with doubly stochastic transition matrices. When $\gamma = r_s$, 
that is the squeezed reflecting propagator, in the same limit $\bm{\lambda}\left(\frac{\pi\kappa}{N-1}\right)^t = \bm{0}$ for $\kappa \neq 0$ 
and $\bm{\lambda}\left(0\right)^t = 2^{-1}\mathbb{J}$. Hence, the steady state in this case can be found as
$\bm{P}^{(r_s)}_{n_{0}}(n, t \to \infty) = (N-1)^{-1}[\alpha_1,\alpha_{2}]^{\intercal}$ ($n \neq \{1, N\}$) while 
$\bm{P}^{(r_s)}_{n_{0}}(1,t\to \infty) = \bm{P}^{(r_s)}_{n_{0}}(N, t \to \infty) = [2(N-1)]^{-1}[\alpha_1,\alpha_{2}]^{\intercal}$. In contrast, when either $f$ or $b$ equals 1, $\bm{\lambda}\left(\pi \mathcal{N}^{(\gamma)}_{\kappa_1}, ..., \pi
\mathcal{N}^{(\gamma)}_{\kappa_d}\right)$ and $\bm{\lambda}^{(\mathcal{H})}\!\big(\bm{\kappa}\big)$ reduce to
generalised permutation matrices, which makes clear that no steady state will be reached. 

When $f= 1$, in the periodically bounded domain, one can show that the
walker will be at $n_0$ with certainty every $t=\mathfrak{c}N$ steps ($\mathfrak{c}\in \mathbb{N} \cup \{0\} $), that is the system is a Markov chain with periodicity $N$. To do so, let us 
consider the one-dimensional case (the analysis is readily extended to arbitrary dimensions), where $\bm{\lambda}\left(\frac{2\pi \kappa}{N}\right) = 
\left(\begin{smallmatrix}e^{\frac{2\pi i \kappa  }{N}} & 0\\ 0 & e^{\frac{-2\pi i\kappa  }{N}}\end{smallmatrix}\right)$. Consequently, \begin{small}$\bm{P}_{n_0}^{(p)}(n, t) = N^{-1}\sum_{\kappa = 0}^{N-1}\exp\left(-\frac{2\pi i \kappa(n-n_0)}{N}\right)\left(\begin{smallmatrix}e^{\frac{2\pi i\kappa  t}{N}} & 0\\ 0 & e^{\frac{-2\pi i\kappa t}{N}}\end{smallmatrix}\right)\cdot \left[\begin{matrix} \alpha_1 \\ \alpha_2 \end{matrix}\right]$\end{small}.
Upon inspection, one clearly sees that when $t=\mathfrak{c}N$ and $n = n_0$, the propagator reduces to \begin{small}$\bm{P}_{n_0}^{(p)}(n, t) = \left[\begin{matrix} \alpha_1 \\ \alpha_2 \end{matrix}\right]$\end{small}, such that 
$P^{(p)}_{n_0}(n_0, \mathfrak{c}N) = 1$.

\section{The Hexagonal Propagator}\label{app: hex}
\begin{figure}[h!]
 \centering
 \includegraphics[width = 0.6\textwidth]{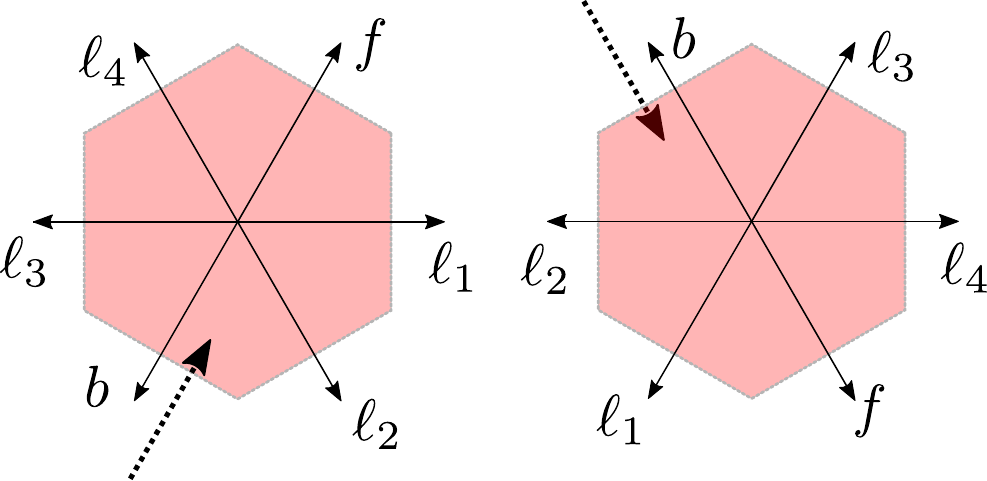}
 \caption{(Colour Online). A schematic representing two of the six permissible movement steps in the correlated
 hexagonal random walk where the solid arrows represent the movement probabilities if the walker had entered the site
 via the dotted arrow. The movements shown on the left side case are shown in the $3^{\text{rd}}$ column of Eq.
 (\ref{eq: hex_structure_func}) of the main text and the movements shows in the right panel are encoded in the  $2^{\text{nd}}$ column
 of Eq. (\ref{eq: hex_structure_func}) of the main text.}
 \label{fig: hex_direction}
\end{figure}

With $Z=6$, the governing equations for correlated motion on the hexagonal lattice comprisies of six coupled equations,
which we write explicitly in the Supplementary Material (ref. \cite{supp_mat}). With the initial condition
$\bm{U}_{m_{0}} = \delta_{n_1, n_{0_1}}\delta_{n_2, n_{0_2}}[\alpha_1, ..., \alpha_6]^{\intercal}$, and following the
same procedure as for the hypercubic lattice (Sec. \ref{app: unbounded_prop} of ref. \cite{supp_mat}), it is possible to
write the lattice Green's function for the hexagonal lattice as
\begin{equation}
\widetilde{\bm{Q}}^{(\mathcal{H})}_{n_0}(\bm{n}, z) = \frac{1}{4\pi^2}\int_{-\pi}^{\pi}\int_{-\pi}^{\pi}e^{-i\bm{\xi}(\bm{n}-\bm{n}_0)}\bm{\lambda}^{(\mathcal{H})}(\xi_1, \xi_2)\cdot \bm{U_{m_0}} \text{d}\xi_1\text{d}\xi_2,
\label{eq: hex_LGF}
\end{equation}
where 
\begin{equation}
 \scriptsize
 \begin{aligned}
&\bm\lambda^{(\mathcal{H})} (\xi_1, \xi_2) = \\ &\bbordermatrix{      & \nwarrow & \searrow & \nearrow & \swarrow & \longrightarrow & \longleftarrow \cr
                                                         \nwarrow  &       fe^{-i \xi_1} + c^{(4)} &    b e^{-i \xi_1} &               \ell_4 e^{-i \xi_1} &              \ell_2 e^{-i \xi_1} &            \ell_3 e^{-i \xi_1} &                  \ell_1 e^{-i \xi_1} \cr
                                                         \searrow &        be^{i \xi_1} &               f e^{i \xi_1} + c^{(4)}&       \ell_2 e^{i \xi_1} &               \ell_4 e^{i \xi_1} &             \ell_1 e^{i \xi_1} &                   \ell_3 e^{i \xi_1} \cr
                                                         \nearrow &        \ell_1 e^{-i \xi_2} &        \ell_3 e^{-i \xi_2} &          f e^{-i \xi_2} + c^{(4)}&          b e^{-i \xi_2} &                 \ell_4 e^{-i \xi_2} &                  \ell_2 e^{-i \xi_2} \cr
                                                         \swarrow &        \ell_3 e^{i \xi_2} &         \ell_1 e^{i \xi_2} &           b e^{i \xi_2} &                    f e^{i \xi_2} + c^{(4)}&         \ell_2 e^{i \xi_2} &                   \ell_4 e^{i \xi_2} \cr
                                                         \longrightarrow & \ell_2 e^{i(\xi_1 - \xi_2)} &  \ell_4 e^{i(\xi_1 - \xi_2)} &    \ell_1 e^{i(\xi_1 - \xi_2)} &        \ell_3 e^{i(\xi_1 - \xi_2)} &      f e^{i(\xi_1 - \xi_2)}+ c^{(4)} &        b e^{i(\xi_1 - \xi_2)}\cr
                                                         \longleftarrow &  \ell_4 e^{-i(\xi_1 - \xi_2)} & \ell_2 e^{-i(\xi_1 - \xi_2)} &   \ell_3 e^{-i(\xi_1 - \xi_2)} &       \ell_1 e^{-i(\xi_1 - \xi_2)} &     b e^{-i(\xi_1 - \xi_2)} &                f e^{-i(\xi_1 - \xi_2)}+ c^{(4)}} ,
  \label{eq: hex_structure_func}
 \end{aligned}
\end{equation}
\normalsize
with the elements representing the different movement options are labeled, in Eq. (\ref{eq: hex_structure_func}), via
the arrows. For clarity we also show pictorially some movement options in Fig. \ref{fig: hex_direction}.

A valid image set for the standard (diffusive) periodic hexagonal propagator is known as \cite{marris2023exact}
\begin{equation}
\label{eq: periodic_images}
  \widetilde{\bm{P}}^{(\mathcal{H})}_{\bm{n}_{0}}(n_1, n_2, z) = \sum_{\kappa_1 = -\infty}^{\infty}\sum_{\kappa_2 = -\infty}^{\infty}\widetilde{\bm{Q}}_{\bm{n}_{0}}^{(\mathcal{H})}(n_1 + \hat{n}_1, n_2 + \hat{n}_2, z) ,
\end{equation}
where 
\begin{equation}
  \begin{bmatrix}
      \widehat{n}_1 \\ \widehat{n}_2 
  \end{bmatrix} = \begin{bmatrix}
      -R\kappa_1 +(2R+1)\kappa_2 \\  -(R+1)\kappa_1 -R\kappa_2
  \end{bmatrix},
  \label{eq: right_shift1}
\end{equation}
and since, like the hypercubic case, the periodic boundary conditions for the correlated random walk are identical to
their diffusive counterpart, Eq. ~(\ref{eq: hex_LGF}) can be solved explicitly following the same procedure as the
solution to the LRW with periodic boundary conditions derived in \cite{marris2023exact}. This leads to Eq. (\ref{eq:
hex_periodic}) of the main text. 

As mentioned in Sec. \ref{sec: hex} of the main text, the non-orthogonality of the coordinates leads to a `shift' in the
position of the walker when traversing the periodic boundary. For a deeper discussion we again refer the interested
reader to \cite{marris2023exact}, and provide here a simple explanation of the dynamics seen in Fig. \ref{fig: hex_occu}
of the main text by describing the path taken by the probability pulse that is near North-East of the domain in the
ballistic case depicted in Fig. \ref{fig: hex_occu}(a) (other pulses may then be understood similarly due to the rotational symmetry
of the system). 

With initial condition at the origin, the pulse in the North-East first travelled due West and reached the West most
corner of the hexagon, at $t = 8$. Crossing the boundary at $t=9$ the hexagonal periodicity brings the walker back into
the domain at the North-East corner. Since the pulse is ballistically travelling due West, it will remain travelling on
this trajectory along the top boundary until $t=17$, passing the point $\bm{n} = (-6, -2, 8)$ at time $t = 11$ as shown
in Fig. \ref{fig: hex_occu}(a).

As mentioned in Sec. \ref{sec: hex} of the main text, the shift in the boundary condition means that the
walker will hit every site in the finite domain with certainty even if $f=1$, something 
that is not the case in the $d > 1$ hypercubic lattice. We show
this formally here by considering the $z$-transfrom of Eq. (\ref{eq: hex_periodic}) may be written in the form
\cite{barry_hughes_book}
\begin{equation}
\widetilde{\bm{P}}^{(\mathcal{H})}_{\bm{n}_{0}}(\bm{n},z) = \frac{1}{\Omega}\left[\mathbb{I} - z\bm{\lambda}^{(\mathcal{H})}(0,0)\right]^{-1} +  \widetilde{\bm{P}}^{(\kappa \neq 0)}_{\bm{n}_0}(\bm{n}, z),
\end{equation}
where $\widetilde{P}^{(\kappa \neq 0)}_{\bm{n}_0}(\bm{n}, z)$ is the generating function of the double summation in Eq.
(\ref{eq: hex_periodic}). Since $\bm{\lambda}^{(\mathcal{H})}(0,0) = \mathbb{I}$ for $(\kappa_1, \kappa_2) = (0,0)$
only, it is clear that at $z=1$, each element $\widetilde{\bm{P}}^{(\kappa \neq 0)}_{\bm{n}_0}(\bm{n}, z)$ is
holomorphic. Therefore, $\widetilde{\bm{P}}^{(\mathcal{H})}_{\bm{n}_{0}}(\bm{n}, z) \to \infty$ as $z\to 1$ showing 
it is certain that all sites will be hit eventually. This is true for all movement parameters except when $b= 1$ or $\ell_i = 1$.

\section{Alternative Derivation of the Single Target First-Passage Probability}\label{app: defect}
We generalise here the approach taken in ref. \cite{larralde2020first}. To do so, we allow 
for an arbitrary initial weighting distribution and make no assumptions about the symmetry of the system 
meaning it is valid for any boundary condition.

For simplicity in the algebra we assume without loss of generality that the system is one-dimensional with two internal states $m = \{1, 2\}$.
The equations relating the state-level occupation probability and state-level first passage are written as:
\begin{equation}
 \begin{aligned}
     \alpha_{m_{0_1}}P_{n_{0}, 1}(s, 1, t) &= \alpha_{m_{0_1}}\delta_{m_0, 1}\delta_{t, 0}\delta_{n, n_0} \\ & \quad \quad\quad + \sum_{t^{\prime} = 0}^{t}F_{n_0, 1}(s, 1, t^{\prime})P_{s, 1}(s, 1, t-t^{\prime})+F_{n_0, 1}(s, 2, t^{\prime})P_{s, 2}(s, 1, t-t^{\prime}),\\
     \alpha_{m_{0_2}}P_{n_{0}, 2}(s, 1, t) &= \alpha_{m_{0_2}}\delta_{m_0, 1}\delta_{t, 0}\delta_{n, n_0}\\ & \quad \quad\quad + \sum_{t^{\prime} = 0}^{t}F_{n_0, 2}(s, 1, t^{\prime})P_{s, 1}(s, 1, t-t^{\prime})+F_{n_0, 2}(s, 2, t^{\prime})P_{s, 2}(s, 1, t-t^{\prime}),\\
     \alpha_{m_{0_1}}P_{n_{0}, 1}(s, 2, t) &= \alpha_{m_{0_1}}\delta_{m_0, 2}\delta_{t, 0}\delta_{n, n_0}\\ & \quad \quad\quad + \sum_{t^{\prime} = 0}^{t}F_{n_0, 1}(s, 1, t^{\prime})P_{s, 1}(s, 2, t-t^{\prime})+F_{n_0, 1}(s, 2, t^{\prime})P_{s, 2}(s, 2, t-t^{\prime}),\\
     \alpha_{m_{0_2}}P_{n_{0}, 2}(s, 2, t) &= \alpha_{m_{0_2}}\delta_{m_0, 2}\delta_{t, 0}\delta_{n, n_0}\\ & \quad \quad\quad + \sum_{t^{\prime} = 0}^{t}F_{n_0, 2}(s, 1, t^{\prime})P_{s, 1}(s, 2, t-t^{\prime})+F_{n_0, 2}(s, 2, t^{\prime})P_{s, 2}(s, 2, t-t^{\prime}).\\
 \end{aligned}
 \label{eq: sup_renewal_coupled}
\end{equation}
Equation (\ref{eq: sup_renewal_coupled}) reduces to Eq. (2) of ref. \cite{larralde2020first} with the choice of $\alpha_{m_{0_1}} = 1$, $\alpha_{m_{0_2}}=0$, which 
implies $F_{n_0, 2}(s, 1, t) =F_{n_0, 2}(s, 2, t) = 0$ as no trajectories start in state 2.

Proceeding in the $z$-domain, and assuming $s \neq n_0$, we find
\begin{equation}
 \begin{aligned}
     \alpha_{m_{0_1}}\widetilde{P}_{n_{0}, 1}(s, 1, z) &= \widetilde{F}_{n_0, 1}(s, 1, z)\widetilde{P}_{s, 1}(s, 1, z)+\widetilde{F}_{n_0, 1}(s, 2, z)\widetilde{P}_{s, 2}(s, 1, z),\\
     \alpha_{m_{0_2}}\widetilde{P}_{n_{0}, 2}(s, 1, z) &= \widetilde{F}_{n_0, 2}(s, 1, z)\widetilde{P}_{s, 1}(s, 1, z)+\widetilde{F}_{n_0, 2}(s, 2, z)\widetilde{P}_{s, 2}(s, 1, z),\\
     \alpha_{m_{0_1}}\widetilde{P}_{n_{0}, 1}(s, 2, z) &= \widetilde{F}_{n_0, 1}(s, 1, z)\widetilde{P}_{s, 1}(s, 2, z)+\widetilde{F}_{n_0, 1}(s, 2, z)\widetilde{P}_{s, 2}(s, 2, z),\\
     \alpha_{m_{0_2}}\widetilde{P}_{n_{0}, 2}(s, 2, z) &= \widetilde{F}_{n_0, 2}(s, 1, z)\widetilde{P}_{s, 1}(s, 2, z)+\widetilde{F}_{n_0, 2}(s, 2, z)\widetilde{P}_{s, 2}(s, 2, z),\\
 \end{aligned}
\end{equation}
which one may formulate as the matrix equation 
\begin{small}
 \begin{equation}
     \!\begin{bmatrix}
         \widetilde{F}_{n_0, 1}(s, 1, z) \\
         \widetilde{F}_{n_0, 1}(s, 2, z) \\
         \widetilde{F}_{n_0, 2}(s, 1, z) \\
         \widetilde{F}_{n_0, 2}(s, 2, z)
     \end{bmatrix} 
     \!=\! 
     \begin{bmatrix}
         \widetilde{P}_{s, 1}(s, 1, z) &\widetilde{P}_{s, 2}(s, 1, z) & 0 & 0\\
         0 & 0 & \widetilde{P}_{s, 1}(s, 1, z) & \widetilde{P}_{s, 2}(s, 1, z) \\
         \widetilde{P}_{s, 1}(s, 2, z) &\widetilde{P}_{s, 2}(s, 2, z) & 0 & 0 \\
         0 & 0 & \widetilde{P}_{s, 1}(s, 2, z) & \widetilde{P}_{s, 2}(s, 2, z)
     \end{bmatrix} ^{-1}
 \! \cdot\!
 \begin{bmatrix}
     \alpha_{m_{0_1}}\widetilde{P}_{n_{0}, 1}(s, 1, z) \\
     \alpha_{m_{0_2}}\widetilde{P}_{n_{0}, 2}(s, 1, z) \\
     \alpha_{m_{0_1}}\widetilde{P}_{n_{0}, 1}(s, 2, z) \\
     \alpha_{m_{0_2}}\widetilde{P}_{n_{0}, 2}(s, 2, z)
 \end{bmatrix}\!.
 \label{eq: renewal2}
 \end{equation}
\end{small}

Upon evaluating Eq. (\ref{eq: renewal2}), one finds Eq. (\ref{eq: one_whole_site_FP}) of the main text by performing
$\sum_{m_0=1}^{2}\sum_{m=1}^{2}\widetilde{F}_{n_0, m_0}(s, m, z)$, namely performing a summation over the column vector
of state level first-passages. 

The procedure detailed here generalises trivially to a target of arbitrary number of states (i.e., it is valid in any dimension), however
the matrix that requires inversion may become large. In such cases, the approach via 
defective sites presented in the main text is more suitable.

Pertinent to the directional first-passage as well 
as the first-passage in the unbounded one-dimensional case (as the directional first-passage equals the 
first-passage in this case) is the situation of one target state at some $\bm{s}, m_{\bm{s}}$.
We focus on the arbitrarily dimensional directional first-passage probability. In such a situation, the renewal equations simplify to   
\begin{equation}
 \begin{aligned}
     \alpha_{m_{0_1}}P_{\bm{n}_{0}, 1}(\bm{s}, m_{\bm{s}}, t) &= \alpha_{m_{0_1}}\delta_{m_0, 1}\delta_{t, 0}\delta_{\bm{s}, \bm{n}_0} + \sum_{t^{\prime} = 0}^{t}F_{\bm{n}_0, 1}(\bm{s}, m_{\bm{s}}, t^{\prime})P_{\bm{s}, m_{\bm{s}}}(\bm{s}, m_{\bm{s}}, t-t^{\prime}),\\
     \alpha_{m_{0_2}}P_{\bm{n}_{0}, 2}(\bm{s}, m_{\bm{s}}, t) &= \alpha_{m_{0_2}}\delta_{m_0, 2}\delta_{t, 0}\delta_{\bm{s}, \bm{n}_0}+ \sum_{t^{\prime} = 0}^{t}F_{\bm{n}_0, 2}(\bm{s}, m_{\bm{s}}, t^{\prime})P_{\bm{s}, m_{\bm{s}}}(\bm{s}, m_{\bm{s}}, t-t^{\prime}),\\
     & \; \; \vdots \\
     \alpha_{m_{0_M}}P_{\bm{n}_{0}, M}(\bm{s}, m_{\bm{s}}, t) &= \alpha_{m_{0_M}}\delta_{m_0, M}\delta_{t, 0}\delta_{\bm{s}, \bm{n}_0}+ \sum_{t^{\prime} = 0}^{t}F_{\bm{n}_0, M}(\bm{s}, m_{\bm{s}}, t^{\prime})P_{\bm{s}, m_{\bm{s}}}(\bm{s}, m_{\bm{s}}, t-t^{\prime}).
 \end{aligned}
 \label{eq: 1_tar_renewal}
\end{equation}

We again assume $\bm{s} \neq \bm{n}_0$ and take the generating function to find 
\begin{equation}
 \begin{aligned}
     \alpha_{m_{0_1}}\widetilde{P}_{\bm{n}_{0}, 1}(\bm{s}, m_{\bm{s}}, z) &= \widetilde{F}_{\bm{n}_0, 1}(\bm{s}, m_{\bm{s}}, z)\widetilde{P}_{\bm{s}, m_{\bm{s}}}(\bm{s}, m_{\bm{s}}, z),\\
     \alpha_{m_{0_2}}\widetilde{P}_{\bm{n}_{0}, 2}(\bm{s}, m_{\bm{s}}, z) &= \widetilde{F}_{\bm{n}_0, 2}(\bm{s}, m_{\bm{s}}, z)\widetilde{P}_{\bm{s}, m_{\bm{s}}}(\bm{s}, m_{\bm{s}}, z),\\
     & \; \; \vdots \\
     \alpha_{m_{0_M}}\widetilde{P}_{\bm{n}_{0}, M}(\bm{s}, m_{\bm{s}}, z) &= \widetilde{F}_{\bm{n}_0, M}(\bm{s}, m_{\bm{s}}, z)\widetilde{P}_{\bm{s}, m_{\bm{s}}}(\bm{s}, m_{\bm{s}}, z).
 \end{aligned}
 \label{eq: 1_tar_renewalz}
\end{equation}
For each equation in Eq. (\ref{eq: 1_tar_renewalz}) one may solve for 
\begin{equation}
 \widetilde{F}_{\bm{n}_0, i}(\bm{s}, m_{\bm{s}}, z) = \frac{\alpha_{m_{0_i}}\widetilde{P}_{\bm{n}_{0}, i}(\bm{s}, m_{\bm{s}}, z)}{\widetilde{P}_{\bm{s}, m_{\bm{s}}}(\bm{s}, m_{\bm{s}}, z)}, 
\end{equation} 
independently of the other equations and upon summation $\widetilde{F}_{\bm{n}_0}(\bm{s}, m_{\bm{s}}, z) =
\sum_{i=1}^{M}\widetilde{F}_{\bm{n}_0, i}(\bm{s}, m_{\bm{s}}, z)$, we find Eq. (\ref{eq: single_tar_FP}) of the main
text. By again setting $\alpha_{m_{0_i}} = 1$ and all other $\alpha_{m_{0_j}} = 0$ ($j \neq i$)  we recover the known cases of refs.
\cite{larralde2020first} and \cite{rws_on_latticesIII}.

For specific cases the first-passage expression, Eq. (\ref{eq: multi_tar_FP}), may actually 
be simplified further. This occurs when $\widetilde{P}_{n, i}(n, i, z) = \widetilde{P}_{n, j}(n, j, z)$ and 
$\widetilde{P}_{n, j}(n, i, z) = \widetilde{P}_{n, i}(n, j, z)$, which is true only when the walk is translationally invariant and 
we demonstrate this explicitly here in the one-dimensional case.

Inserting the above identities into Eq. (\ref{eq: one_whole_site_FP}) we find
\begin{small}
 \begin{align}
     & \widetilde{F}_{n_0}(s, z) =  \nonumber \\ & \frac{\left\{\!\alpha_{1}\!\left[\!\widetilde{P}_{n_0, 1}(s, 1,z)\!+\!\widetilde{P}_{n_0, 1}(s, 2,z)\right]\!+\!\alpha_{2}\left[\widetilde{P}_{n_0, 2}(s, 1,z)\!+\!\widetilde{P}_{n_0, 2}(s, 2,z)\!\right]\right\}\!\left[\!\widetilde{P}_{s, 2}(s, 2,z)\!-\!\widetilde{P}_{s, 1}(s, 2,z)\!\right]}{\widetilde{P}_{s, 2}(s, 2,z)^2-\widetilde{P}_{s, 1}(s, 2,z)^2}\!.
        \label{eq: pmvis1}
 \end{align}
\end{small}
Since, $\alpha_{1}\left[\widetilde{P}_{n_0, 1}(s, 1,z)+\widetilde{P}_{n_0, 1}(s, 2,z)\right]+\alpha_{2}\left[\widetilde{P}_{n_0, 2}(s, 1,z)+\widetilde{P}_{n_0, 2}(s, 2,z)\right] = \widetilde{P}_{n_0}(s, z)$, Eq. (\ref{eq: pmvis1})
may be rewritten as
\begin{align}
 \widetilde{F}_{n_0}(s, z) = \frac{\widetilde{P}_{n_0}(s, z)\left[\widetilde{P}_{s, 2}(s, 2,z)-\widetilde{P}_{s, 1}(s, 2,z)\right]}{\widetilde{P}_{s, 2}(s, 2,z)^2-\widetilde{P}_{s, 1}(s, 2,z)^2},
   \label{eq: pmvis2}
\end{align}
and then upon using the differences of two squares property on the denominator we find 
\begin{align}
 \widetilde{F}_{n_0}(s, z) = \frac{\widetilde{P}_{n_0}(s, z)}{\widetilde{P}_{s, 2}(s, 2,z)+\widetilde{P}_{s, 1}(s, 2,z)}.
   \label{eq: pmvis4}
\end{align}
To perform the final simplification we note that with $\alpha_1 + \alpha_2 = 1$, the denominator may be 
rewritten as $(\alpha_1 +\alpha_2)\left[\widetilde{P}_{s, 2}(s, 2,z)+\widetilde{P}_{s, 1}(s, 2,z)\right]$. 
By multiplying this expression out and once again using the properties of the translationally 
invariant propagators given above, we see that
\begin{align}
 (\alpha_1 +\alpha_2)&\left[\widetilde{P}_{s, 2}(s, 2,z)+\widetilde{P}_{s, 1}(s, 2,z)\right] = \nonumber \\ & \alpha_{1}\left[\widetilde{P}_{s, 1}(s, 1,z)+\widetilde{P}_{s, 1}(s, 2,z)\right]+\alpha_{2}\left[\widetilde{P}_{s, 2}(s, 1,z)+\widetilde{P}_{s, 2}(s, 2,z)\right],
 \label{eq: denom_simp}
\end{align}
where it is clear from the LHS of Eq. (\ref{eq: denom_simp}) that we may re-write the denominator of Eq. (\ref{eq: pmvis4}) as 
$\widetilde{P}_s(s,z)$ leading to 
\begin{equation}
 \widetilde{F}_{n_0}(s, z) = \frac{\widetilde{P}_{n_0}(s, z)}{\widetilde{P}_{s}(s, z)}.
\end{equation}

\bibliography{references}

\begin{thebibliography}{100}

\bibitem{nathan2022big}
R.~Nathan, C.~T. Monk, R.~Arlinghaus, T.~Adam, J.~Al{\'o}s, M.~Assaf,
  H.~Baktoft, C.~E. Beardsworth, M.~G. Bertram, A.~I. Bijleveld, et~al.
\newblock Big-data approaches lead to an increased understanding of the ecology
  of animal movement.
\newblock {\em Science}, 375(6582):eabg1780, 2022.

\bibitem{meijering2012methods}
E.~Meijering, O.~Dzyubachyk, and I.~Smal.
\newblock Methods for cell and particle tracking.
\newblock {\em Method Enzymol.}, 504:183--200, 2012.

\bibitem{chenouard2014objective}
N.~Chenouard, I.~Smal, F.~De~Chaumont, M.~Ma{\v{s}}ka, I.~F. Sbalzarini,
  Y.~Gong, J.~Cardinale, C.~Carthel, S.~Coraluppi, M.~Winter, et~al.
\newblock Objective comparison of particle tracking methods.
\newblock {\em Nature methods}, 11(3):281--289, 2014.

\bibitem{rws_on_latticesII}
E.~W. Montroll and G.~H. Weiss.
\newblock Random walks on lattices. {II}.
\newblock {\em J. {M}ath. {P}hys.}, 6(2):167--181, 1965.

\bibitem{weiss1994aspects}
G.~H. Weiss.
\newblock {\em Aspects and applications of the random walk}.
\newblock Elsevier Science \& Technology, 1994.

\bibitem{LucaPRX}
L.~Giuggioli.
\newblock Exact spatiotemporal dynamics of confined lattice random walks in
  arbitrary dimensions: A century after {S}moluchowski and {P}{\'o}lya.
\newblock {\em {Phys. Rev. X}}, 10:021045, 2020.

\bibitem{morters2010brownian}
P.~M{\"o}rters and Y.~Peres.
\newblock {\em Brownian motion}, volume~30.
\newblock Cambridge University Press, 2010.

\bibitem{bressloff2014stochastic}
P.~C. Bressloff.
\newblock {\em Stochastic processes in cell biology}, volume~41.
\newblock Springer, 2014.

\bibitem{kenkre2021theory}
V.M. Kenkre and L.~Giuggioli.
\newblock {\em Theory of the spread of epidemics and movement ecology of
  animals: an interdisciplinary approach using methodologies of physics and
  mathematics}.
\newblock Cambridge University Press, 2021.

\bibitem{bejan2007constructal}
A.~Bejan and G.~W. Merkx.
\newblock {\em Constructal theory of social dynamics}.
\newblock Springer, 2007.

\bibitem{embrechts2013modelling}
P.~Embrechts, C.~Kl{\"u}ppelberg, and T.~Mikosch.
\newblock {\em Modelling extremal events: for insurance and finance},
  volume~33.
\newblock Springer, 2013.

\bibitem{selmeczi2005cell}
D.~Selmeczi, S.~Mosler, P.~H. Hagedorn, N.~B. Larsen, and H.~Flyvbjerg.
\newblock Cell motility as persistent random motion: theories from experiments.
\newblock {\em Biophys. J.}, 89(2):912--931, 2005.

\bibitem{prasad2006searching}
B.~R.~G. Prasad and R.~M. Borges.
\newblock Searching on patch networks using correlated random walks: space
  usage and optimal foraging predictions using {M}arkov chain models.
\newblock {\em J. Theor. Biol.}, 240(2):241--249, 2006.

\bibitem{echeverria2021estimating}
I.~Echeverr{\'\i}a-Huarte, A.~Garcimart{\'\i}n, R.C. Hidalgo,
  C.~Mart{\'\i}n-G{\'o}mez, and I.~Zuriguel.
\newblock Estimating density limits for walking pedestrians keeping a safe
  interpersonal distancing.
\newblock {\em Sci. Rep.}, 11(1):1534, 2021.

\bibitem{korabel2022hemocytes}
N.~Korabel, G.~D. Clemente, D.~Han, F.~Feldman, T.~H. Millard, and T.~A. Waigh.
\newblock Hemocytes in drosophila melanogaster embryos move via heterogeneous
  anomalous diffusion.
\newblock {\em Commun. Phys.}, 5(1):269, 2022.

\bibitem{speckner2021single}
K.~Speckner and M.~Weiss.
\newblock Single-particle tracking reveals anti-persistent subdiffusion in cell
  extracts.
\newblock {\em Entropy}, 23(7):892, 2021.

\bibitem{benelli2022probing}
R.~Benelli and M.~Weiss.
\newblock Probing local chromatin dynamics by tracking telomeres.
\newblock {\em Biophys. J.}, 121(14):2684--2692, 2022.

\bibitem{gabel2012random}
A.~Gabel and S.~Redner.
\newblock Random walk picture of basketball scoring.
\newblock {\em J. Quant. Anal. Sports}, 8(1), 2012.

\bibitem{kenkre1983coherence}
V.M. Kenkre, V.~Ern, and A.~Fort.
\newblock Coherence effects in triplet-exciton transport via time-dependent
  delayed fluorescence.
\newblock {\em Phys. Rev. B}, 28(2):598, 1983.

\bibitem{kenkre1981effect}
V.M. Kenkre and Y.M. Wong.
\newblock Effect of transport coherence on trapping: Quantum-yield calculations
  for excitons in molecular crystals.
\newblock {\em Phys. Rev. B}, 23(8):3748, 1981.

\bibitem{rudnick2004elements}
J.~Rudnick and G.~Gaspari.
\newblock {\em Elements of the Random Walk}, volume~1.
\newblock Cambridge University Press, 2004.

\bibitem{jose2022active}
S.~Jose, D.~Mandal, M.~Barma, and K.~Ramola.
\newblock Active random walks in one and two dimensions.
\newblock {\em Phys. Rev. E}, 105(6):064103, 2022.

\bibitem{martens2012probability}
K.~Martens, L.~Angelani, R.~Di~Leonardo, and L.~Bocquet.
\newblock Probability distributions for the run-and-tumble bacterial dynamics:
  An analogy to the lorentz model.
\newblock {\em Eur. Phys. J. E}, 35:1--6, 2012.

\bibitem{malakar2018steady}
K.~Malakar, V.~Jemseena, A.~Kundu, K.~V. Kumar, S.~Sabhapandit, S.~N. Majumdar,
  S.~Redner, and A.~Dhar.
\newblock Steady state, relaxation and first-passage properties of a
  run-and-tumble particle in one-dimension.
\newblock {\em J. Stat. Mech-Theory E.}, 2018(4):043215, 2018.

\bibitem{redner2001guide}
S.~Redner.
\newblock {\em A guide to first-passage processes}.
\newblock Cambridge university press, 2001.

\bibitem{metzler2014first}
R.~Metzler, S.~Redner, and G.~Oshanin.
\newblock {\em First-passage phenomena and their applications}, volume~35.
\newblock World Scientific, 2014.

\bibitem{bressloff2022encounter}
P.~C. Bressloff.
\newblock Encounter-based model of a run-and-tumble particle.
\newblock {\em J. of Stat. Mech. Theory Exp.}, 2022(11):113206, 2022.

\bibitem{angelani2015run}
L.~Angelani.
\newblock Run-and-tumble particles, telegrapher’s equation and absorption
  problems with partially reflecting boundaries.
\newblock {\em J. Phys. A-Math. Theor.}, 48(49):495003, 2015.

\bibitem{guerin2016mean}
T.~Gu{\'e}rin, N.~Levernier, O.~B{\'e}nichou, and R.~Voituriez.
\newblock Mean first-passage times of non-{M}arkovian random walkers in
  confinement.
\newblock {\em Nature}, 534(7607):356--359, 2016.

\bibitem{larralde1997transport}
H.~Larralde.
\newblock Transport properties of a two-dimensional “chiral” persistent
  random walk.
\newblock {\em Phys. Rev. E}, 56(5):5004, 1997.

\bibitem{masoliver2017three}
J.~Masoliver.
\newblock Three-dimensional telegrapher's equation and its fractional
  generalization.
\newblock {\em Phys. Rev. E}, 96(2):022101, 2017.

\bibitem{furth1920brownian}
R.~Furth.
\newblock The brownian motion with consideration of the longevity of the
  direction of movement.
\newblock {\em Z. Phys.}, 2:244--256, 1920.

\bibitem{taylor1922diffusion}
G.~I. Taylor.
\newblock Diffusion by continuous movements.
\newblock {\em Proc. London Math. Soc.}, 2(1):196--212, 1922.

\bibitem{larralde2020first}
H.~Larralde.
\newblock First-passage probabilities and mean number of sites visited by a
  persistent random walker in one-and two-dimensional lattices.
\newblock {\em Phys. Rev. E}, 102(6):062129, 2020.

\bibitem{rws_on_latticesIII}
E.~W. Montroll.
\newblock Random walks on lattices. {III}. {C}alculation of first-passage times
  with application to exciton trapping on photosynthetic units.
\newblock {\em J. {M}ath. {P}hys.}, 10(4):753--765, 1969.

\bibitem{tejedor2012optimizing}
V.~Tejedor, R.~Voituriez, and O.~B{\'e}nichou.
\newblock Optimizing persistent random searches.
\newblock {\em Phys. Rev. Lett.}, 108(8):088103, 2012.

\bibitem{montroll1955effect}
E.~W. Montroll and R.~B. Potts.
\newblock Effect of defects on lattice vibrations.
\newblock {\em Physical Review}, 100(2):525, 1955.

\bibitem{szabo1984localized}
A.~Szabo, G.~Lamm, and G.~H. Weiss.
\newblock Localized partial traps in diffusion processes and random walks.
\newblock {\em Journal of statistical physics}, 34:225--238, 1984.

\bibitem{giuggioli2022spatio}
L.~Giuggioli and S.~Sarvaharman.
\newblock Spatio-temporal dynamics of random transmission events: from
  information sharing to epidemic spread.
\newblock {\em J. Phys. A-Math. Theor.}, 55(37):375005, 2022.

\bibitem{kenkre2021memory}
V.~M. Kenkre.
\newblock {\em Memory functions, projection operators, and the defect
  technique: some tools of the trade for the condensed matter physicist},
  volume 982.
\newblock Springer Nature, 2021.

\bibitem{shum2015hydrodynamic}
H.~Shum and E.~A. Gaffney.
\newblock Hydrodynamic analysis of flagellated bacteria swimming in corners of
  rectangular channels.
\newblock {\em Phys. Rev. E}, 92(6):063016, 2015.

\bibitem{van1998remarks}
N.~G. van Kampen.
\newblock Remarks on non-{M}arkov processes.
\newblock {\em Braz. J. Phys.}, 28:90--96, 1998.

\bibitem{masoliver1992solutions}
J.~Masoliver, J.~M. Porra, and G.~H. Weiss.
\newblock Solutions of the telegrapher’s equation in the presence of traps.
\newblock {\em Phys. Rev. A}, 45(4):2222, 1992.

\bibitem{landman1979stochastic2}
U.~Landman and M.~F. Shlesinger.
\newblock Stochastic theory of multistate diffusion in perfect and defective
  systems. ii. case studies.
\newblock {\em Phys. Rev. B}, 19(12):6220, 1979.

\bibitem{supp_mat}
See supplementary material for further details on the derivations.

\bibitem{barry_hughes_book}
B.~D. Hughes.
\newblock {\em Random walks and random environments Volume 1: Random Walks}.
\newblock Clarendon Press Oxford ; New York, 1995.

\bibitem{ernst1988random}
M.~H. Ernst.
\newblock Random walks with short memory.
\newblock {\em J. Stat. Phys.}, 53:191--201, 1988.

\bibitem{bechinger2016active}
C.~Bechinger, R.~Di~Leonardo, H.~L{\"o}wen, C.~Reichhardt, G.~Volpe, and
  G.~Volpe.
\newblock Active particles in complex and crowded environments.
\newblock {\em Rev. Mod. Phys.}, 88(4):045006, 2016.

\bibitem{kalay2012effects}
Z.~Kalay.
\newblock Effects of confinement on the statistics of encounter times: exact
  analytical results for random walks in a partitioned lattice.
\newblock {\em J. Phys. A-Math. Theor.}, 45(21):215001, 2012.

\bibitem{keller1953scope}
J.~B. Keller.
\newblock The scope of the image method.
\newblock {\em Commun. Pur. Appl. Math.}, 6(4):505--512, 1953.

\bibitem{masoliver1993solution}
J.~Masoliver, J.~M. Porra, and G.~H. Weiss.
\newblock Solution to the telegrapher’s equation in the presence of
  reflecting and partly reflecting boundaries.
\newblock {\em Phys. Rev. E}, 48(2):939, 1993.

\bibitem{montroll1956random}
E.~W. Montroll.
\newblock Random walks in multidimensional spaces, especially on periodic
  lattices.
\newblock {\em J. Soc. Ind. Appl. Math.}, 4(4):241--260, 1956.

\bibitem{beylkin2008fast}
G.~Beylkin, C.~Kurcz, and L.~Monz{\'o}n.
\newblock Fast algorithms for {H}elmholtz {G}reen's functions.
\newblock {\em P. Roy. Soc. A-Math. Phy.}, 464(2100):3301--3326, 2008.

\bibitem{chandrasekhar1943stochastic}
S.~Chandrasekhar.
\newblock Stochastic problems in physics and astronomy.
\newblock {\em Rev. Mod. Phys.}, 15(1):1, 1943.

\bibitem{proesmans2020phase}
K.~Proesmans, R.~Toral, and C.~Van~den Broeck.
\newblock Phase transitions in persistent and run-and-tumble walks.
\newblock {\em Physica A Stat.}, 552:121934, 2020.

\bibitem{smith2022exact}
N.~R. Smith, P.~Le~Doussal, S.~N. Majumdar, and G.~Schehr.
\newblock Exact position distribution of a harmonically confined run-and-tumble
  particle in two dimensions.
\newblock {\em Phys. Rev. E}, 106(5):054133, 2022.

\bibitem{miri2003persistent}
M.~Miri and H.~Stark.
\newblock Persistent random walk in a honeycomb structure: Light transport in
  foams.
\newblock {\em Phys. Rev. E}, 68(3):031102, 2003.

\bibitem{her1995geometric}
I.~Her.
\newblock Geometric transformations on the hexagonal grid.
\newblock {\em {IEEE Trans. Imag. Process.}}, 4(9):1213--1222, 1995.

\bibitem{marris2023exact}
D.~Marris, S.~Sarvaharman, and L.~Giuggioli.
\newblock Exact spatiotemporal dynamics of lattice random walks in hexagonal
  and honeycomb domains.
\newblock {\em Phys. Rev. E}, 107(5):054139, 2023.

\bibitem{spitzer1976principles}
F.~Spitzer.
\newblock {\em Principles of random walk}, volume~34.
\newblock SpringerVerlag, second edition, 1976.

\bibitem{larralde2015three}
H.~Larralde and F.~Leyvraz.
\newblock Three-dimensional diffusion with helical persistence.
\newblock {\em J. Phys. A-Math. Theor.}, 48(26):265001, 2015.

\bibitem{masoliver1993some}
J.~Masoliver, J.~M. Porra, and G.~H. Weiss.
\newblock Some two and three-dimensional persistent random walks.
\newblock {\em Physica A: Statistical Mechanics and its Applications},
  193(3-4):469--482, 1993.

\bibitem{hargus2021odd}
C.~Hargus, J.~M. Epstein, and K.~K. Mandadapu.
\newblock Odd diffusivity of chiral random motion.
\newblock {\em Phys. Rev. Lett.}, 127(17):178001, 2021.

\bibitem{kummel2013circular}
F.~K{\"u}mmel, B.~Ten~Hagen, R.~Wittkowski, I.~Buttinoni, R.~Eichhorn,
  G.~Volpe, H.~L{\"o}wen, and C.~Bechinger.
\newblock Circular motion of asymmetric self-propelling particles.
\newblock {\em Phys. Rev. Lett.}, 110(19):198302, 2013.

\bibitem{riedel2005self}
I.~H. Riedel, K.~Kruse, and J.~Howard.
\newblock A self-organized vortex array of hydrodynamically entrained sperm
  cells.
\newblock {\em Science}, 309(5732):300--303, 2005.

\bibitem{fuchter2023three}
P.~Fuchter and H.~Bloomfield-Gad{\^e}lha.
\newblock The three-dimensional coarse-graining formulation of interacting
  elastohydrodynamic filaments and multi-body microhydrodynamics.
\newblock {\em J. R. Soc. Interface}, 20(202):20230021, 2023.

\bibitem{abate1992numerical}
J.~Abate and W.~Whitt.
\newblock Numerical inversion of probability generating functions.
\newblock {\em Oper. Res. Lett.}, 12(4):245--251, 1992.

\bibitem{abate2000introduction}
J.~Abate, G.~L. Choudhury, and W.~Whitt.
\newblock An introduction to numerical transform inversion and its application
  to probability models.
\newblock In {\em Computational probability}, pages 257--323. Springer, 2000.

\bibitem{sarvaharman2020closed}
S.~Sarvaharman and L.~Giuggioli.
\newblock Closed-form solutions to the dynamics of confined biased lattice
  random walks in arbitrary dimensions.
\newblock {\em Phys. Rev. E}, 102(6):062124, 2020.

\bibitem{holhouse2023first}
J.~Holhouse and S.~Redner.
\newblock First-passage on disordered intervals.
\newblock {\em arXiv preprint arXiv:2307.08879}, 2023.

\bibitem{bonomo2021first}
O.~L. Bonomo and A.~Pal.
\newblock First passage under restart for discrete space and time: application
  to one-dimensional confined lattice random walks.
\newblock {\em Phys. Rev. E}, 103(5):052129, 2021.

\bibitem{giuggioli2023multi}
L.~Giuggioli, S.~Sarvaharman, D.~Das, D.~Marris, and T.~Kay.
\newblock Multi-target search in bounded and heterogeneous environments: a
  lattice random walk perspective.
\newblock {\em arXiv preprint arXiv:2311.00464}, 2023.

\bibitem{kac1947random}
M.~Kac.
\newblock Random walk and the theory of brownian motion.
\newblock {\em Am. Math. Mon.}, 54(7):369--391, 1947.

\bibitem{kulkarni2023first}
M.~Kulkarni and S.~N. Majumdar.
\newblock First detection probability in quantum resetting via random
  projective measurements.
\newblock {\em arXiv preprint arXiv:2305.15123}, 2023.

\bibitem{verechtchaguina2006first}
T.~Verechtchaguina, I.M. Sokolov, and L.~Schimansky-Geier.
\newblock First passage time densities in non-{M}arkovian models with
  subthreshold oscillations.
\newblock {\em Europhys. Lett.}, 73(5):691, 2006.

\bibitem{giuggioli2012predicting}
L.~Giuggioli, J.R. Potts, and S.~Harris.
\newblock Predicting oscillatory dynamics in the movement of territorial
  animals.
\newblock {\em J. R. Soc. Interface.}, 9(72):1529--1543, 2012.

\bibitem{kenkre2006exciton}
V.~M. Kenkre and P.~Reineker.
\newblock {\em Exciton dynamics in molecular crystals and aggregates}.
\newblock Springer, 2006.

\bibitem{das2022discrete}
D.~Das and L.~Giuggioli.
\newblock Discrete space-time resetting model: Application to first-passage and
  transmission statistics.
\newblock {\em J. Phys. A-Math. Theor.}, 55(42):424004, 2022.

\bibitem{das2023dynamics}
D.~Das and L.~Giuggioli.
\newblock Dynamics of lattice random walk within regions composed of different
  media and interfaces.
\newblock {\em J. Stat. Mech-Theory E.}, 2023(1):013201, 2023.

\bibitem{sarvaharman2022particle}
S.~Sarvaharman and L.~Giuggioli.
\newblock Particle-environment interactions in arbitrary dimensions: a unifying
  analytic framework to model diffusion with inert spatial heterogeneities.
\newblock {\em Phys. Rev. Res.}, 5(4):043281, 2023.

\bibitem{benichou2014first}
O.~B{\'e}nichou and R.~Voituriez.
\newblock From first-passage times of random walks in confinement to
  geometry-controlled kinetics.
\newblock {\em Phys. Rep.}, 539(4):225--284, 2014.

\bibitem{condamin2005first}
S.~Condamin, O.~B{\'e}nichou, and M.~Moreau.
\newblock First-passage times for random walks in bounded domains.
\newblock {\em Phys. Rev. Lett.}, 95(26):260601, 2005.

\bibitem{godec2016first}
A.~Godec and R.~Metzler.
\newblock First passage time distribution in heterogeneity controlled kinetics:
  going beyond the mean first passage time.
\newblock {\em Sci. Rep-UK}, 6(1):20349, 2016.

\bibitem{godec2016universal}
A.~Godec and R.~Metzler.
\newblock Universal proximity effect in target search kinetics in the
  few-encounter limit.
\newblock {\em Phys. Rev. X}, 6(4):041037, 2016.

\bibitem{schultz2014aging}
J.~H.~P. Schulz, E.~Barkai, and R.~Metzler.
\newblock Aging renewal theory and application to random walks.
\newblock {\em Phys. Rev. X}, 4:011028, 2014.

\bibitem{godoy1992reflection}
S.~Godoy and S.~Fujita.
\newblock Reflection principles for biased correlated walks. simple
  applications.
\newblock {\em J. Math. Phys.}, 33(9):2998--3003, 1992.

\bibitem{rossetto2018one}
V.~Rossetto.
\newblock The one-dimensional asymmetric persistent random walk.
\newblock {\em J. Stat. Mech-Theory E.}, 2018(4):043204, 2018.

\bibitem{garcia2007solution}
R.~Garc{\'\i}a-Pelayo.
\newblock Solution of the persistent, biased random walk.
\newblock {\em Physica A}, 384(2):143--149, 2007.

\bibitem{kay2022diffusion}
T.~Kay and L.~Giuggioli.
\newblock Diffusion through permeable interfaces: fundamental equations and
  their application to first-passage and local time statistics.
\newblock {\em Phys. Rev. Res.}, 4(3):L032039, 2022.

\bibitem{bressloff2022probabilistic}
P.C. Bressloff.
\newblock A probabilistic model of diffusion through a semi-permeable barrier.
\newblock {\em Proc. R. Soc. A}, 478(2268):20220615, 2022.

\bibitem{kosztolowicz2001random}
T.~Koszto{\l}owicz.
\newblock Random walk in a discrete and continuous system with a thin membrane.
\newblock {\em Phys. A: Stat. Mech. Appl.}, 298(3-4):285--296, 2001.

\bibitem{szasz1984persistent}
D.~Sz{\'a}sz and B.~T{\'o}th.
\newblock Persistent random walks in a one-dimensional random environment.
\newblock {\em J. Stat. Phys.}, 37(1-2):27--38, 1984.

\bibitem{hill1980discrete}
J.~M. Hill.
\newblock A discrete random walk model for diffusion in media with double
  diffusivity.
\newblock {\em ANZIAM J.}, 22(1):58--74, 1980.

\bibitem{hill1985general}
J.~M. Hill and B.D. Hughes.
\newblock On the general random walk formulation for diffusion in media with
  diffusivities.
\newblock {\em ANZIAM J.}, 27(1):73--87, 1985.

\bibitem{landman1979stochastic}
U.~Landman and M.~F. Shlesinger.
\newblock Stochastic theory of multistate diffusion in perfect and defective
  systems. i. mathematical formalism.
\newblock {\em Phys. Rev. B}, 19(12):6207, 1979.

\bibitem{zhang2019persistent}
S.~Zhang, A.~Chong, and B.~D. Hughes.
\newblock Persistent exclusion processes: Inertia, drift, mixing, and
  correlation.
\newblock {\em Phys. Rev. E}, 100(4):042415, 2019.

\bibitem{teomy2019transport}
E.~Teomy and R.~Metzler.
\newblock Transport in exclusion processes with one-step memory: density
  dependence and optimal acceleration.
\newblock {\em J. Phys. A-Math. Theor.}, 52(38):385001, 2019.

\bibitem{gavagnin2018modeling}
E.~Gavagnin and C.~A. Yates.
\newblock Modeling persistence of motion in a crowded environment: The
  diffusive limit of excluding velocity-jump processes.
\newblock {\em Phys. Rev. E}, 97(3):032416, 2018.

\bibitem{regnier2023universal}
L.~R{\'e}gnier, M.~Dolgushev, S.~Redner, and O.~B{\'e}nichou.
\newblock Universal exploration dynamics of random walks.
\newblock {\em Nat. Comms.}, 14(1):618, 2023.

\bibitem{regnier2023record}
L.~R{\'e}gnier, M.~Dolgushev, and O.~B{\'e}nichou.
\newblock Record ages of non-{M}arkovian scale-invariant random walks.
\newblock {\em Nat. Commun.}, 14(1):6288, 2023.

\bibitem{cressoni2007amnestically}
J.~C. Cressoni, M.~A.~A. da~Silva, and G.M. Viswanathan.
\newblock Amnestically induced persistence in random walks.
\newblock {\em Phys, Rev. Lett.}, 98(7):070603, 2007.

\bibitem{schutz2004elephants}
G.~M. Sch{\"u}tz and S.~Trimper.
\newblock Elephants can always remember: Exact long-range memory effects in a
  non-{M}arkovian random walk.
\newblock {\em Phys. Rev. E}, 70(4):045101, 2004.

\bibitem{kenkre2007analytic}
V.M. Kenkre.
\newblock Analytic formulation, exact solutions, and generalizations of the
  elephant and the alzheimer random walks.
\newblock {\em arXiv preprint arXiv:0708.0034}, 2007.

\bibitem{boettcher2013renormalization}
S.~Boettcher, S.~Falkner, and R.~Portugal.
\newblock Renormalization group for quantum walks.
\newblock In {\em J. Phys. Conf. Ser.}, volume 473, page 012018. IOP
  Publishing, 2013.

\bibitem{klafter2011first}
J.~Klafter and I.~M. Sokolov.
\newblock {\em First steps in random walks: from tools to applications}.
\newblock OUP Oxford, 2011.

\bibitem{bernstein2009matrix1}
D.~S. Bernstein.
\newblock {\em Matrix mathematics: theory, facts, and formulas}.
\newblock Princeton University Press, 2009.

\end{thebibliography}
\bibliographystyle{unsrt}

\pagebreak
\begin{center}
\textbf{\large Supplemental Materials: Persistent and anti-Persistent Motion in Bounded Space: Resolution of the
First-Passage Problem}

\end{center}
%%%%%%%%%% Merge with supplemental materials %%%%%%%%%% %%%%%%%%% Prefix a "S" to all equations, figures, tables and
%reset the counter %%%%%%%%%%
\setcounter{equation}{0}
\setcounter{figure}{0}
\setcounter{table}{0}
\setcounter{page}{1}
\makeatletter
\renewcommand{\theequation}{S\arabic{equation}} \renewcommand{\thefigure}{S\arabic{figure}}

%%%%%%%%%% Prefix a "S" to all equations, figures, tables and reset the counter %%%%%%%%

\section*{Continuum Limit of the Master Equation}\label{app: limit} 
It is straightforward to reduce the two state Master equation i.e., Eq. (\ref{eq: Master_eq}) in the main text to the telegrapher's equation when $c=0$ \cite{weiss1994aspects},
that is when the walker has no option to stay. Here we show the limiting procedure when $c\not=0$. 

We define the time between successive steps as $\Delta t$ and the spacing between the lattice sites as $\Delta x$ and
following standard procedures, see e.g., \cite{weiss1994aspects}, we expand Eq. (\ref{eq: Master_eq}) to first order in
$\Delta t$ and $\Delta x$
\begin{equation}
 \begin{aligned}
     Q_1 + \Delta t \frac{\partial Q_1}{\partial t} &= f\left(Q_1 - \Delta x \frac{\partial Q_1}{\partial x}\right)+ b\left(Q_2 - \Delta x \frac{\partial Q_2}{\partial x}\right) + cQ_1, \\
     Q_2 + \Delta t \frac{\partial Q_2}{\partial t} &= f\left(Q_2 + \Delta x \frac{\partial Q_2}{\partial x}\right)+ b\left(Q_1 + \Delta x \frac{\partial Q_1}{\partial x}\right) + cQ_2,
 \end{aligned}
\end{equation}
where $Q_1$ and $Q_2$ represent the two states. We define a scaling limit for the probabilities, namely 
\begin{equation}
 f = \alpha - \frac{\Delta t}{2T},  \quad b = \frac{\Delta t}{2T} \quad \text{and} \quad c = 1 - \alpha,
\end{equation}
where $0 \leq \alpha \leq 1$, $T$ has the dimensions of time and $\lim_{\Delta x, \Delta t \to 0}\frac{\Delta x}{\Delta
t}=v$. After some algebra, dividing through by $\Delta t$ and letting $\Delta t$ and $\Delta x$ tend to zero, we find
\begin{equation}
 \begin{aligned}
     \frac{\partial Q_1}{\partial t} &= \frac{1}{2T}(Q_2 - Q_1) - \alpha v \frac{\partial Q_1}{\partial x} \\
     \frac{\partial Q_2}{\partial t} &= \frac{1}{2T}(Q_1 - Q_2) + \alpha v \frac{\partial Q_2}{\partial x}.
 \end{aligned}  
 \label{eq: coupled_eqs}
\end{equation}
Defining $U = Q_1 + Q_2$ and $W = Q_1 - Q_2$ and summing and subtracting the two lines of Eq. (\ref{eq: coupled_eqs}),
one obtains
\begin{equation}
 \frac{\partial ^2 U}{\partial t^2} = -\frac{1}{T} \frac{\partial U}{\partial t} + \alpha^2 v^2\frac{\partial^2 U}{\partial x^2},
 \label{eq: telegraphers}
\end{equation}
which shows that $\alpha=1-c$ is simply a rescaling of the velocity $v$.

\section*{Derivation of the Unbounded Propagator}\label{app: unbounded_prop} 

Analogously to the $d=1$ case, in higher dimensions one writes the Master equation for the dynamics into each state. As
an example we show explicitly the $d=2$ ($Z = 4$) case, which gives the following set of four coupled equations
\begin{equation}
    \begin{aligned}
        Q(\bm{n}, 1, t \!+\! 1) &= fQ(n_1-1, n_2, 1, t) \!+\! bQ(n_1-1, n_2, 2, t)\!+\! \ell Q(n_1-1, n_2, 3, t)\!+\! \ell Q(n_1-1, n_2, 4, t) \!+\! c^{(2)}Q(n_1, n_2, 1, t),\\
        Q(\bm{n}, 2, t \!+\! 1) &= bQ(n_1+1, n_2, 1, t) \!+\! fQ(n_1+1, n_2, 2, t)\!+\! \ell Q(n_1+1, n_2, 3, t)\!+\! \ell Q(n_1+1, n_2, 4, t) \!+\! c^{(2)}Q(n_1, n_2, 2, t),\\
        Q(\bm{n}, 3, t \!+\! 1) &= \ell Q(n_1, n_2-1, 1, t) \!+\! \ell Q(n_1, n_2-1, 2, t)\!+\! fQ(n_1, n_2-1, 3, t)\!+\! bQ(n_1, n_2-1, 4, t) \!+\! c^{(2)}Q(n_1, n_2, 3, t),\\
        Q(\bm{n}, 4, t \!+\! 1) &= \ell Q(n_1, n_2+1, 1, t) \!+\! \ell Q(n_1, n_2+1, 2, t)\!+\! bQ(n_1, n_2+1, 3, t)\!+\! fQ(n_1, n_2+1, 4, t) \!+\! c^{(2)}Q(n_1, n_2, 4, t),
    \end{aligned}
    \label{eq: 2d_ME_app}
\end{equation}
where the movement parameters are defined in the first paragraph of Sec. \ref{sec: 2D} of the main text. 

From Eq. (\ref{eq: 2d_ME_app}) it is straightforward to write a Master equation in matrix form as  
\begin{equation}
    \begin{aligned}
    &\bm{Q}(\bm{n}, t + 1) = \\ & 
    \mathbb{A}^{(1)}_4 \cdot \bm{Q}(n_1-1, n_2, t) + \mathbb{B}^{(1)}_4 \cdot \bm{Q}(n_1+1, n_2, t)+
    \mathbb{A}^{(2)}_4 \cdot \bm{Q}(n_1, n_2-1, t) + \mathbb{B}^{(2)}_4 \cdot \bm{Q}(n_1, n_2+1, t) + \mathbb{C}_4 \cdot \bm{Q}(n_1, n_2, t), 
    \end{aligned}
\end{equation} 
where 
\begin{equation}
    \mathbb{A}^{(1)}_4 = \begin{bmatrix}
        f  & b & \ell & \ell \\
        0 & 0 & 0 & 0 \\
        0 & 0 & 0 & 0 \\
        0 & 0 & 0 & 0 \\
    \end{bmatrix}, \; \mathbb{B}^{(1)}_4 = \begin{bmatrix}
        0 & 0 & 0 & 0 \\
        b & f & \ell & \ell \\
        0 & 0 & 0 & 0 \\
        0 & 0 & 0 & 0 \\
    \end{bmatrix}, \;
    \mathbb{A}^{(2)}_4 = \begin{bmatrix}
        0 & 0 & 0 & 0 \\
        0 & 0 & 0 & 0 \\
        \ell & \ell & f & b \\
        0 & 0 & 0 & 0 \\
    \end{bmatrix}, \;  \mathbb{B}^{(2)}_4 = \begin{bmatrix}
        0 & 0 & 0 & 0 \\
        0 & 0 & 0 & 0 \\
        0 & 0 & 0 & 0 \\
        \ell & \ell & b & f \\
    \end{bmatrix} \; \text{and} \; \mathbb{C}_4 = c^{(2)} \mathbb{I}.
\end{equation}

In fact, this procedure can be generalised to arbitrary dimension $d$, and one can write a matricial Master equation, also
reported in Eq. (\ref{eq: Dd_ME_matrix}) of the main text, of the form
\begin{equation}
    Q(\bm{n}, t+1) = \sum_{i=1}^{d}\left\{\mathbb{A}_{2d}^{(i)}\cdot Q(\bm{n}-\bm{e}_i, t) + \mathbb{B}_{2d}^{(i)}\cdot Q(\bm{n}+\bm{e}_i, t)\right\}  + \mathbb{C}_{2d}\cdot Q(\bm{n}, t), 
    \label{eq: app_Dd_ME_matrix}
\end{equation} 
where $\bm{e}_i$ is a unit vector along dimension $i$.

While the dynamics across the states is dependent on the internal degree of freedom, the site dynamics are
translationally invariant. This allows one to apply the discrete Fourier transform $\widehat{f}(\xi) =
\sum_{n=-\infty}^{\infty}e^{i n \xi}f(n)$, and we also introduce the $z$-transform $\widetilde{f}(z) =
\sum_{t=0}^{\infty}z^tf(t)$, to find the unbounded solution in Fourier and $z$-domain as
\cite{barry_hughes_book}
\begin{equation}
 \widehat{\widetilde{\bm{Q}}}(\xi_1, ..., \xi_d, z) = \big[\mathbb{I} - z\bm{\lambda}(\xi_1, ..., \xi_d)\big]^{-1}\cdot \widehat{\bm{Q}}(\xi_1, ..., \xi_d, 0).
 \label{eq: unbounded_fz}
\end{equation}
Introducing the Fourier transform of the initial condition, namely $\sum_{n_1=-\infty}^{\infty} ...
\sum_{n_d=-\infty}^{\infty}e^{i \bm{\xi}^{\intercal}\cdot \bm{n}}\left(\prod_{i=1}^{d}\delta_{n,
n_{0_i}}\right)\bm{U_{m_0}} =  e^{i\bm{\xi}^{\intercal}\cdot \bm{n}_0}\bm{U_{m_0}}$, with $\bm{\xi}=(\xi_1,...,\xi_d)$,
into Eq. (\ref{eq: unbounded_fz}), we find
\begin{equation}
    \widehat{\widetilde{\bm{Q}}}_{\bm{n}_0}(\xi_1, ..., \xi_d, z) = e^{i\bm{\xi}^{\intercal}\cdot \bm{n}_0}\left[\mathbb{I} - z\bm{\lambda}(\xi_1, ..., \xi_d)\right]^{-1}\cdot \bm{U_{m_0}},
    \label{eq: unbounded_fz1}
   \end{equation}
which gives through an inverse Fourier transformation the formal lattice Green's function 
\begin{equation}
    \widetilde{\bm{Q}}_{\bm{n}_0}(\bm{n}_1, ..., \bm{n}_d, z) = \frac{1}{(2\pi)^d}\int_{-\pi}^{\pi} ... \int_{-\pi}^{\pi} e^{-i \bm{\xi}^{\intercal}\cdot(\bm{n} - \bm{n}_0)}\left[\mathbb{I} - z\bm{\lambda}(\xi_1, ..., \xi_d)\right]^{-1}\cdot \bm{U_{m_0}} \text{d}^{d}\xi,
    \label{eq: unbounded_z}
\end{equation}
where the structure of the matrix $\bm{\lambda}(\xi_1, ..., \xi_d)$ is discussed in \ref{app: s_func} of the
main text. As noted in ref. \cite{barry_hughes_book}, to obtain the state level probability one needs only to integrate
over the specific element in the matrix in Eq. (\ref{eq: unbounded_z}), e.g., $\widetilde{Q}_{\bm{n}_0, m_0}(\bm{n}_1,
..., \bm{n}_d, m, z) = (2\pi)^{-d}\int_{-\pi}^{\pi} ...
\int_{-\pi}^{\pi}\bm{e}_m^{\intercal}\cdot\widetilde{\bm{Q}}_{\bm{n}_0}(\bm{n}_1, ..., \bm{n}_d,
z)\cdot\bm{e}_{m_0}\text{d}^{d}\xi$.

We note here that other representations of the unbounded occupation probability viz. Eqs. (\ref{eq: unbounded_fz1}) and
(\ref{eq: unbounded_z}) are known. For example, in ref. \cite{ernst1988random}, a form of the propagator is given where
the walk is decomposed into the movement steps along each spatial dimension and the full occupation probability is
obtained via a summation over each spatial direction. Such a summation yields the same outcome as performing
$\bm{Q}_{\bm{n}_0}(\bm{n}_1, ..., \bm{n}_d, z) = \sum_{i=1}^{2d}\widetilde{\bm{Q}}_{\bm{n}_0}(\bm{n}_1, ..., \bm{n}_d,
z)_i$ in the present context. However, the present approach is more flexible as it is able to discern exactly the state
level  occupation probability, easily include a sojourn probability $c$ and an arbitrary distribution over
$\bm{U_{m_0}}$, with very little algebraic burden. 

\subsection*{Closed-form solution for $d=1$ case:} 
In one dimension it is possible to calculate explicitly the integral in Eq. (\ref{eq: unbounded_z}). With $\bm{U_{m_0}}=
[\alpha_1, \alpha_2]^{\intercal}$ and $\bm{\lambda}(\xi)$ defined in Eq. (\ref{eq: 1D_sf}) of the main text one has
\begin{equation}
 \widetilde{\bm{Q}}_{n_0}(n, z) = \int_{-\pi}^{\pi} \frac{e^{-i\xi(n- n_0)}}{1-2z\big[c - f \cos(\xi)\big] + z^2 \big[f^2 - b^2 + c^2 + 2fc\cos(\xi)\big]} 
 \begin{bmatrix}
     1-z\left(c + fe^{-i\xi}\right) & bze^{i\xi} \\
     bze^{-i\xi} & 1-z\left(c + fe^{i\xi}\right)
 \end{bmatrix}\cdot\begin{bmatrix}
     \alpha_1 \\
     \alpha_2
 \end{bmatrix}\text{d}\xi.
 \label{eq: unbounded1}
\end{equation}
which is the explicit version Eq. (\ref{eq: internal_states_LGF}) in the main text. By rewriting Eq. (\ref{eq:
unbounded1}) as 
\begin{equation}
    \widetilde{\bm{Q}}_{n_0}(n, z) = \int_{-\pi}^{\pi} \frac{e^{-i\xi(n- n_0)}}{(1-2c z)\big[1-z\varepsilon\left(e^{i \xi}+e^{-i \xi}\right) + \delta z^2 \big]} 
    \begin{bmatrix}
        1-z\left(c + fe^{-i\xi}\right) & bze^{i\xi} \\
        bze^{-i\xi} & 1-z\left(c + fe^{i\xi}\right)
    \end{bmatrix}\cdot\begin{bmatrix}
        \alpha_1 \\
        \alpha_2
    \end{bmatrix}\text{d}\xi,
    \label{eq: unbounded2}
   \end{equation}
and defining  
\begin{align}
r_{\pm}(z)=
\frac{1+\delta z^2}{2\varepsilon z}\left(1\pm\sqrt{1-\frac{4\varepsilon^2z^2}{[1+\delta z^2]^2}}\right),
\label{eqn:r_pm}
\end{align}
where $r_+=r_-^{-1}$, $|r_-|<1<|r_+|$, $\varepsilon = f(1-cz)(1-2cz)^{-1}$ and $\delta = (f^2-b^2+c^2)(1-2cz)^{-1}$ one
can follow closely to the solution of a similar integral \cite{godoy1992reflection} and find the explicitly generating
functions for $\widetilde{Q}_{n_0}(n, 1, z)$ and $\widetilde{Q}_{n_0}(n, 2, z)$. Namely, 
\begin{equation}
\begin{bmatrix}
\widetilde{Q}_{n_0, 1}(n,1, z) \\
\widetilde{Q}_{n_0, 1}(n, 2, z)
\end{bmatrix} = \frac{\alpha_1}{(1-2cz)\sqrt{\left(1+\delta z^2\right)^2-4\varepsilon^2z^2}}
\begin{bmatrix}
 \frac{1-cz}{r_+(z)^{|n-n_0|}}-\frac{fz}{r_+(z)^{|n-n_0+1|}}\\
 \frac{bz}{r_+(z)^{|n-n_0+1|}}
\end{bmatrix}
\label{eqn:exact_sol_za}
\end{equation}
when the walker is assumed to have reached $n_0$ from the left at $t=0$, and 
\begin{equation}
\begin{bmatrix}
\widetilde{Q}_{n_0, 2}(n, 1, z) \\
\widetilde{Q}_{n_0, 2}(n, 2, z)
\end{bmatrix}
=\frac{\alpha_2}{(1-2cz)\sqrt{\left(1+\delta z^2\right)^2-4\varepsilon^2z^2}}
\begin{bmatrix}
\frac{bz}{r_+(z)^{|n-n_0-1|}} \\
\frac{1-cz}{r_+(z)^{|n-n_0|}}-\frac{fz}{r_+(z)^{|n-n_0-1|}},
\end{bmatrix}.
\label{eqn:exact_sol_zb}
\end{equation}
when it reached from the right.

The solution at each state is irrespective of how the walker arrived at $n_0$ at time $t=0$ and we sum the rows in Eqs.
(\ref{eqn:exact_sol_za}) and (\ref{eqn:exact_sol_zb}) to find
\begin{equation}
    \begin{aligned}
            \begin{bmatrix}
                \widetilde{Q}_{n_0}(n, 1, z) \\
                \widetilde{Q}_{n_0}(n, 2, z) 
            \end{bmatrix} = \frac{1}{(1-2cz)\sqrt{\left(1+\delta z^2\right)^2-4\varepsilon^2z^2}}\begin{bmatrix}
                    \frac{\alpha_1(1-cz)}{r(z)^{|n-n_0|}} - \frac{\alpha_1 fz}{r(z)^{|n-n_0+1|}} + \frac{\alpha_2 bz}{r(z)^{|n-n_0-1|}} \\
                    \frac{\alpha_2(1-cz)}{r(z)^{|n-n_0|}} - \frac{\alpha_2 fz}{r(z)^{|n-n_0-1|}} + \frac{\alpha_1 bz}{r(z)^{|n-n_0+1|}} 
            \end{bmatrix},
    \end{aligned}
    \label{eq: unbounded_propagator}
\end{equation}
where for simplicity we have dropped the subscript + from the definition of $r_+$. A summation of the two states then
gives Eq. (\ref{eqn:sol_1d}) of the main text. 

\section*{Periodic Boundary Conditions}\label{app: dD_periodic} The periodic images in arbitrary dimensions are written
as \cite{barry_hughes_book}
\begin{equation}
 \begin{aligned}
         \widetilde{\bm{P}}^{(p)}_{n_{0_1}, ..., n_{0_d}}(n_1, ..., n_d, z) = \sum_{\kappa_1 = -\infty}^{\infty} ...\sum_{\kappa_d = -\infty}^{\infty} \widetilde{\bm{Q}}_{n_{0_1} + \kappa_1 N_1, ...,n_{0_d} + \kappa_d N_d }(n_1, ..., n_d, z),
 \end{aligned}
\end{equation}
inserting this into Eq. (\ref{eq: unbounded_z}) we find that
\begin{equation}
 \widetilde{\bm{P}}^{(p)}_{n_{0_1}, ..., n_{0_d}}(n_1, ..., n_d, z) = \frac{1}{(2\pi)^d}\int_{-\pi}^{\pi}...\int_{-\pi}^{\pi} \sum_{\kappa_1=-\infty}^{\infty}...\sum_{\kappa_d=-\infty}^{\infty}e^{-i (\bm{n}^{\intercal}-\bm{n}_0^{\intercal}+N\bm{\kappa}^{\intercal})\cdot \bm{\xi}}\left[ \mathbb{I}-z\bm{\lambda}(\xi_1, ..., \xi_d)\right]^{-1}\cdot\bm{U_{m_0}}\text{d}^{d}\xi.
\end{equation}

Using the $d$-dimensional Poisson summation formula
\begin{equation}
 \sum_{\kappa_1= -\infty}^{\infty} ... \sum_{\kappa_d= -\infty}^{\infty}\mu^d\delta(\bm{\xi} - \bm{\kappa}^{\intercal}\cdot\bm{\mu}) =   \sum_{\kappa_1= -\infty}^{\infty} ... \sum_{\kappa_d= -\infty}^{\infty} e^{-i\bm{\kappa}^{\intercal} \cdot \bm{N}\bm{\xi}}
 \label{eq: psm}
\end{equation}
where we define $\mu^d = \prod_{j=1}^{d}2\pi/N_j$ and $\bm{\mu} = \left[2\pi/N_1, ..., 2\pi/N_d\right]$, one finds
\begin{equation}
 \widetilde{\bm{P}}^{(p)}_{n_{0_1}, ..., n_{0_d}}(n_1, ..., n_d, z) \!=\! \frac{1}{N^d}\int_{-\varepsilon_1}^{2\pi-\varepsilon_1}...\int_{-\varepsilon_d}^{2\pi-\varepsilon_d} \!\sum_{\kappa_1=-\infty}^{\infty}...\sum_{\kappa_d=-\infty}^{\infty}\!e^{-i (\bm{n}^{\intercal}-\bm{n_0}^{\intercal})\cdot \bm{\xi}}\delta(\bm{\xi} - \bm{\kappa}^{\intercal}\cdot\bm{\mu})\left[ \mathbb{I}-z\bm{\lambda}(\xi_1, ..., \xi_d)\right]^{-1}\cdot\bm{U_{m_0}}\text{d}^{d}\xi,
    \label{eq: periodic_integral}
\end{equation}
where $N^d = \prod_{j=1}^{d}N_j$ and the periodicity of the function has allowed us to replace the region of integration from $V = [-\pi, \pi]^d$ to $[-\varepsilon_d, 2\pi -\varepsilon_d]$ along
each dimension, where 
$0 < \varepsilon_d < 2\pi/N_d$. The singularities of $\delta(\bm{\xi} -
\bm{\kappa}^{\intercal}\cdot\bm{\mu})$ lie in the integral region if and only if $\bm{\kappa}$ is a site in the lattice
(for a general discussion of solving integrals of this type see \cite{barry_hughes_book}). We may use this 
property to perform the intergrals in Eq.
(\ref{eq: periodic_integral}) and obtain
\begin{equation}
 \widetilde{\bm{P}}^{(p)}_{n_{0_1}, ..., n_{0_d}}(n_1, ..., n_d, z) = \frac{1}{N^d}\sum_{\kappa_1=0}^{N-1}...\sum_{\kappa_d=0}^{N-1}e^{-i (\bm{n}^{\intercal}-\bm{n_0}^{\intercal})\cdot \frac{2\pi}{N}\bm{\kappa}}\left[ \mathbb{I}-z\bm{\lambda}\left(\frac{2\pi\kappa_1}{N}, ..., \frac{2\pi\kappa_d}{N}\right)\right]^{-1}\cdot\bm{U_{m_0}},
 \label{eq: per_int} 
\end{equation}
which after taking the inverse $z$-transform via the identity $\sum_{t=0}^{\infty}z^t\bm{\lambda}(\bm{\xi})^t=
\left[\mathbb{I}-z\bm{\lambda}(\bm{\xi})\right]^{-1}$, one finds Eq. (\ref{eq: high_dim}) of the main text.

In the $1d$ case, one finds, explicitly, the probability over the entire site as
\begin{equation}
 \begin{aligned}
 P^{(p)}_{n_0}(n, t) =& \frac{1}{N}\sum_{\kappa=0}^{N-1}\exp\left(\frac{-2\pi i\kappa(n-n_0)}{N}\right)\frac{be^{-\frac{2\pi i \kappa}{N}}}{2\sqrt{b^2-f^2+f^2\cos^2\left(\frac{2\pi \kappa}{N}\right)}}\times \\ &
 \left[(\lambda_-)^t\left(1+v_-\left(\frac{2\pi\kappa}{N}\right)\right)\left(\alpha_2 v_+\left(\frac{2\pi\kappa}{N}\right)-\alpha_1\right)+(\lambda_+)^t\left(1+v_+\left(\frac{2\pi\kappa}{N}\right)\right)\left(\alpha_2 v_-\left(\frac{2\pi\kappa}{N}\right)-\alpha_1\right) \right], 
 \end{aligned}
 \label{eq: scalar_periodic1}
\end{equation}
where $v_{\pm}\left(\frac{2\pi\kappa}{N}\right) = \frac{1}{2b}\left[-f-fe^{\frac{4\pi i \kappa}{N}} \pm
2e^{\frac{2\pi i \kappa}{N}}\sqrt{b^2-f^2+f^2\cos^{2}\left(\frac{2\pi\kappa}{N}\right)}\right]$, and $\lambda_{\pm}$ are
defined in Eq. (\ref{eq: lambdapm}) of the main text. Equation (\ref{eq: scalar_periodic1}) reduces to Eq. (\ref{eq:
scalar_periodic}) of the main text when $\alpha_1 = \alpha_2$.

\section*{Reflective Boundary Conditions}\label{app: supp_dD_reflective} We detail here the corresponding procedure for
the reflective boundary condition. 
\subsection*{Derivation of the $1d$ Semi-bounded Propagator}\label{app: semi-bounded}
The image set used, for the semi-infinite propagator, is dependent on whether the reflection is to the left or the right
of the initial condition. We assume, without loss of generality, that the barrier is placed between sites $n=1$ and $n=0$ (with $n_0 \geq 1$)
meaning $P(1, \{2, 1\}, t) = P(0, \{1, 2\}, t)$. 

Therefore, in the symmetric $\alpha_1 = \alpha_2 = \frac{1}{2}$ case, the semi-bounded propagator,
$\widetilde{\bm{P}}^{(\mathdutchcal{r})}_{n_0}(n, z)$, can be written as
\begin{equation}
 \widetilde{\bm{P}}^{(\mathdutchcal{r})}_{n_0}(n, z) = \widetilde{\bm{Q}}_{n_0}(n, z) + \widetilde{\bm{Q}}_{-n_0 + 1}(n, z),
\end{equation}
which is written, after summing the components from Eq. (\ref{eq: unbounded_propagator}), as 
\begin{equation}
 \begin{aligned}
     \widetilde{P}^{(\mathdutchcal{r})}_{n_0}(n, z) = \frac{1}{(1-2cz)\sqrt{\left(1+\delta z^2\right)^2-4\varepsilon^2z^2}}\bigg[&\frac{1-cz}{r(z)^{|n-n_0|}} + \frac{(b-f)z}{2r(z)^{|n-n_0+1|}} + \frac{(b-f)z}{2r(z)^{|n-n_0-1|}} + \\
     &\frac{1-cz}{r(z)^{|n+n_0-1|}} + \frac{(b-f)z}{2r(z)^{|n+n_0|}} + \frac{(b-f)z}{2r(z)^{|n+n_0-1|}}\bigg].
 \end{aligned}
     \label{eq: semi-bounded_ref}
\end{equation}
In the ($b=0$) ballistic limit, the boundary condition can be simplified to $P(0, 1, t) = P(1, 2, t)$ as there will
never be an instance of the walker, entering state $m=1$ and then crossing the boundary at a later time. In fact, the
trajectories in state $m=1$, will remain in $m=1$ for all time, meaning the image of this part of the dynamics will
deterministically travel towards minus infinity and never interact with the domain of interest. As such, one can write
the propagator in this case as 
\begin{align}
 \widetilde{P}^{(\mathdutchcal{r})}_{n_0}(n, z) &= \widetilde{Q}_{n_0}(n, 1, z) + \widetilde{Q}_{n_0}(n, 2, z) + \widetilde{Q}_{-n_0+1}(n, 2, z)=\frac{1}{(1-2cz)\sqrt{\left(1+\delta z^2\right)^2-4\varepsilon^2z^2}}\bigg[\frac{1-cz}{r(z)^{|n-n_0|}} \nonumber \\
&+ \frac{1-cz}{r(z)^{|n+n_0-1|}} - \frac{zf}{2}\bigg(\frac{1}{r(z)^{|n-n_0+1|}} + \frac{1}{r(z)^{|n-n_0-1|}} + \frac{1}{r(z)^{|n+n_0|}}\bigg)\bigg].
 \label{eq: ballistic_semi_bound}
\end{align}

\subsection*{Fully Bounded Propagator}
We derive here the propagator in arbitrary dimensions. The method of images required to satisfy the 
`bouncing' reflective boundary condition along the $j^{\text{th}}$ dimension may be written as
\cite{LucaPRX},
\begin{equation}
    \begin{aligned}
            \widetilde{\bm{P}}^{(r)}_{n_{0_1}, ..., n_{0_j}, n_{0_d}}(n_{1}, ..., n_{j}, ..., n_{d}, z) = \sum_{\kappa_j = -\infty}^{\infty} & \bigg\{\widetilde{ \bm{Q}}_{n_{0_1}, ...,n_{0_j} + 2\kappa_j N_j, ..., n_{0_d}}(n_{1}, ..., n_{j}, ..., n_{d}, z) \\
            & + \widetilde{ \bm{Q}}_{n_{0_1}, ...,-n_{0_j} +1+ 2\kappa_j N_j, ..., n_{0_d}}(n_{1}, ..., n_{j}, ..., n_{d}, z)\bigg \} ,
    \end{aligned}
    \label{eq: reflective_images}
   \end{equation}
where we have related the position of the walker in any dimension to its image set along that axis independently of any other dimension, since 
the nearest-neighbour transition probabilities dictate the walker may never interact with two boundaries simultaneously.

Writing the unbounded propagator, Eq. (\ref{eq: unbounded_z}) (also given in Eq. (\ref{eq: internal_states_LGF}) of the main text), as 
\begin{equation}
    \widetilde{\bm{Q}}_{n_{0_1}, ..., n_{0_d}}(n_1, ..., n_d, z) = \frac{1}{(2\pi)^d}\int_{-\pi}^{\pi} ... \int_{-\pi}^{\pi} \left[\prod_{j=1}^{d}e^{-i \xi_j\left(n_j - n_{0_j}\right)}\right]\left[\mathbb{I} - z\bm{\lambda}(\xi_1, ..., \xi_d)\right]^{-1}\cdot \bm{U_{m_0}} \text{d}^{d}\xi,
    \label{eq: unbounded_z}
\end{equation}
and applying the method of images along each dimension independently, we find the reflective propagator as
\begin{equation}
 \begin{aligned}
     &\widetilde{\bm{P}}^{(r)}_{n_{0_1}, ..., n_{0_d}}(n_1, ..., n_d, z) = \\
     &\frac{1}{(2 \pi)^d}\int_{-\pi}^{\pi} ... \int_{-\pi}^{\pi}\sum_{\kappa_1 = -\infty}^{\infty} ... \sum_{\kappa_d = -\infty}^{\infty}\left[\prod_{j=1}^{d}e^{-i(n_j-n_{0_j})\xi_j} + e^{-i(n_j+n_{0_j}-1)\xi_j}\right]e^{i\bm{\kappa}^{\intercal}\cdot 2\bm{N}\bm{\xi}}\left[\mathbb{I}-z\bm{\lambda}\left(\xi_1, ..., \xi_d\right)\right]^{-1}\cdot\bm{U_{m_0}}\text{d}^{d}\xi,
 \end{aligned}
\end{equation}
where we have swapped the order of integration and summation as is permissible in theory of generalised functions \cite{barry_hughes_book}.

Shifting the region of integration from $V = [-\pi, \pi]^d$ to $[-\pi-\varepsilon_j, \pi-\varepsilon_j]$ along each dimension and applying Eq. (\ref{eq: psm}) where now $\mu^d = \prod_{j=1}^{d}\frac{\pi}{N_j}$ and $\bm{\mu} = \left[\pi/N_1, ..., \pi/N_d\right]$, one obtains
\begin{equation}
 \begin{aligned}
     &\widetilde{\bm{P}}^{(r)}_{n_{0_1}, ..., n_{0_d}}(n_1, ..., n_d, z) = \\ 
     &\frac{1}{(2 N)^d}\int_{-\pi-\varepsilon_1}^{\pi-\varepsilon_1} \!...\! \int_{-\pi-\varepsilon_d}^{\pi-\varepsilon_d}\!\sum_{\kappa_1 = -\infty}^{\infty}\! ... \!\sum_{\kappa_1 = -\infty}^{\infty}\!\delta\left(\bm{\xi}-\bm{\kappa}^{\intercal}\cdot\bm{\mu}\right)\!  \left[\prod_{j=1}^{d}e^{-i(n_j-n_{0_j})\xi_j} \!+\! e^{-i(n_j+n_{0_j}-1)\xi_j}\right]\!\left[\mathbb{I}\!-\!z\bm{\lambda}\left(\xi_1, ..., \xi_d\right)\right]^{-1}\!\cdot\bm{U_{m_0}}\!\text{d}\xi.
 \end{aligned}
 \label{eq: reflect_integral1}
\end{equation}
We may now solve the integral via the same procedure as Eq. (\ref{eq: per_int}), that is performing a summation over the singularities of Eq. (\ref{eq: reflect_integral1}), leading to
\begin{equation}
    \begin{aligned}
        &\widetilde{\bm{P}}^{(r)}_{n_{0_1}, ..., n_{0_d}}(n_1, ..., n_d, z)  =\frac{1}{N^d}\left[\mathbb{I}-z\bm{\lambda}\left(0, ...,0 \right)\right]^{-1}\cdot\bm{U_{m_0}} + 
        \\ & \frac{1}{(2N)^d}\mathop{\sum_{\kappa_1 = -(N_1-1)}^{N_1-1}...\sum_{\kappa_d = -(N_d-1)}^{N_d-1}}_{\kappa_1+...+\kappa_d \neq 0}\left[\prod_{j=1}^{d}e^{-\frac{i\pi\kappa_j (n_j-n_{0_j})}{N_j}} + e^{-\frac{i\pi\kappa_j (n_j+n_{0_j}-1)}{N_j}}\right]\left[\mathbb{I}-z\bm{\lambda}\left(\frac{\pi\kappa_1}{N_1}, ...,\frac{\pi\kappa_d}{N_d} \right)\right]^{-1}\cdot\bm{U_{m_0}},
    \end{aligned}
    \label{eq: reflect_11}
\end{equation}
where we have omitted any terms containing the dummy variable $\kappa_j=-N_j$ as there is clearly no contribution to the summation and 
have also isolated the $(\kappa_1, ..., \kappa_d) = (0, ..., 0)$ term.

Applying Euler's formula to each term in the product and noticing that the $\sin$ terms give no contribution allows us to write 
Eq. (\ref{eq: reflect_11}) as 
\begin{equation}
    \begin{aligned}
        &\widetilde{\bm{P}}^{(r)}_{n_{0_1}, ..., n_{0_d}}(n_1, ..., n_d, z)  =\frac{1}{N^d}\left[\mathbb{I}-z\bm{\lambda}\left(0, ...,0 \right)\right]^{-1}\cdot\bm{U_{m_0}} + 
        \\ & \frac{1}{(2N)^d}\!\mathop{\sum_{\kappa_1 = -(N_1-1)}^{N_1-1}\!...\!\sum_{\kappa_d = -(N_d-1)}^{N_d-1}}_{\kappa_1+...+\kappa_d \neq 0}\left[\prod_{j=1}^{d}\cos\left(\frac{\pi\kappa_j (n_j-n_{0_j})}{N_j}\right) \!+\! \cos\left(\frac{\pi\kappa_j (n_j+n_{0_j}-1)}{N_j}\right)\right] \\ &
        \quad \quad \quad \quad \times \left[\mathbb{I}-z\bm{\lambda}\left(\frac{\pi\kappa_1}{N_1}, ...,\frac{\pi\kappa_d}{N_d} \right)\right]^{-1}\cdot\bm{U_{m_0}}.
    \end{aligned}
    \label{eq: reflect_11}
\end{equation}
One may simplify this expression by noticing that the periodic nature of the function ensures that for each summation
the part in the range $\kappa_j \in [-N_j +1, -1] $ gives the same contribution as the part in the range 
$\kappa_j \in [1, N_j -1] $, and sequentially working from the inside summation outwards, we can write 
\begin{equation}
    \begin{aligned}
        &\widetilde{\bm{P}}^{(r)}_{n_{0_1}, ..., n_{0_d}}(n_1, ..., n_d, z)  =\frac{1}{N^d}\left[\mathbb{I}-z\bm{\lambda}\left(0, ...,0 \right)\right]^{-1}\cdot\bm{U_{m_0}} + 
        \\ & \frac{1}{N^d}\!\mathop{\sum_{\kappa_1 = 0}^{N_1-1}\!...\!\sum_{\kappa_d = 0}^{N_d-1}}_{\kappa_1+...+\kappa_d \neq 0}\left[\prod_{j=1}^{d}\cos\left(\frac{\pi\kappa_j (n_j-n_{0_j})}{N_j}\right) \!+\! \cos\left(\frac{\pi\kappa_j (n_j+n_{0_j}-1)}{N_j}\right)\right]  \left[\mathbb{I}-z\bm{\lambda}\left(\frac{\pi\kappa_1}{N_1}, ...,\frac{\pi\kappa_d}{N_d} \right)\right]^{-1}\cdot\bm{U_{m_0}}.
    \end{aligned}
    \label{eq: reflect_12}
\end{equation}
Finally, some using the well-known trigonomentric identity $2\cos(a)\cos(b) = \cos(a+b)+\cos(a-b)$ on each term in the product, taking the inverse $z$-transform 
and recombining the $(\kappa_1, ..., \kappa_d) = (0, ..., 0)$ term, we find 
\begin{equation}
    \begin{aligned}
        &\widetilde{\bm{P}}^{(r)}_{n_{0_1}, ..., n_{0_d}}(n_1, ..., n_d, z)  = 
        \\ & \frac{1}{N^d}\!\sum_{\kappa_1 = 0}^{N_1-1}\!...\!\sum_{\kappa_d = 0}^{N_d-1}\left[\prod_{j=1}^{d}\beta_{\kappa_j}\cos\left(\frac{\pi\kappa(n_j-\frac{1}{2})}{N}\right) \cos\left(\frac{\pi\kappa(n_{0_j}-\frac{1}{2})}{N}\right)\right]  \left[\mathbb{I}-z\bm{\lambda}\left(\frac{\pi\kappa_1}{N_1}, ...,\frac{\pi\kappa_d}{N_d} \right)\right]^{-1}\cdot\bm{U_{m_0}},
    \end{aligned}
    \label{eq: reflect_12}
\end{equation}
where $\beta_{\kappa_j} = 2$ if $\kappa_j=0$ otherwise $\beta_{\kappa_j} = 1$, as reported in Eq. (14) of the main text.
%%%%%%%%%%%%%%%%%%%%%%%%%%%%%%%%%%%%%%%%%%%%%%%%%%%%%%%%%%%%%%%%%%%%%%%%%%
\section*{The propagator in a two-dimensional Square lattice}\label{app: 2D_special_cases} 
The generating function of Eq. (\ref{eq: high_dim}) when $d=2$ is given by
\begin{equation}
    \widetilde{\bm{P}}^{(\gamma)}_{n_{01}, n_{02}}(n_1, n_2, t) = \frac{1}{N^2}\sum_{\kappa_1=0}^{N-1}\sum_{\kappa_2=0}^{N-1}\left[\prod_{i=1}^{2}g_{\kappa_i}^{(\gamma)}(n_i, n_{0_i})\right]\left[\mathbb{I}-z\bm{\lambda}\left(\pi \mathcal{N}^{(\gamma)}_{\kappa_1}, \pi \mathcal{N}^{(\gamma)}_{\kappa_2}\right)\right]^{-1}\cdot\bm{U}_{\bm{m_{0}}}.
   \label{eq: high_dim1}
\end{equation}
In general, one can invert the matrix in Eq. (\ref{eq: high_dim1}), however with arbitrary parameters it is a fourth
order polynomial, and writing the roots explicitly may not be convenient. There are, however, two non-trivial cases for
which the matrix inversion is greatly simplified. They occur when $c^{(2)} = 0$ and with a uniform localised initial
condition, that is $\bm{U}_{\bm{m}_{0}}=[1/4, 1/4, 1/4, 1/4]^{\intercal}$, and we present them below.

\subsection*{The case with $b=\ell$}
By inverting the matrix in Eq. (\ref{eq: high_dim1}), one finds
\begin{equation}
    \begin{aligned}
    &\widetilde{P}^{(\gamma)}_{n_{0_1}, n_{0_2}}(n_1, n_2, z)  =\frac{1}{2N^2}\sum_{\kappa_1 = 0}^{N-1}\sum_{\kappa_2 = 0}^{N-1}\left[\prod_{i=1}^{2}g_{\kappa_i}^{(\gamma)}(n_i, n_{0_i})\right]
    \\ &\times\frac{2+3z(4\ell-1)\psi(\kappa_1, \kappa_2)+2z^2(1-4\ell)^2\sigma(\kappa_1, \kappa_2)-z^3(1-4\ell)^3\psi(\kappa_1, \kappa_2)}{1-2z(1-3\ell)\psi(\kappa_1, \kappa_2)+2z^2(1-6\ell+8\ell^2)\sigma(\kappa_1, \kappa_2)+2z^3(1-4\ell)^2(\ell-1)\psi(\kappa_1, \kappa_2)-z^4(4\ell-1)^3},
    \end{aligned}    
\end{equation}
where $\psi(\kappa_1, \kappa_2)  = \cos\left(\pi \mathcal{N}_{\kappa_1}^{(\gamma)}\right)+\cos\left(\pi
\mathcal{N}_{\kappa_2}^{(\gamma)}\right)$, and $\sigma(\kappa_1, \kappa_2) = 1+2\cos\left(\pi
\mathcal{N}_{\kappa_1}^{(\gamma)}\right)\cos\left(\pi \mathcal{N}_{\kappa_2}^{(\gamma)}\right)$.

\subsection*{The case with $f=\ell$}
To study this case, we make use of the theory of matrix resolvent \cite{bernstein2009matrix1}, for which
\begin{equation}
    \det\left[z\mathbb{I}-\bm{\lambda}\left(\pi \mathcal{N}^{(\gamma)}_{\kappa_1}, \pi \mathcal{N}^{(\gamma)}_{\kappa_2}\right)\right] = \big[(1 - 4 \ell)^2 - z^2\big] \Big\{1 - 4 \ell - z^2 + 2 \ell z \left[\cos\left(\pi \mathcal{N}_{\kappa_1}^{(\gamma)}\right)+\cos\left(\pi \mathcal{N}_{\kappa_2}^{(\gamma)}\right)\right]\Big\},
\end{equation}
to find the eigenvalues of $\bm{\lambda}\left(\pi \mathcal{N}^{(\gamma)}_{\kappa_1}, \pi \mathcal{N}^{(\gamma)}_{\kappa_2}\right)$
as
\begin{equation}
    \lambda_{1}\! =\! 1\!-\!4\ell, \quad \lambda_{2}\! =\! -1\!+\!4\ell, \quad \lambda_{\{3,4\}}\! =\! \ell \left[\cos\left(\pi \mathcal{N}_{\kappa_1}^{(\gamma)}\right)\!+\!\cos\left(\pi \mathcal{N}_{\kappa_2}^{(\gamma)}\right)\right] \pm \sqrt{1\! -\! 4 \ell \!+\! \ell^2 \left[\cos\left(\pi \mathcal{N}_{\kappa_1}^{(\gamma)}\right)\!+\!\cos\left(\pi \mathcal{N}_{\kappa_2}^{(\gamma)}\right)\right]^2}.
\end{equation}
By finding the corresponding eigenvectors, the matrix power in Eq. (\ref{eq: high_dim1}) can be computed exactly and after some
laborious algebra, one arrives at Eq. (\ref{eqn:2D_corr_final}) of the main text. 

\section*{The Governing Equations for the Correlated Hexagonal Propagator}\label{app: hex_deriv} 
With the sojourn probability, $c^{(4)} = 1 - (f + b+ \sum_{i=1}^{4}\ell_i)$, we write the Master equation as
\begin{equation}
\label{eq: hex-corr-master-eq1}
 \begin{aligned}
     Q(&n_1, n_2, 1, t+1) = \bigg[fQ(n_1 + 1, n_2, 1, t) +  bQ(n_1 + 1, n_2, 2, t) + \ell_4 Q(n_1 + 1, n_2, 3, t) +  \\ &\ell_2 Q(n_1 + 1, n_2, 4, t) + \ell_3 Q(n_1 +1 , n_2, 5, t) + \ell_1 Q(n_1 + 1, n_2, 6, t)\bigg] + c^{(4)}Q(n_1, n_2, 1, t),
 \end{aligned}
\end{equation}
\begin{equation}
\label{eq: hex-corr-master-eq2}
 \begin{aligned}
     Q(&n_1, n_2, 2, t+1) = \bigg[bQ(n_1 - 1, n_2, 1, t) +  fQ(n_1 - 1, n_2, 2, t) + \ell_2 Q(n_1 - 1, n_2, 3, t) +  \\ &\ell_4 Q(n_1 - 1, n_2,4, t) + \ell_1 Q(n_1 - 1 , n_2,5, t) + \ell_3 Q(n_1 - 1, n_2, 6, t)\bigg] + c^{(4)}Q(n_1, n_2, 2, t),
 \end{aligned}
\end{equation}
\begin{equation}
\label{eq: hex-corr-master-eq3}
 \begin{aligned}
     Q(&n_1, n_2, 3, t+1) = \bigg[\ell_1 Q(n_1, n_2 + 1, 1, t) +  \ell_3 Q(n_1, n_2 + 1, 2,t) + fQ(n_1, n_2 + 1,3, t) +  \\ &bQ(n_1, n_2 + 1,4, t) + \ell_4 Q(n_1, n_2 + 1,5, t) + \ell_2 Q(n_1, n_2 + 1, 6,t)\bigg] + c^{(4)}Q(n_1, n_2, 3, t),
 \end{aligned}
\end{equation}
\begin{equation}
\label{eq: hex-corr-master-eq4}
 \begin{aligned}
     Q(&n_1, n_2,4, t+1) = \bigg[\ell_3 Q(n_1, n_2 - 1,1, t) +  \ell_1 Q(n_1, n_2 - 1,2, t) + bQ(n_1, n_2 - 1,3, t) +  \\ &fQ(n_1, n_2 - 1, 4,t) + \ell_2 Q(n_1, n_2 - 1, 5, t) + \ell_4 Q(n_1, n_2 - 1, 6, t)\bigg] + c^{(4)}Q(n_1, n_2, 4, t),
 \end{aligned}
\end{equation}
\begin{equation}
\label{eq: hex-corr-master-eq5}
 \begin{aligned}
     Q(&n_1, n_2, 5, t+1) = \bigg[\ell_2 Q(n_1 - 1, n_2 + 1, 1,t) +  \ell_4 Q(n_1 -1, n_2 + 1,2, t) + \ell_1 Q(n_1 - 1, n_2 + 1,3, t) +  \\ &\ell_3Q(n_1 - 1, n_2 + 1,4, t) + fQ(n_1 - 1, n_2 + 1, 5,t) + bQ(n_1 - 1, n_2 + 1,6, t)\bigg] + c^{(4)}Q(n_1, n_2, 5, t),
 \end{aligned}
\end{equation}
\begin{equation}
\label{eq: hex-corr-master-eq6}
 \begin{aligned}
     Q(&n_1, n_2,6, t+1) = \bigg[\ell_4 Q(n_1 + 1, n_2 - 1,1, t) +  \ell_2 Q(n_1 + 1, n_2 - 1,2, t) + \ell_3 Q(n_1 + 1, n_2 - 1,3, t) +  \\ &\ell_1 Q(n_1 + 1, n_2 - 1,4, t) + bQ(n_1 + 1, n_2 - 1, 5, t) + fQ(n_1 + 1, n_2 - 1, 6, t)\bigg] + c^{(4)}Q(n_1, n_2, 6, t).
 \end{aligned}
\end{equation}

\section*{Derivations of Specific First Passage Cases}\label{sec: sup_matFP}
\subsection*{One State Case}

With a single target we have a set $\mathcal{S}$ of targets with cardinality one, namely $\mathdutchcal{S} =
\left\{(\bm{s}, m_{\bm{s}})\right\}$. Starting from Eq. (\ref{eq: defect_sol2}) of Appendix \ref{app: defect} in the
main text, we find 
\begin{equation}
    \begin{aligned}
    \widetilde{P}^{(a)}_{\bm{n}_0, m_0}(\bm{s}, m, z) = \alpha_{m_0}\widetilde{P}_{\bm{n}_0, m_0}(\bm{s}, m, z) + \frac{\rho}{\rho-1}\alpha_{m_{\bm{s}}}\widetilde{P}_{\bm{s}, m_{\bm{s}}}(\bm{s}, m, z)\widetilde{P}^{(a)}_{\bm{n}_0, m_0}(\bm{s},m_{\bm{s}}, z).
    \end{aligned}
    \label{eq: defect_sol_1tar}
\end{equation} 
To proceed we set here $(\bm{s}, m) = (\bm{s}, m_{\bm{s}})$, solve for the unknown function $
\widetilde{P}^{(a)}_{\bm{n}_0, m_0}(\bm{s},m_{\bm{s}}, z)$ in terms of the known functions,
\begin{equation}
    \widetilde{P}^{(a)}_{\bm{n}_0, m_0}(\bm{s},m_{\bm{s}}, z) = \frac{\alpha_{m_0}\widetilde{P}_{\bm{n}_0, m_0}(\bm{s}, m_{\bm{s}}, z)}{1 + \frac{\rho}{\rho-1}\alpha_{m_{\bm{s}}}\widetilde{P}_{\bm{s}, m_{\bm{s}}}(\bm{s}, m_{\bm{s}}, z)}
    \label{eq: solo1}
\end{equation}
and then substitute Eq. (\ref{eq: solo1}) into Eq. (\ref{eq: defect_sol_1tar}), which leads to 
\begin{equation}
    \widetilde{P}^{(a)}_{\bm{n}_0, m_0}(\bm{s}, m, z) = \alpha_{m_0}\widetilde{P}_{\bm{n}_0, m_0}(\bm{s}, m, z) - \rho\alpha_{m_{\bm{s}}}\widetilde{P}_{\bm{s}, m_{\bm{s}}}(\bm{s}, m, z)\frac{\alpha_{m_0}\widetilde{P}_{\bm{n}_0, m_0}(\bm{s}, m_{\bm{s}}, z)}{1-\rho + \rho\alpha_{m_{\bm{s}}}\widetilde{P}_{\bm{s}, m_{\bm{s}}}(\bm{s}, m_{\bm{s}}, z)}.
\end{equation}
Performing a summation over $m$ and $m_0$, we find the occupation probability with one partially absorbing trap as
\begin{equation}
    \widetilde{P}^{(a)}_{\bm{n}_0}(\bm{s}, z) = \widetilde{P}_{\bm{n}_0}(\bm{s}, z) - \rho\alpha_{m_{\bm{s}}}\sum_{m=1}^{M}\widetilde{P}_{\bm{s}, m_{\bm{s}}}(\bm{s}, m, z)\frac{\sum_{j=1}^{M}\alpha_{m_{0_j}}\widetilde{P}_{\bm{n}_0, m_{0_j}(\bm{s}, m_{\bm{s}}, z)}}{1-\rho + \rho\alpha_{m_{\bm{s}}}\widetilde{P}_{\bm{s}, m_{\bm{s}}}(\bm{s}, m_{\bm{s}}, z)}.
\end{equation}
Following the same procedure outlined in Sec. \ref{sec: first-passage} to pass from Eq. (\ref{eq: correlated_defect}) to
Eq. (\ref{eq: multi_tar_FP}) of the main text we find the first-\textit{absorption} probability with one absorbing state
as 
\begin{equation}
    \widetilde{F}_{\bm{n}_0}(\bm{s}, m_{\bm{s}}, z) = \rho\alpha_{m_{\bm{s}}}\frac{\sum_{j=1}^{M}\alpha_{m_{0_j}}\widetilde{P}_{\bm{n}_0, m_{0_j}}(\bm{s}, m_{\bm{s}}, z)}{1-\rho + \rho\alpha_{m_{\bm{s}}}\widetilde{P}_{\bm{s}, m_{\bm{s}}}(\bm{s}, m_{\bm{s}}, z)},
    \label{eq: pa_single_tar_FP}
\end{equation} 
which, in the $\rho \to 1$ limit, gives Eq. (\ref{eq: single_tar_FP}) of the main text.

\subsection*{Two State Case}

As explained in Sec. \ref{sec: Absorbing} of the main text, the first-passage probability can be found from (\ref{eq:
multi_tar_FP}) as
\begin{equation}
   \widetilde{F}_{\bm{n}_0}(\mathdutchcal{S}, z) = \sum_{j=1}^{M}\sum_{i = 1}^{S}\alpha_{m_{\bm{s}_i}}\frac{\det[\mathbb{H}^{(i)}(\bm{n}_0, m_{0_j}, z)]}{\det[\mathbb{H}(z)]}.
 \label{eq: multi_tar_FP_app}
\end{equation}
For the cases with low numbers of targets, Eq. (\ref{eq: multi_tar_FP_app}) can be expanded explicitly. An important
case is when there are two targets, which is pertinent to the one-dimensional case with a target in both states of $s$.
With two targets, one finds 
\begin{equation}
 \widetilde{F}_{n_0}(\mathdutchcal{S}, z)\! =\! \frac{\alpha_{m_{\bm{s}_1}}\det[\mathbb{H}^{(1)}(\bm{n}_0, 1, z)]\!+\!\alpha_{m_{\bm{s}_2}}\det[\mathbb{H}^{(2)}(\bm{n}_0, 1, z)]}{\det[\mathbb{H}(z)]} \!+\! \frac{\alpha_{m_{\bm{s}_1}}\det[\mathbb{H}^{(1)}(\bm{n}_0, 2, z)]\!+\!\alpha_{m_{\bm{s}_2}}\det[\mathbb{H}^{(2)}(\bm{n}_0, 2, z)]}{\det[\mathbb{H}(z)]}.
 \label{eq: multi_tar_FP_app_ex}
\end{equation}
where $\mathbb{H}(z)_{l,k} = \alpha_{m_{\bm{s}_k}}\widetilde{P}_{\bm{s}_k, m_{\bm{s}_k}}(\bm{s}_l, m_{\bm{s}_l}, z)$,
$\mathbb{H}(z)_{k,k} = \alpha_{m_{\bm{s}_k}}\widetilde{P}_{\bm{s}_k, m_{\bm{s}_k}}(\bm{s}_k, m_{\bm{s}_k}, z)$ and
$\mathbb{H}^{(i)}(\bm{n}_0, j, z)$ is the same, but with the $i^{\text{th}}$ column replaced with
$\alpha_{m_{0_j}}\left[\widetilde{P}_{\bm{n}_0, m_{0_j}}(\bm{s}_1, m_{\bm{s}_1}, z), ..., \widetilde{P}_{\bm{n}_0,
m_{0_j}}(\bm{s}_S, m_{\bm{s}_S}, z) \right]^{\intercal}$. One may write this more explicitly as 
\begin{equation}
    \begin{aligned}
    &\widetilde{F}_{n_0}(\mathdutchcal{S}, z)= \nonumber \\  & \alpha_{m_{\bm{s}_1}}\frac{\begin{vmatrix}
        \alpha_{m_{0_1}}\widetilde{P}_{\bm{n}_0, m_{0_1}}(\bm{s}_1, m_{\bm{s}_1},z) & \alpha_{m_{\bm{s}_2}}\widetilde{P}_{\bm{s}_2, m_{\bm{s}_2}}(\bm{s}_1, m_{\bm{s}_1},z) \\ 
        \alpha_{m_{0_1}}\widetilde{P}_{\bm{n}_0, m_{0_1}}(\bm{s}_2, m_{\bm{s}_2},z) & \alpha_{m_{\bm{s}_2}}\widetilde{P}_{\bm{s}_2, m_{\bm{s}_2}}(\bm{s}_2, m_{\bm{s}_2},z) \\ 
\end{vmatrix} + \begin{vmatrix}
    \alpha_{m_{0_2}}\widetilde{P}_{\bm{n}_0, m_{0_2}}(\bm{s}_1, m_{\bm{s}_1},z) & \alpha_{m_{\bm{s}_2}}\widetilde{P}_{\bm{s}_2, m_{\bm{s}_2}}(\bm{s}_1, m_{\bm{s}_1},z) \\ 
    \alpha_{m_{0_2}}\widetilde{P}_{\bm{n}_0, m_{0_2}}(\bm{s}_2, m_{\bm{s}_2},z) & \alpha_{m_{\bm{s}_2}}\widetilde{P}_{\bm{s}_2, m_{\bm{s}_2}}(\bm{s}_2, m_{\bm{s}_2},z) \\ 
\end{vmatrix}}{\begin{vmatrix}
                \alpha_{m_{\bm{s}_1}}\widetilde{P}_{\bm{s}_1, m_{\bm{s}_1}}(\bm{s}_1, m_{\bm{s}_1},z) & \alpha_{m_{\bm{s}_2}}\widetilde{P}_{\bm{s}_2, m_{\bm{s}_2}}(\bm{s}_1, m_{\bm{s}_1},z) \\ 
                \alpha_{m_{\bm{s}_1}}\widetilde{P}_{\bm{s}_1, m_{\bm{s}_1}}(\bm{s}_2, m_{\bm{s}_2},z) & \alpha_{m_{\bm{s}_2}}\widetilde{P}_{\bm{s}_2, m_{\bm{s}_2}}(\bm{s}_2, m_{\bm{s}_2},z) \\ 
        \end{vmatrix} } 
    \\ & + \alpha_{m_{\bm{s}_2}}\frac{\begin{vmatrix}
        \alpha_{m_{\bm{s}_1}}\widetilde{P}_{\bm{s}_1, m_{\bm{s}_1}}(\bm{s}_1, m_{\bm{s}_1},z) & \alpha_{m_{0_1}}\widetilde{P}_{\bm{n}_0, m_{0_1}}(\bm{s}_1, m_{\bm{s}_1},z) \\ 
        \alpha_{m_{\bm{s}_1}}\widetilde{P}_{\bm{s}_1, m_{\bm{s}_1}}(\bm{s}_2, m_{\bm{s}_2},z) & \alpha_{m_{0_1}}\widetilde{P}_{\bm{n}_0, m_{0_1}}(\bm{s}_2, m_{\bm{s}_2},z) \\ 
\end{vmatrix} \!+\! \begin{vmatrix}
            \alpha_{m_{\bm{s}_1}}\widetilde{P}_{\bm{s}_1, m_{\bm{s}_1}}(\bm{s}_1, m_{\bm{s}_1},z) & \alpha_{m_{0_2}}\widetilde{P}_{\bm{n}_0, m_{0_2}}(\bm{s}_1, m_{\bm{s}_1},z) \\ 
            \alpha_{m_{\bm{s}_1}}\widetilde{P}_{\bm{s}_1, m_{\bm{s}_1}}(\bm{s}_2, m_{\bm{s}_2},z) & \alpha_{m_{0_2}}\widetilde{P}_{\bm{n}_0, m_{0_2}}(\bm{s}_2, m_{\bm{s}_2},z) \\ 
    \end{vmatrix}}{\begin{vmatrix}
                \alpha_{m_{\bm{s}_1}}\widetilde{P}_{\bm{s}_1, m_{\bm{s}_1}}(\bm{s}_1, m_{\bm{s}_1},z) & \alpha_{m_{\bm{s}_2}}\widetilde{P}_{\bm{s}_2, m_{\bm{s}_2}}(\bm{s}_1, m_{\bm{s}_1},z) \\ 
                \alpha_{m_{\bm{s}_1}}\widetilde{P}_{\bm{s}_1, m_{\bm{s}_1}}(\bm{s}_2, m_{\bm{s}_2},z) & \alpha_{m_{\bm{s}_2}}\widetilde{P}_{\bm{s}_2, m_{\bm{s}_2}}(\bm{s}_2, m_{\bm{s}_2},z) \\ 
        \end{vmatrix} } .
    \end{aligned}
    \label{eq: multi_tar_FP_app_ex1}
\end{equation}
Performing the determinants and some trivial simplification leads to Eq. (\ref{eq: one_whole_site_FP}) of the main text.

\section*{The Mean First-Passage Time}\label{app: MFPT}
\subsection*{Derivation of the Multi Target Mean First-Passage Time}\label{sec: multiMFPT}
As mentioned in the main text, the general procedure of finding the multi-target mean first-passage time (MFPT) may be
found in ref. \cite{giuggioli2022spatio} and we refer the interested reader there. Here, we present the extension
from that work that is required to find the MFPT with correlated random variables. 

We start from Eq. (\ref{eq: multi_tar_FP}) of the main text and since we assume here that the $\bm{U_{m_0}}$ is uniformly
distributed, one may factor out $\alpha = 1/2d$ as it is a constant factor across all elements of each determinant and rewrite the first passage as 
\begin{equation}
    \widetilde{F}_{\bm{n}_0}(\mathdutchcal{S}, z) = \frac{1}{2d}\sum_{j=1}^{M}\sum_{i = 1}^{S}\frac{\det[\mathbb{H}^{(i)}(\bm{n}_0, m_{0_j}, z)]}{\det[\mathbb{H}(z)]}, 
\end{equation}
where $\mathbb{H}(z)_{l,k} = \widetilde{P}_{\bm{s}_k, m_{\bm{s}_k}}(\bm{s}_l, m_{\bm{s}_l}, z)$, $\mathbb{H}(z)_{k,k} =
\widetilde{P}_{\bm{s}_k, m_{\bm{s}_k}}(\bm{s}_k, m_{\bm{s}_k}, z)$ and $\mathbb{H}^{(i)}(\bm{n}_0, m_0, z)$ is the same,
but with the $i^{\text{th}}$ column replaced with $\left[\widetilde{P}_{\bm{n}_0, m_{0_j}}(\bm{s}_1, m_{\bm{s}_1}, z),
..., \widetilde{P}_{\bm{n}_0, m_{0_j}}(\bm{s}_S, m_{\bm{s}_S}, z) \right]^{\intercal}$. Using the linearity of the outer
summation, one may perform $\mathdutchcal{F}_{\bm{n_0}\to \{\mathdutchcal{S}\}} = \left.\frac{\partial
\widetilde{F}_{\bm{n_0}}(\mathdutchcal{S}, z)}{\partial z}\right|_{z=1}$ on each element and following ref.
\cite{giuggioli2022spatio}, we find 
\begin{equation}
    \mathdutchcal{F}_{\bm{n_0}\to \{\mathdutchcal{S}\}} = \frac{1}{2d}\sum_{j=1}^{2d}\frac{\det[\mathbb{T}_{m_{0_j}}]}{\det[\mathbb{T}_{1}]-\det[\mathbb{T}]},
    \label{eq: app_MFPT}
   \end{equation}
where $\mathbb{T}_{i,i} = 0$, $\mathbb{T}_{i,k} = \mathdutchcal{F}_{\bm{s}_k, m_{\bm{s}_k}\to\bm{s}_i, m_{\bm{s}_i}}$
($i\neq k$),
$\mathbb{T}_{{1}_{i,k}} = \mathbb{T}_{i,k} - 1$ and the $(i,k)^{\text{th}}$ element of $\mathbb{T}_{m_{0_j}}$ is $ \mathbb{T}_{i,k} -\mathdutchcal{F}_{\bm{n}_0, m_{0_j}\to\bm{s}_i, m_{\bm{s}_i}}$.
Repeated use of the multi-linearity property of the determinant allows one to bring inside the summation over the initial states 
allowing us to arrive at Eq. (\ref{eq: MFPT}) of the main text. 

\subsection*{Derivation of the State Level Mean First-Passage Time}
We derive here the MFPT from one state to another state, as needed to use Eq. (\ref{eq: MFPT}) of the main text. Namely,
we are interested in calculating the quantity $ \mathdutchcal{F}_{\bm{n}_0, m_{0}\to\bm{s}_i, m_{\bm{s}_i}} =
\frac{\partial}{\partial z}\left.\frac{\widetilde{P}_{\bm{n}_0, m_0}(\bm{s}, m_{\bm{s}}, z)}{\widetilde{P}_{\bm{s},
m_{\bm{s}}}(\bm{s}, m_{\bm{s}}, z)}\right|_{z=1}$. We give the procedure here for the periodic propagator but note that
the derivation is general as long as three assumptions are held. Namely, (i) one may use the propagator as the
defect-free dynamics on which to place traps (see Sec. \ref{sec: Absorbing} of the main text for discussion), (ii) the
walk is recurrent and (iii) is not confined to a genuine sublattice. 

We begin by writing the first-passage as
\begin{equation}
 \widetilde{F}^{(p)}_{\bm{n}_0, m_{0}}(\bm{s}_i, m_{\bm{s}_i},z) =  \frac{\bm{e}_{m}^{\intercal}\cdot \left\{\left[\mathbb{I}-z\bm{\lambda}\left(0, ..., 0\right)\right]^{-1}+\sum_{\bm{\kappa}\neq \bm{0}}\left(\prod_{i=1}^{d}g^{(p)}_{\kappa_i}(s_i, n_{0_i})\right)\left[\mathbb{I}-z\bm{\lambda}\left(\frac{2\pi\kappa_1}{N_1}, ..., \frac{2\pi\kappa_d}{N_d}\right)\right]^{-1}\right\}\cdot\bm{e}_{m_0} }{\bm{e}_{m}^{\intercal}\cdot \left\{\left[\mathbb{I}-z\bm{\lambda}\left(0, ..., 0\right)\right]^{-1}+\sum_{\bm{\kappa}\neq \bm{0}}\left[\mathbb{I}-z\bm{\lambda}\left(\frac{2\pi\kappa_1}{N_1}, ..., \frac{2\pi\kappa_d}{N_d}\right)\right]^{-1}\right\}\cdot\bm{e}_{m}^{\intercal}},
 \label{eq: MFPT1}
\end{equation}
and multiplying the top and bottom of Eq. (\ref{eq: MFPT1}) by $\det[\mathbb{I}-z\bm{\lambda}\left(0, ..., 0\right)]$
leads to 
\begin{equation}
    \begin{aligned}
        &\widetilde{F}^{^{(p)}}_{\bm{n}_0,
        m_{0}}(\bm{s}_i, m_{\bm{s}_i},z) =\\&  \frac{\bm{e}_{m}^{\intercal}\cdot \left\{\text{Adj}\left[\mathbb{I}-z\bm{\lambda}\left(0, ..., 0\right)\right]+\det[\mathbb{I}-z\bm{\lambda}\left(0, ..., 0\right)]\sum_{\bm{\kappa}\neq \bm{0}}\left(\prod_{i=1}g^{(p)}_{\kappa_i}(s_i, n_{0_i})\right)\left[\mathbb{I}-z\bm{\lambda}\left(\frac{2\pi\kappa_1}{N_1}, ..., \frac{2\pi\kappa_d}{N_d}\right)\right]^{-1}\right\}\cdot\bm{e}_{m_0} }{\bm{e}_{m}^{\intercal}\cdot \left\{\text{Adj}\left[\mathbb{I}-z\bm{\lambda}\left(0\right)\right]+\det[\mathbb{I}-z\bm{\lambda}\left(0\right)]\sum_{\bm{\kappa}\neq\bm{0}}\left[\mathbb{I}-z\bm{\lambda}\left(\frac{2\pi\kappa_1}{N_1}, ..., \frac{2\pi\kappa_d}{N_d}\right)\right]^{-1}\right\}\cdot\bm{e}_{m}^{\intercal}},       
    \end{aligned}
\label{eq: MFPT1}
\end{equation} 
where $\text{Adj}\left[\mathbb{A}\right]$ denotes the adjoint of $\mathbb{A}$. The recurrence of the walk ensures
$\det[\mathbb{I}-\bm{\lambda}\left(\bm{0}\right)]=0$, and we utilise this property upon differentiating to find
\begin{equation}
 \begin{aligned}
      &\mathdutchcal{F}^{^{(p)}}_{\bm{n}_0, m_{0}\to \bm{s}_i, m_{\bm{s}_i}}\! =\! \frac{1}{\left[\bm{e}_{m}^{\intercal}\cdot\text{Adj}\left[\mathbb{I}-\bm{\lambda}\left(0, ..., 0\right)\right]\cdot\bm{e}_{m}\right]^2} \bigg\{\bm{e}_{m}^{\intercal}\cdot\text{Adj}\left[\mathbb{I}\!-\!\bm{\lambda}\left(0, ..., 0\right)\right]\cdot\bm{e}_{m}\bigg[\bm{e}_{m}^{\intercal}\cdot\bigg( \left.\frac{\partial \text{Adj}[\mathbb{I}\!-\!z\bm{\lambda}(0, ..., 0)]}{\partial z}\right|_{z=1}  \\
      &\!+\! \left.\frac{\partial \det[\mathbb{I}-z\bm{\lambda}(0, ..., 0)]}{\partial z}\right|_{z=1} \sum_{\bm{\kappa}\neq\bm{0}}\left(\prod_{i=1}^{d}g^{(p)}_{\kappa_i}(s_i, n_{0_i})\right)\!\left[\mathbb{I}\!-\!\bm{\lambda}\left(\frac{2\pi\kappa_1}{N_1}, ..., \frac{2\pi \kappa_d}{N_d}\right)\right]^{-1}\!\bigg)\!\cdot \!\bm{e}_{m_0} \bigg] \!- \!\bm{e}_{m}^{\intercal}\!\cdot\!\text{Adj}\left[\mathbb{I}-\bm{\lambda}\left(0, ..., 0\right)\right]\!\cdot\!\bm{e}_{m_0}
      \\ &\times\bigg[\bm{e}_{m}^{\intercal}\cdot\bigg( \left.\frac{\partial \text{Adj}[\mathbb{I}-z\bm{\lambda}(0, ..., 0)]}{\partial z}\right|_{z=1}  + \left.\frac{\partial \det[\mathbb{I}-z\bm{\lambda}(0, ..., 0)]}{\partial z}\right|_{z=1} \sum_{\bm{\kappa}\neq\bm{0}}\left[\mathbb{I}-\bm{\lambda}\left(\frac{2\pi\kappa_1}{N_1}, ..., \frac{2\pi\kappa_d}{N_d}\right)\right]^{-1}\bigg)\cdot \bm{e}_{m}^{\intercal} \bigg]\bigg\}.
 \end{aligned}
 \label{eq: MFPT2}
\end{equation}

It is the knowledge of the structure function $\bm{\lambda}(0, ...., 0)$ that allows simplification of Eq. (\ref{eq:
MFPT2}). In one dimension we have
\begin{equation}
 \text{Adj}[\mathbb{I}-\bm{\lambda}(0)]= b\mathbb{J}, \quad \left.\frac{\partial \text{Adj}[\mathbb{I}-z\bm{\lambda}(0)]}{\partial z}\right|_{z=1}= \begin{bmatrix} b-1 & b \\ b & b-1 \end{bmatrix} \; \; \text{and} \left.\frac{\partial \det[\mathbb{I}-z\bm{\lambda}(0))}{\partial z}\right|_{z=1}= -2b,
\end{equation}
where $\mathbb{J}$ is the all ones matrix, and substitution into Eq. (\ref{eq: MFPT2}) leads to, with
$g^{(p)}_{\kappa_i}(s_i, n_{0_i})$ taken from the main text,
\begin{equation}
 \begin{aligned}
 \mathdutchcal{F}^{(p)}_{n_0, m_{0}\to s, m_{s}} =&  \bm{e}_{m_{s}}^{\intercal}\cdot\begin{bmatrix}0 & \frac{1}{b} \\ \frac{1}{b} & 0 \end{bmatrix}\cdot \bm{e}_{m_0} + \\& 
  2
 \sum_{\kappa=1}^{N-1} \left[\bm{e}_{m_{s}}^{\intercal}\cdot\left[\mathbb{I}-\bm{\lambda}\left(\frac{2\pi\kappa}{N}\right)\right]^{-1}\cdot \bm{e}_{m_{s}}-\exp\left(\frac{-2\pi i\kappa(s-n_0)}{N}\right)\bm{e}_{m_{s}}^{\intercal}\cdot\left[\mathbb{I}-\bm{\lambda}\left(\frac{2\pi\kappa}{N}\right)\right]^{-1}\cdot \bm{e}_{m_0}\right],
 \end{aligned}
 \label{eq: appsingle_tarMFPT}
\end{equation}
where 
\begin{equation}
 \begin{aligned}
     &\big[\mathbb{I}-\bm{\lambda}(x)\big]^{-1} = \frac{1}{2 f (b + f) (1 - \cos(x))} 
 \begin{bmatrix} b+f-fe^{-i x} & be^{i x} \\
     be^{-i x}& b+f-fe^{i x}
 \end{bmatrix}.
 \end{aligned}
\end{equation} 
The first term in Eq. (\ref{eq: appsingle_tarMFPT}) corresponds to the dynamics within the site and leads to, in the
degenerate $N = 1$ case, a MFPT of zero if $m=m_0$ and $b$ if $m \neq m_0$, while the summation corresponds to the
dynamics between the sites.

To simplify Eq. (\ref{eq: appsingle_tarMFPT}) further, one can exploit the structure of the matrices and the property
that when $m_0 = m_s$ ($m_0 \neq m_s$) a diagonal (anti-diagonal) element will be chosen. To illustrate, consider the
four options, namely (I) $m_0 = m_s = 1$, (II) $m_0 = m_s = 2$, (III) $m_0 = 1, m_s = 2$ and (IV) $m_0 = 2, m_s =1$. In
turn, these options give \\ \\
(I)
\begin{equation}
    \mathdutchcal{F}^{(p)}_{n_0, 1 \to s, 1} =  
    \sum_{\kappa=1}^{N-1}\frac{\left(b+f-f e^{\frac{-2 \pi i\kappa}{N}}\right)\left(1-e^{\frac{-2\pi i\kappa(s-n_0)}{N}}\right)}{f (b + f) \left(1 - \cos\left( \frac{2\pi \kappa}{N}\right)\right)}  ,
\end{equation}
(II)
\begin{equation}
    \mathdutchcal{F}^{(p)}_{n_0, 2\to s, 2}= 
    \sum_{\kappa=1}^{N-1} \frac{\left(b+f- f e^{\frac{2 \pi i\kappa}{N}}\right)\left(1-e^{\frac{-2\pi i\kappa(s-n_0)}{N}}\right)}{ f (b + f) \left(1 - \cos\left( \frac{2\pi \kappa}{N}\right)\right)} ,
\end{equation}
(III)
\begin{equation}
    \mathdutchcal{F}^{(p)}_{n_0, 2\to s, 1} =  \frac{1}{b}+ 
    \sum_{\kappa=1}^{N-1}\frac{\left(b+f- f e^{\frac{2 \pi i\kappa}{N}}\right) -b e^{\frac{-2\pi i\kappa(s-n_0)}{N}} e^{\frac{-2\pi i \kappa}{N}}}{f (b + f) \left(1 - \cos\left( \frac{2\pi \kappa}{N}\right)\right)},
\end{equation}
and (IV)
\begin{equation}
    \mathdutchcal{F}^{(p)}_{n_0, 1\to s, 2} = \frac{1}{b}+ 
    \sum_{\kappa=1}^{N-1}\frac{\left(b+f-f e^{\frac{-2 \pi i\kappa}{N}}\right) -b e^{\frac{-2\pi i\kappa(s-n_0)}{N}} e^{\frac{2\pi i \kappa}{N}}}{f (b + f) \left(1 - \cos\left( \frac{2\pi \kappa}{N}\right)\right)}.
\end{equation}

Upon combining into one equation, one finds
\begin{equation}
    \begin{aligned}
       &\mathdutchcal{F}_{n_0, m_0 \rightarrow s, m_s}^{(p)} = \frac{1\! -\! \delta_{m_s,m_0}}{b} \\
       & + \sum_{\kappa=1}^{N-1} \frac{\left(b\!+\!f\!-\! f e^{\frac{ 2(\delta_{m_s,2} - \delta_{m_s,1})\pi i\kappa}{N}}\right) \!-\! e^{\frac{-2\pi i\kappa(s-n_0)}{N}}\left[\delta_{m_s, m_0}\left(b\!+\!f\!-\! f e^{\frac{2 (\delta_{m_s,2} \!- \! \delta_{m_s,1}) \pi i\kappa}{N}}\right) \!+\! (1 \!-\! \delta_{m_s, m_0}) b e^{\frac{2\pi i \kappa(m_s - m_0)}{N}} \right] }{f (b + f) (1 - \cos\left(\frac{2\pi \kappa}{N}\right))}.
    \end{aligned}
    \label{eq: single_tarMFPT}
\end{equation} 

The degenerate $N=1$ case, with $f=b=\frac{q}{2}$ here, corresponds exactly to the symmetric walk in an $N=2$ reflective
domain, where the MFPT is precisely zero if $n=n_0$ and $\frac{q}{2}$ if $n\neq n_0$ \cite{LucaPRX}. 

Since the reflective propagator, namely Eq. (\ref{eq: 1d_reflective_kalay}) in the main text, is simply a sum of
periodic propagators, one follows the same procedure used to obtain Eq. (\ref{eq: appsingle_tarMFPT}) making it
straightforward to find the MFPT in the reflective case as
\begin{equation}
    \begin{aligned}
    \mathdutchcal{F}^{(r_s)}_{n_0, m_{0}\to s, m_{s}}&= \frac{1 - \delta_{m_s,m_0}}{b}+ 2\sum_{\kappa=1}^{2N-3} \left\{\bm{e}_m^{\intercal}\cdot\left[1+\mu(s)\exp\left(\frac{2\pi i \kappa(N-s)}{N-1}\right)\right]\left[\mathbb{I}-\bm{\lambda}\left(\frac{\pi\kappa }{N-1}\right)\right]^{-1}\cdot \bm{e}_{m}\right. \\&\left.- \bm{e}_m^{\intercal}\cdot\left[\exp\left(\frac{\pi i \kappa(s-n_0)}{N-1}\right)+\mu(s)\exp\left(\frac{\pi \kappa(2N-s-n_0)}{N-1}\right)\right]\left[\mathbb{I}-\bm{\lambda}\left(\frac{\pi\kappa }{N-1}\right)\right]^{-1}\cdot \bm{e}_{m_0}\right\}.
    \end{aligned}
    \label{eq: single_tarMFPT_ref}
\end{equation}

In the square lattice,
\begin{equation}
 \begin{aligned}
     &\mathdutchcal{F}^{^{(p)}}_{\bm{n}_0, m_{0}\to\bm{s}_i, m_{\bm{s}_i}} = \frac{1}{\left[\bm{e}_{m}^{\intercal}\cdot\text{Adj}\left[\mathbb{I}-\bm{\lambda}\left(0, 0\right)\right]\cdot\bm{e}_{m}\right]^2} \bigg\{ \bm{e}_{m}^{\intercal}\cdot\text{Adj}\left[\mathbb{I}-\bm{\lambda}\left(0, 0\right)\right]\cdot\bm{e}_{m}\bigg[\bm{e}_{m}^{\intercal}\cdot\bigg( \left.\frac{\partial \text{Adj}[\mathbb{I}-z\bm{\lambda}(0, 0)]}{\partial z}\right|_{z=1}  \\
     & + \left.\frac{\partial \det[\mathbb{I}-z\bm{\lambda}(0, 0))}{\partial z}\right|_{z=1} \sum_{\bm{\kappa}\neq\bm{0}}\exp\left(-\frac{2\pi i[\kappa_1(n_1-n_{0_1})+\kappa_2(n_2 - n_{0_2})]}{N}\right)\left[\mathbb{I}-\bm{\lambda}\left(\frac{2\pi\kappa_1}{N}, \frac{2\pi\kappa_2}{N}\right)\right]^{-1}\bigg)\cdot \bm{e}_{m_0} \bigg]  \\
     &-\bm{e}_{m}^{\intercal}\cdot\text{Adj}\left[\mathbb{I}-\bm{\lambda}\left(0, 0\right)\right]\cdot\bm{e}_{m_0}\bigg[\bm{e}_{m}^{\intercal}\cdot\bigg( \left.\frac{\partial \text{Adj}[\mathbb{I}-z\bm{\lambda}(0,0)]}{\partial z}\right|_{z=1}  + \\
     &\; \; \; \; \; \; \; \; \; \; \; \; \; \; \; \; \; \; \; \; \; \; \;\; \; \; \; \; \; \; \; \; \;\; \; \; \; \; \; \; \; \; \; \; \; \; \; \; \; \; \; \; \;\; \; \; \; \; \; \; \left.\frac{\partial \det[\mathbb{I}-z\bm{\lambda}(0, 0))}{\partial z}\right|_{z=1} \sum_{\bm{\kappa}\neq\bm{0}}\left[\mathbb{I}-\bm{\lambda}\left(\frac{2\pi\kappa_1}{N}, \frac{2\pi\kappa_2}{N}\right)\right]^{-1}\bigg)\cdot \bm{e}_{m}^{\intercal} \bigg]\bigg\},
\end{aligned}
 \label{eq: MFPT_2d1}
\end{equation}
and we find 
\begin{equation}
 \begin{aligned}
 &\text{Adj}[\mathbb{I}-\bm{\lambda}(0, 0)]= 4\ell(b+\ell)^2\mathbb{J}, \left.\frac{\partial \det[\mathbb{I}-z\bm{\lambda}(0,0))}{\partial z}\right|_{z=1}= -16\ell (b+\ell)^2 \; \; \text{and}\quad \left.\frac{\partial \text{Adj}[\mathbb{I}-z\bm{\lambda}(0,0)]}{\partial z}\right|_{z=1}\\
 &=\left[\begin{smallmatrix} -2(b+\ell)[b-\ell+6(sf+\ell^2-c\ell)] & 2 (b+\ell)^2 (-1 + 6\ell) &-4\ell (b+\ell) (-2 + 3 c + 3 f + 3\ell) &-4\ell (b+\ell) (-2 + 3 c + 3 f + 3\ell) \\ 2 (b+\ell)^2 (-1 + 6\ell) & -2(b+\ell)[b-\ell+6(sf+\ell^2-c\ell)] & -4\ell (b+\ell) (-2 + 3 c + 3 f + 3\ell)& -4\ell (b+\ell) (-2 + 3 c + 3 f + 3\ell)\\-4\ell (b+\ell) (-2 + 3 c + 3 f + 3\ell) & -4\ell (b+\ell) (-2 + 3 c + 3 f + 3\ell) &-2(b+\ell)[b-\ell+6(sf+\ell^2-c\ell)] & 2 (b+\ell)^2 (-1 + 6\ell)\\-4\ell (b+\ell) (-2 + 3 c + 3 f + 3\ell) & -4\ell (b+\ell) (-2 + 3 c + 3 f + 3\ell) & 2 (b+\ell)^2 (-1 + 6\ell)&-2(b+\ell)[b-\ell+6(sf+\ell^2-c\ell)]\end{smallmatrix}\right],
 \end{aligned}
\end{equation}
which upon substitution and simplification, gives
\begin{equation}
 \begin{aligned}
 &\mathdutchcal{F}^{(p)}_{\bm{n}_0, m_{0}\to\bm{s}_i, m_{\bm{s}_i}} = \delta_{|m-m_0|>1}\frac{3\ell+b}{2\ell(b+\ell)}+ \delta_{|m-m_0|,1}\left((1-\delta_{m_0 \in \{2,3\}}\delta_{m \in \{2,3\}})\frac{2}{b+\ell}\! +\! \delta_{m_0 \in \{2,3\}}\delta_{m \in \{2,3\}}\frac{3\ell+b}{2\ell(b+\ell)}\right)
 \\ &+4\underset{\kappa_1 + \kappa_2 > 0}{\sum_{\kappa_1=0}^{N-1}\sum_{\kappa_2=0}^{N-1}}\left[ \!\bm{e}_{m}^{\intercal}\cdot \left[\mathbb{I}\!-\!\bm{\lambda} \left(\frac{2\pi\kappa_1}{N}, \frac{2\pi\kappa_2}{N}\right) \right]^{-1}\cdot \bm{e}_{m}- \exp\left(\frac{-i 2\pi[\bm{\kappa}\!\cdot\!(\bm{s}-\bm{n}_0)]}{N}\right)\bm{e}_{m}^{\intercal}\cdot\left[\mathbb{I}\!-\!\bm{\lambda}\left(\frac{2\pi\kappa_1}{N}, \frac{2\pi\kappa_2}{N}\right)\!\right]^{-1}\!\cdot \!\bm{e}_{m_0}\!\right].
 \end{aligned}
 \label{eq: appsingle_tarMFPT2D}
\end{equation}

In the hexagonal case, we find
\begin{equation}
 \begin{aligned}
 &\text{Adj}[\mathbb{I}-\bm{\lambda}(0, 0)]= 48\ell^2(b+2\ell)^3\mathbb{J}, \left.\frac{\partial \det[\mathbb{I}-z\bm{\lambda}(0,0))}{\partial z}\right|_{z=1}= -288 \ell^2 (b+2\ell)^3,
 \end{aligned}
\end{equation}
and we omit explicitly giving $\left.\frac{\partial \text{Adj}[\mathbb{I}-z\bm{\lambda}(0,0)]}{\partial
z}\right|_{z=1}$, owing to its large size. However, upon simplification, many terms cancel out and one obtains 
\begin{equation}
 \begin{aligned}
 &\mathdutchcal{F}^{(\mathcal{H})}_{\bm{n}_0, m_{0}\to\bm{s}_i, m_{\bm{s}_i}} = \delta_{|m-m_0|>1}\frac{5\ell+b}{2\ell(b+2\ell)}+ \delta_{|m-m_0|,1}\left( \delta^{(m,m_0)} \frac{5\ell+b}{2\ell(b+2\ell) }+(1-\delta^{(m,m_0)})\frac{3}{b+2\ell}\right)\\ &+ 6 \sum_{r=0}^{R-1}\sum_{s=0}^{3r+2} \Bigg\{\bm{e}_{m}^{\intercal}\cdot\left(\left[\mathbb{I}-\bm{\lambda}^{(\mathcal{H})}\left(\frac{2\pi \kappa_1}{\Omega}, \frac{2\pi \kappa_2}{\Omega}\right)\right]^{-1} +
\left[\mathbb{I}-\bm{\lambda}^{(\mathcal{H})}\left(\frac{-2\pi \kappa_1}{\Omega}, \frac{-2\pi \kappa_2}{\Omega}\right)\right]^{-1}\right) \cdot \bm{e}_{m} \\
& - \bm{e}_{m}^{\intercal}\cdot\left(e^{\frac{2\pi i\bm{k}\cdot(\bm{s} -\bm{n}_{0})}{\Omega}}\left[\mathbb{I}-\bm{\lambda}^{(\mathcal{H})}\left(\frac{2\pi \kappa_1}{\Omega}, \frac{2\pi \kappa_2}{\Omega}\right)\right]^{-1} +
e^{\frac{-2\pi i\bm{\kappa}\cdot(\bm{s} -\bm{n}_{0})}{\Omega}}\left[\mathbb{I}-\bm{\lambda}^{(\mathcal{H})}\left(\frac{-2\pi \kappa_1}{\Omega}, \frac{-2\pi \kappa_2}{\Omega}\right)\right]^{-1}\right) \cdot \bm{e}_{m_0} \Bigg\},
\end{aligned}    
\label{eq: appsingle_tarMFPThex}
\end{equation}
where $\delta^{(m,m_0)} =\delta_{m \in \{2,3\}}\delta_{m_0 \in \{2,3\}}+\delta_{m \in \{5,6\}}\delta_{m_0 \in \{5,6\}}$, where 
we have lightened the notation from $\kappa_i(r,s)$ (defined just below Eq. (\ref{eq: hex_periodic}) in the main text) to $\kappa_i$.

We note here that the procedure can be carried out in arbitrary dimensions, however numerical procedures may be required
to find $\text{Adj}[\mathbb{I}-\bm{\lambda}(0, ..., 0)], \quad \left.\frac{\partial
\text{Adj}[\mathbb{I}-z\bm{\lambda}(0, ..., 0)]}{\partial z}\right|_{z=1} \; \; \text{and} \left.\frac{\partial
\det[\mathbb{I}-z\bm{\lambda}(0, ..., 0))}{\partial z}\right|_{z=1}$ in higher dimensions.

\section*{Details on the Monte Carlo Simulations}
Discrete space-time stochastic simulations are fairly straightforward, and the inclusion of an internal variable, which
keeps track of the short term memory of the system, requires only a marginal extension of standard computational
routines. To simulate the movement process, one must keep track of the internal variable, which is updated based on the
movement direction of the previous step. Depending on this internal variable, and the global movement variables ($f, b,
\ell, c$, etc), a new movement direction is chosen. The position of the walker is then updated and, if necessary, one
imposes a boundary condition. We note that when we have considered the `squeezed' boundary condition, the simulations are
independent of any periodic dynamics, and we simply impose the Chandrasekhar boundary condition described in the main
text.

Then, disregarding the internal variable, we check whether the walker is located at a target site $\bm{s}$, and if so,
the first passage event is recorded with some probability $\rho_{\bm{s}}$, where for the fully absorbing targets
$\rho_{\bm{s}} = 1$. As an exception, when one simulates the directional first-passage, the internal variable is
required to check whether a first-passage event has occurred, since one must know the direction the walker entered the
site from.
\end{document}